\documentclass[aps,twocolumn,amsmath,amssymb,superscriptaddress,prb]{revtex4-2}
\usepackage{xcolor}

\usepackage{physics}
\usepackage{graphicx}
\usepackage{dcolumn}
\usepackage{bm}

\usepackage{amsmath,amssymb}

\begin{document}


\title{Interplay between charge and spin  noise in the near-surface theory of  decoherence and relaxation of $C_{3v}$ symmetry qutrit spin-1  centers}

\author{Denis R. Candido}
 \email{denisricardocandido@gmail.com}
\affiliation{Department of Physics and Astronomy, University of Iowa, Iowa City, Iowa 52242,
USA}

\author{Michael E. Flatt\'e}%
 \email{michaelflatte@quantumsci.net}
\affiliation{Department of Physics and Astronomy, University of Iowa, Iowa City, Iowa 52242,
USA}
\affiliation{Department of Applied Physics, Eindhoven University of Technology, P.O. Box 513, 5600 MB, Eindhoven, The Netherlands}

\date{\today}

\begin{abstract}

Decoherence and relaxation of solid-state defect qutrits near a crystal surface, where they are commonly used as quantum sensors, originates from charge and magnetic field noise. A complete theory requires a  formalism for  decoherence and relaxation that includes all Hamiltonian terms allowed by the defect's point-group symmetry. This formalism, presented here for the $C_{3v}$ symmetry of a spin-1 defect in a diamond, silicon cardide, or similar host, relies on a Lindblad dynamical equation and clarifies the relative contributions of charge and spin noise to relaxation and decoherence, along with their dependence on the defect spin's depth and resonant frequencies. The calculations agree with the experimental measurements of Sangtawesin~\textit{et al.}, Phys. Rev. X \textbf{9}, 031052 (2019) and point to an unexpected importance of charge noise.


\end{abstract}

\maketitle


\section{Introduction}

Coupling of a spin-1 center in a solid, usually associated with a dopant or defect, to electric and magnetic fields provides a direct method of sensing nanoscale fields~\cite{taylor2008high,electric-magnetic1,dolde2014,schirhagl2014nitrogen,van2015nanometre,degen2017quantum,flebus2018quantum,casola2018probing,mittiga2018imaging,PhysRevX.10.011003,lee2020nanoscale,rustagi2020,candido2021theory}, of tuning the optical emission linewidth for optically-active defects~\cite{tamarat2006stark,anderson2019electrical,de2017stark}, and of coupling to electric or magnetic excitations to realize hybrid quantum coherent systems~\cite{lukaprx,li2015hybrid,li2016hybrid,andrich2017long,lemonde2018phonon,flebus2019entangling,muhlherr2019magnetic,zou2020tuning,candido2020predicted,neumanprl2020,doi:10.1021/acs.jpcc.0c11536,solanki,fukami2021}.
Conversely this coupling also makes the defect spin dynamics very susceptible to charge and magnetic noise,  contributing to  decoherence and  relaxation of the spin qubit states~\cite{electric-magnetic1,electric-magnetic2,electric-magnetic3,electric-magnetic4,electricnoise1,electricnoise2,electricnoise3,magneticnoise1,magneticnoise2,PhysRevB.87.235436,magneticnoise3,magneticnoise4,magneticnoise5,magneticnoise6,magneticnoise7,Kolkowitz1129}, and increasing the photoluminescence linewidth~\cite{tamarat2006stark,anderson2019electrical,de2017stark,candido-pin}.
Many approaches have been explored to diminish the effect of charge noise on defects, {\it e.g.,} controlling the termination of the diamond surface~\cite{electric-magnetic3}, embedding diamonds in materials with a high dielectric constant~\cite{electricnoise1}, covering the diamond surface with an extra layer~\cite{electricnoise2}, and placing the spin center in the depletion region of a \textit{p}--\textit{n} diode~\cite{anderson2019electrical,candido-pin}. 
Nevertheless, surfaces present a useful laboratory for the study of noise sources, as the nature of these fluctuations can be specific to the surface type and also to the depth below the surface. 
For spins acting as quantum sensors for nanoscale fields, the surface noise limits how near to the surface a spin can be placed while still retaining experimentally resolvable coherent dynamics, and thus the spatial resolution achievable with the sensor.
Thus a complete formalism for decoherence and relaxation will permit the surface properties to be optimized within practical parameters and will enable the optimal depth of a spin to be determined when it is acting as a quantum sensor for nanoscale fields.

\begin{figure}[t!]
\begin{center}
\includegraphics[clip=true,width=1\columnwidth]{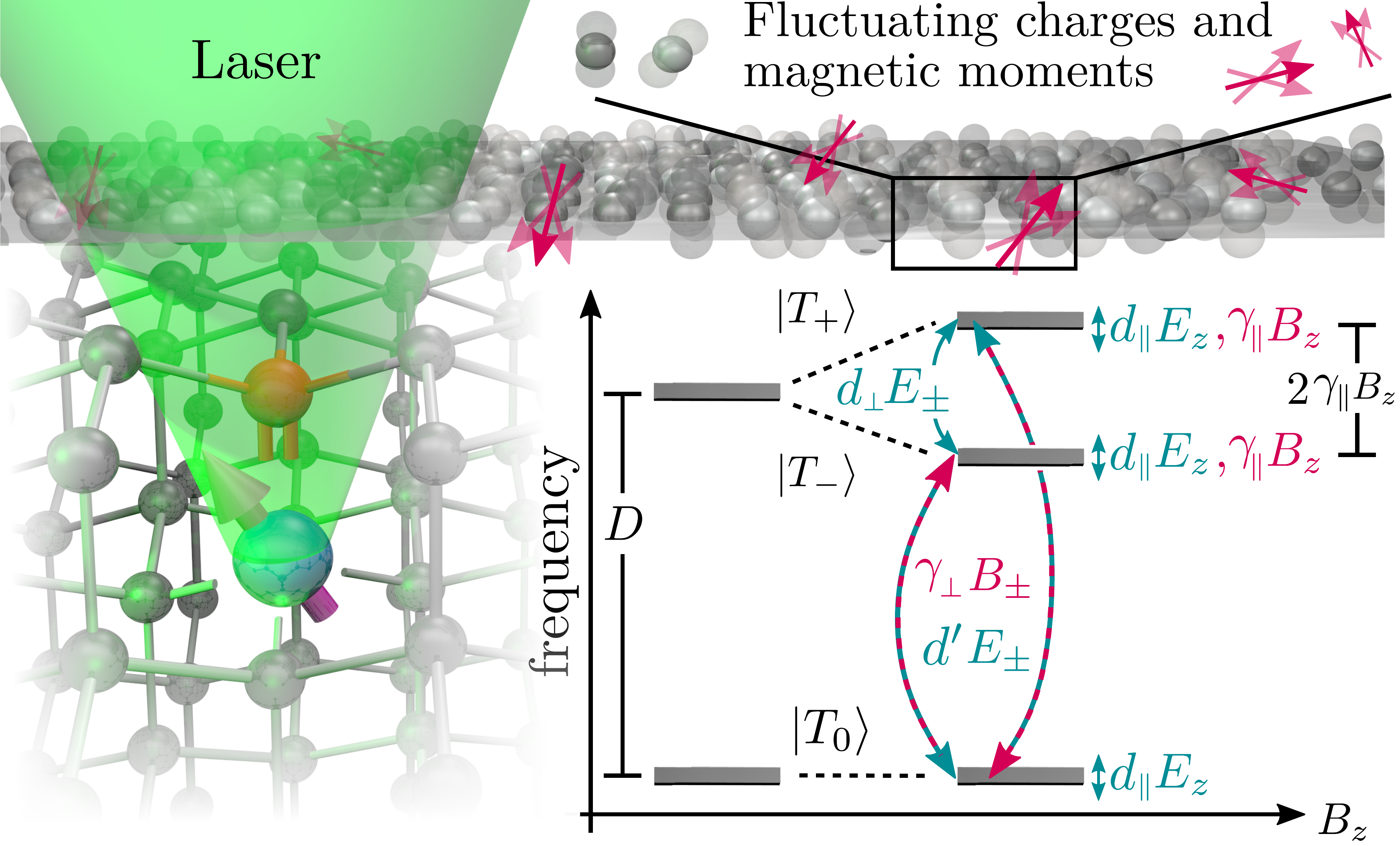}
\caption{Schematic plot of a spin-1 negatively-charged nitrogen-vacancy (NV$^{-}$) center within diamond in the presence of fluctuating surface charges (gray spheres) and magnetic moments (ruby arrows). (Lower right)  NV$^{-}$ center's spin levels and  response to the magnetic field. Dephasing and relaxation processes are indicated schematically  by double arrows associated with the electric dipole terms $d_{\perp}E_{\pm}$, $d_{\parallel}E_z$ and $d'E_{\pm}$ and the magnetic dipole terms $\gamma_{\perp}B_{\pm}$ and $\gamma_{\parallel}B_z$.} 
\label{fig1}
\end{center}
\end{figure}

In this work we provide a complete quantitative theory for the influence of the charge and magnetic noise on the dephasing and relaxation processes of the three states ($\ket{T_{+}}$,  $\ket{T_{0}}$, and $\ket{T_{-}}$, corresponding to the spin-1 projection along the symmetry axis) of  shallow spin-1 (qutrit) solid-state spin centers with $C_{3v}$ symmetry, embedded in hosts such as diamond and silicon carbide (Fig.~\ref{fig1}). Our work includes all electric ($d_{\perp}E_z$, $d_{\parallel}E_{\pm}$ and $d'E_{\pm}$) and magnetic ($\gamma_{\perp}B_{\pm}$ and $\gamma_{\parallel}B_{\pm}$) terms allowed by symmetry (See Fig.~\ref{fig1}). We derive a Lindblad dynamical equation~\cite{lindblad1976} for our qutrit containing eight different Lindblad operators, which captures the resulting  dephasing and relaxation processes. We then  calculate  the population and  dephasing dynamics of the spin, and describe several different regimes and scenarios for these processes, suggesting  improved conditions for the experimental utilization of qutrits. For example, the three-state character of our spin-1 qutrit can improve the ability to probe ``flat'' (frequency-independent) regions of the spectral noise density.  We also show that the relaxation between the spin-defect states $\ket{T_{\pm}}$ and $\ket{T_{0}}$, usually attributed to  magnetic noise (via the $\gamma_{\perp}B_{\pm}$ term),  also has a significant contribution from the charge noise via the commonly-ignored dipole term $d'E_{\pm}$.

We then apply this theory to study the surface noise arising from the fluctuations of charges and magnetic moments on the diamond surface. For both hydrogen (H) and oxygen (O) terminated diamond, the bonds between  carbon atoms of the diamond and these other atoms effectively create either acceptor (hydrogen) or donor (oxygen) levels at the diamond surface~\cite{PhysRevB.73.085313,sussmann2009cvd}. These acceptor (donor) levels are occupied by the electrons (holes) provided by nitrogen dopants (so-called P$_1$ centers) in the diamonds which also  contain negatively-charged nitrogen-vacancy (NV$^{-}$) centers. The electrostatic effects of this charge transfer bends the host bands  and  creates an effective, very low mobility, surface two-dimensional (2D) hole (electron) gas~\cite{KAWARADA1996205,PhysRevLett.85.3472,PhysRevB.68.041304,sussmann2009cvd,surfacediamondreview,https://doi.org/10.1002/admi.201801449,PhysRevB.105.205304,candido2021theory}. Similar effects also emerge from imperfections in the crystal termination \cite{https://doi.org/10.1002/admi.201801449}. The hopping motion of  trapped electrons (holes) and the charge motion within the confined 2D surface hole (electron) gas produce fluctuating electric and magnetic noise that influences our shallow defects, causing relaxation and decoherence of the spin center's quantum state.

The distinct character of the sources of charge and their fluctuations require different theoretical descriptions of their effect on the defect spin. The trapped charges can be modeled as electric dipole fluctuators~\cite{electricnoise2,magneticnoise6}, whereas the confined surface charged gas should be treated as the fluctuation of point-like charges~\cite{electric-magnetic1,electric-magnetic2,electric-magnetic3,magneticnoise6,candido2021theory} (with charge neutrality maintained by the fixed charges of donors). We derive analytical formulas for both types of fluctuating electric fields as a function of the areal density of  fluctuating dipoles or fluctuating charge densities, of dipole length and  defect depth. We  analyze the competition between these two sources of charge noise, and compare them with their bulk noise counterparts~\cite{candido-pin,electricnoise3}. 
For completeness, we include as well the magnetic noise produced by both the fluctuations of the spins' magnetic moments and the movement of charged particles (Biot-Savart law) ~\cite{magneticnoise1,magneticnoise2,magneticnoise3,PhysRevB.87.235436,magneticnoise4,magneticnoise5,electricnoise3}. 
We also identify the scenarios for which the magnetic noise dominates. Our quantitative theory for both magnetic and charge noise enables the study and analysis of the competition between electric (charge) and magnetic noise in different scenarios and environmental conditions.

Finally, combining our quantitative theory for the surface charge and magnetic noise with our complete formalism for relaxation and dephasing of a spin-defect with $C_{3v}$ point-group symmetry, we report calculations of the decoherence and relaxation of the spin center as a function of the surface charge density, the defect depth and the frequency separation between the spin center's energy levels. The dependence of the decoherence and relaxation on the spin center's energy levels will thus allow us to identify the dominant source of noise if these quantities are studied as a function of magnetic field. Our results show good agreement with experimental reports of the dependence of the decoherence time on the defect depth~\cite{electric-magnetic3}. Thus we propose such studies will enable the dominant noise sources to be assigned for various  surface treatments.

Section~\ref{defect-sec} presents the ground state Hamiltonian for a spin-1 defect with $C_{3v}$ point group symmetry, along with its coupling to  external electric and magnetic fields. Assuming that the magnetic and charge noise will manifest as classical electric and magnetic fields, we derive the Lindblad operators followed by the Lindblad dynamical equation for this Hamiltonian. The general expressions produced for the spin population dynamics yield the different  relaxation times and   decoherence times associated with the loss of information among different spin-1 subspaces. Section~\ref{secIII} focuses on the specific case of point-like and dipole charge noise, and explores the competition between these two sources. Different sources of magnetic noise are also calculated, and their contribution compared to that of charge noise. From this we clarify the charge and magnetic noise dependence of the relaxation and decoherence rates on the frequency separation between  the spin center's levels.  Section~\ref{seciv} compares our theoretical findings with  experimental results for the decoherence of shallow NV$^{-}$ centers.

\section{Spin-1 (qutrit) decoherence for $C_{3v}$ point group symmetry}
\label{defect-sec}

Here we establish the general features required for a calculation of the decoherence and relaxation of the quantum state of a spin-1 qutrit  due to electric and magnetic noise. We first present the complete Hamiltonian for qutrits with $C_{3v}$ point group symmetry in the presence of electric and magnetic fields. We then find eight Lindblad operators that produce decoherence and relaxation of the three states of the qutrit. The dynamics of the spin in the presence of fluctuating electric and magnetic fields are obtained from a   Lindblad~\cite{lindblad1976} dynamical equation. Finally, we identify new features of the population dynamics of our spin-1 qutrit, including relaxation rates which have been previously interpreted as due to magnetic noise dominance, but which may be due to electric noise.

\subsection{Spin center Hamiltonian}

Semiconductor spin-1 centers with $C_{3v}$ point group symmetry~\cite{loubser1978electron,van1990electric,PhysRevB.53.13441,tamarat2006stark,hossain2008ab,de2010universal,togan2010quantum,bassett2011electrical,maze2011,electric-magnetic1,doherty2011negatively,acosta2012dynamic,doherty2012,doherty2013nitrogen,dolde2014,schirhagl2014nitrogen,dolde2014nanoscale,rogers2015singlet,ivady2015theoretical,seo2016} do not possess inversion symmetry and therefore permit  linear  coupling of the spin's energy levels to an electric field (Stark effect\cite{tamarat2006stark,electric-magnetic1,bassett2011electrical,acosta2012dynamic,klimov2014electrically,christle2017,de2017stark,miao2019electrically,anderson2019electrical}) and to strain. The spin is also coupled to a magnetic field via the Zeeman effect with an anisotropic  gyromagnetic ratio. The  ground state (GS) Hamiltonian, with all these terms, in the triplet basis $\left|T_{-}\right\rangle ,\left|T_{0}\right\rangle ,\left|T_{+}\right\rangle$ (where $+$, $0$ and $-$ are defined along the symmetry axis), for spin-1 centers with a $C_{3v}$ point group symmetry is, from a group theory analysis~\cite{PhysRevB.5.803,van1990electric,maze2011,doherty2012,PhysRevB.98.075201}
\begin{align} 
\frac{{\cal H}}{h} & = \textbf{B}\cdot \gamma \cdot \textbf{S}+\left(D +d_{\parallel}E_{z}  \right) \left(S_{z}^{2}-\frac{2}{3}\right)+d_{\perp} E_{x} \left(S_{y}^{2}-S_{x}^{2}\right) \nonumber \\
+ & d_{\perp} E_{y}\left\{S_x,S_y \right\}+ d'E_{x}\left\{S_x,S_z \right\} +d'E_{y}\left\{S_y,S_z \right\} 
 \label{hgs},
\end{align}
where $h$ is  Planck's constant, $\gamma$ is the gyromagnetic ratio tensor, $\textbf{S}$ are the triplet spin-1 matrices, 
 ${\textbf{E}=(E_x,E_y,E_z)}$ is the electric field, $\textbf{B}=(B_x,B_y,B_z)$ is the magnetic field, $\{ A,B\}=AB+BA$, $D$ is the zero energy splitting between the triplet states $\left|T_{0}\right\rangle$ and $\left|T_{\pm}\right\rangle$, and $d_{\parallel}$, $d_{\perp}$ and $d'$ are electric dipole constants. The $z$ direction here corresponds to the defect symmetry axis. 

The temporal fluctuations of  magnetic and electric fields generate decoherence and relaxation of the quantum state of the spin. To understand the role of  individual terms within the corresponding decoherence and relaxation processes, we rewrite the Hamiltonian in matrix form, 
\begin{widetext}
\begin{equation}
\frac{{\cal H}}{h} = \left(\begin{array}{ccc}
\frac{D}{3}+\frac{d_{\parallel}}{3}E_{z}+\gamma_{\parallel} B_z & \frac{d'}{\sqrt{2}}E_{-}+\frac{\gamma_{\perp}}{\sqrt{2}}{B_{-}} & -d_{\perp}E_{+}\\
\frac{d'}{\sqrt{2}}E_{+}+\frac{\gamma_{\perp}}{\sqrt{2}}B_{+} & -\frac{2D}{3}-\frac{2d_{\parallel}}{3}E_{z} & -\frac{d'}{\sqrt{2}}E_{-} +\frac{\gamma_{\perp}}{\sqrt{2}}B_{-}\\
-d_{\perp}E_{-}  & -\frac{d'}{\sqrt{2}}E_{+} +\frac{\gamma_{\perp}}{\sqrt{2}}B_{+} & \frac{D}{3}+\frac{d_{\parallel}}{3}E_{z} -\gamma_{\parallel} B_z
\end{array}\right),   
\label{H-matrix}
\end{equation}
\end{widetext}
where $E_{\pm}=E_x\pm iE_y$ and $B_{\pm}=B_{x}\pm i B_{y}$. From the Hamiltonian, Eq.~(\ref{H-matrix}), the magnetic field produces a frequency split  $\propto \gamma_{\parallel}B_z$ between the $\left|T_{\pm}\right\rangle$ states, in addition to a coupling $\propto \gamma_{\perp}B_{\pm}$ between the $\left|T_{0}\right\rangle$ and $\left|T_{\pm}\right\rangle$ states. Similarly, the electric field yields a frequency splitting $\propto d_{\parallel}E_z$ between $\left|T_{\pm}\right\rangle$ and $\left|T_{0}\right\rangle$, in addition to a coupling $\propto d' E_{\pm}$ between the $\left|T_{0}\right\rangle$ and $\left|T_{\pm}\right\rangle$ states. However, unlike  the magnetic field, the electric field also couples the $\left|T_{-}\right\rangle$ and $\left|T_{+}\right\rangle$ subspaces, with a strength proportional to $d_{\perp}E_{\pm}$. These different terms appear  schematically in Fig.~\ref{fig1}. As relaxation processes ($1/T_1$) occur  when different levels are coupled to each other through random temporal fluctuations, the $d'$, $d_{\perp}$ and $\gamma_{\perp}$ terms will contribute to relaxation processes. Conversely, the dephasing processes can also occur due to any terms responsible for relative fluctuations of the energy of the levels, namely $d_{\parallel}$ and $\gamma_{\parallel}$ in addition to  $d'$, $d_{\perp}$ and $\gamma_{\perp}$.

We stress that although prior work has neglected the presence of the $d'$ electric dipole terms within the spin center's Hamiltonian, these are important when  charge noise dominates. They are also important
 to characterize correctly processes involving photoluminescence and spin dynamics near the level anticrossing of the electronic ground state (GSLAC)~\cite{PhysRevLett.108.197601,wang2013sensitive,PhysRevLett.114.247603,doi:10.1063/1.4960171,PhysRevApplied.6.064001,https://doi.org/10.1002/pssb.201700258,PhysRevB.100.075204,PhysRevB.102.224101,PhysRevApplied.13.044023,PhysRevB.103.035307}, as well as for acoustical driving experiments of the $\left| T_{0}\right\rangle  \leftrightarrow \left| T_{\pm}\right\rangle$ spin transition~\cite{PhysRevB.98.075201,PhysRevApplied.13.054068}. Moreover, although up to this point there is no precise experimental verification for the value of $d'$,  Ref.~\cite{PhysRevApplied.13.054068} suggests $d'/d_{\perp} \approx \sqrt{2}/2$, whereas an {\it ab-initio} calculation finds $d'\approx d_{\perp}$~\cite{doherty2012}.

\subsection{Lindblad formaism for $C_{3v}$ spin-1 (qutrit) decoherence and relaxation}

To obtain the Lindblad dynamical equation describing decoherence and relaxation of our qutrit we begin by considering that the only nonfluctuating external field is the dc magnetic field $B_z$, which controls the frequency separation between $\left|T_{+}\right\rangle$ and $\left|T_{-}\right\rangle$ subspaces [See Fig.~\ref{fig1}]. The other fields can fluctuate, so we rewrite our Hamiltonian, Eq.~(\ref{H-matrix}), as the sum of a time independent part and a time-dependent one, i.e., ${\cal H}= {\cal H}_0+{\cal V}(t)$ with 
${\cal H}_0/h= (\gamma_{\parallel}/2\pi) B_z S_z + D (S_{z}^2-2/3)$, and ${\cal V}(t)$ produced by the remaining terms of Eq.~(\ref{H-matrix}). 
In the absence of these (weak, relative to the value of $D$) fluctuating fields the spin center's frequencies are $\omega_0=0$ and $\omega_\pm/2\pi=D\pm\gamma_{\parallel} B_z$.

In order to solve the dynamics of the spin center, we move to the interaction picture with respect to ${\cal H}_0$, which leads to
\begin{equation}
    {\cal V}_{\rm{I}}\left(t\right)\left|\psi\left(t\right)\right\rangle _{\rm{I}}=i\hbar\frac{\partial}{\partial t}\left|\psi\left(t\right)\right\rangle _{\rm{I}},
    \label{int-eq}
\end{equation}
where $\left|\psi\left(t\right)\right\rangle _{\rm{I}}\equiv e^{i\frac{{\cal H}_{0}}{\hbar}t}\left|\psi\left(t\right)\right\rangle  $ and 
\begin{equation}
{{\cal V}_{\rm{I}}(t)=e^{i \frac{{\cal H}_0}{\hbar}t} {\cal V}(t) e^{-i \frac{{\cal H}_0}{\hbar}t}}.
\end{equation}
 The evolution of the density matrix in the interaction picture associated with Eq.~(\ref{int-eq}) is 
\begin{equation}
\hat{\rho}_{\rm{I}}\left(t\right)  ={\cal T}e^{-i\int_{t_{0}}^{t}d\tau\thinspace {\cal V}_{\rm{I}}\left(\tau\right)/\hbar}\hat{\rho}_{\rm{I}}\left(t_{0}\right){\cal T}e^{i\int_{t_{0}}^{t}d\tau\thinspace {\cal V}_{\rm{I}}\left(\tau\right)/\hbar},\label{rhoform}
\end{equation}
where ${\cal T}$ is the time-ordering operator. The solution to Eq.~(\ref{rhoform}) can be obtained through a perturbative (Dyson) expansion of the propagator $\Pi(t,t_0)$, namely,
\begin{equation}
\hat{\rho}_{\rm{I}}\left(t\right)=\Pi\left(t,t_{0}\right)\hat{\rho}_{\rm{I}}\left(t_{0}\right),
\end{equation}
with $\Pi=\Pi_{0}+\Pi\Sigma\Pi_{0}$ where $\Sigma$ is the self-energy. This yields the general dynamical equation for $\hat{\rho}_{\rm{I}}(t)$~\cite{MAKHLIN2004315,doi:10.1063/1.2162537}
\begin{equation}
    \frac{d\hat{\rho}_{\rm{I}}\left(t\right)}{dt}=\frac{1}{i\hbar}\left[{\cal V}_{I}(t),\hat{\rho}_{\rm{I}}\left(t\right)\right]+\int_{t_0}^{t}d\tau\Sigma\left(t-\tau\right)\hat{\rho}_{\rm{I}}\left(\tau\right).
\end{equation}
Mapping this equation onto a Lindblad  equation~\cite{lindblad1976} requires some  assumptions and approximations. 
The  equation is expanded in a diagrammatic perturbation series  to second order, and  the average is taken over different realizations, namely, $\langle \cdots \rangle$. We further assume the  correlation time of the noise, $\tau_c$, is much smaller than the time interval $t-t_0$, {\it i.e.,} $t-t_0\gg \tau_c$. A further Markovian approximation yields the result 
\begin{equation}
    \frac{d\langle\hat{\rho}_{\rm{I}} \left(t\right)\rangle }{dt} = -\frac{1}{\hbar^2} \int_{0}^{\infty} d\tau \langle \left[{\cal V}_{I}\left(\tau\right) , \left[{\cal V}_{I}\left(0\right),\hat{\rho}_{\rm{I}}\left(t\right)\right]\right]\rangle.
\end{equation}
The identity 
\begin{align}
\left[{\cal A},\left[{\cal B},{\cal C}\right]\right] &= \frac{1}{2}\left[\left[{\cal A},{\cal B}\right],{\cal C}\right]\\\nonumber
&-\left( {\cal A}{\cal C}{\cal B}+{\cal B}{\cal C}{\cal A}-\frac{1}{2}\left\{{\cal A}{\cal B},{\cal C}\right\}-\frac{1}{2}\left\{{\cal B}{\cal A},{\cal C}\right\}\right)
\end{align}
 is of great use; the first right hand side term  produces new contributions to the coherent evolution {\it e.g.,} Stark shifts and Lamb shifts, however the remaining ones produce the terms associated with Lindblad operators. Defining ${{\cal H}_{{\rm eff},\rm{I}}=(i/2\hbar)\int_{0}^{\infty}d\tau\left\langle \left[{\cal V}_{\rm{I}}\left(\tau\right),{\cal V}_{I}\left(0\right)\right]\right\rangle }$, we obtain
{\small {} \begin{align}
\frac{d\langle\hat{\rho}_{\rm{I}}\left(t\right)\rangle}{dt} & =\frac{1}{i\hbar}\left[{\cal H}_{{\rm eff},\rm{I}},\langle \hat{\rho}_{\rm{I}} \left(t\right)\rangle\right] \nonumber \\
- & \frac{1}{\hbar^{2}}\int_{0}^{\infty}d\tau\left[\left\langle {\cal V}_{\rm{I}}\left(\tau\right)\hat{\rho}_{\rm{I}}\left(t\right){\cal V}_{\rm{I}}\left(0\right)\right\rangle +\left\langle {\cal V}_{\rm{I}}\left(0\right)\hat{\rho}_{\rm{I}}\left(t\right){\cal V}_{\rm{I}}\left(\tau\right)\right\rangle \right]\nonumber \\ 
+ & \frac{1}{2\hbar^{2}}\int_{0}^{\infty}d\tau\left\langle \left\{ {\cal V}_{\rm{I}}\left(\tau\right){\cal V}_{\rm{I}}\left(0\right),\hat{\rho}_{\rm{I}}\left(t\right)\right\} \right\rangle \nonumber \\
+ & \frac{1}{2\hbar^{2}}\int_{0}^{\infty}d\tau\left\langle \left\{ {\cal V}_{\rm{I}}\left(0\right){\cal V}_{\rm{I}}\left(\tau\right),\hat{\rho}_{\rm{I}}\left(t\right)\right\} \right\rangle. \label{drhodt-int}
\end{align}}
The rotating wave approximation with respect to the frequency separations between the spin center's energy levels, $\omega_{\mu\nu}=\omega_{\mu}-\omega_{\nu}$ with $\mu,\nu=\{0,\pm \}$ simplifies Eq.~(\ref{drhodt-int}). This approximation cannot be employed if two of the states become degenerate, and hence the results are only valid for lifted degeneracies.
We use Novikov's theorem~\cite{novikov1965functionals,PhysRevA.64.052110,PhysRevA.63.012106,costaquantum,PhysRevA.95.052126} together with the weak coupling between our spin center states and the fluctuating fields~\cite{PhysRevA.64.052110,PhysRevA.63.012106,costaquantum,PhysRevA.95.052126}.  We further assume temporal
translational symmetry for the fluctuating fields (stationary regime), $\left\langle E_{i}\left(t\right)E_{j}\left(t'\right)\right\rangle =\left\langle E_{j}\left(t-t'\right)E_{j}\left(0\right)\right\rangle $ and $\left\langle B_{i}\left(t\right)B_{j}\left(t'\right)\right\rangle =\left\langle B_{j}\left(t-t'\right)B_{j}\left(0\right)\right\rangle $, 
and also $\left\langle E_{i}\left(\tau\right)E_{j}\left(0\right)\right\rangle =\delta_{ij} f\left(\tau\right)$ with $\left\langle B_{i}\left(\tau\right)B_{j}\left(0\right)\right\rangle = \delta_{ij}\bar{f}\left(\tau\right)$
for $i=x,y,z$, which follows for fluctuating fields lacking a preferential direction.
The corresponding noise spectral densities are
\begin{align}
    S_{{E}_i}\left(\omega\right)&=\int_{-\infty}^{\infty}d\tau\left\langle E_{i}\left(\tau\right)E_{i}\left(0\right)\right\rangle e^{i\omega\tau},\label{specE} \\
    S_{{B}_i}\left(\omega\right)&=\int_{-\infty}^{\infty}d\tau\left\langle B_{i}\left(\tau\right)B_{i}\left(0\right)\right\rangle e^{i\omega\tau}, \label{specB}
\end{align}
where due to the classical character of our fluctuating fields we have $S_{{E(B)}}\left(\omega\right)=S_{{E(B)}}\left(-\omega\right)$. This holds for $\hbar \omega \ll k_B T$ since $S_{{E(B)}}(\omega)/S_{{E(B)}}(-\omega)=e^{\frac{\hbar \omega}{k_B T}}$. For the NV$^{-}$  the largest frequency split is  $\approx 2.5$~GHz, so the approximation holds for $T\gtrsim 1$~K. Another consequence of the frequency-symmetric noise spectral density is the absence of an effective coherent Hamiltonian arising from the noise, i.e., no Stark nor Lamb shift, so ${\cal H}_{\rm{eff,I}}=0$. Three rates are usefully associated with the charge noise spectral density, namely 
\begin{align}
\Gamma_{d_{\perp}}\left(\omega\right)&={\tilde{d}_{\perp}^{2}}\left[S_{{E_x}}\left(\omega\right)+S_{{ E}_y}\left(\omega\right)\right], \label{Gammadperp} \\
\Gamma_{d'}\left(\omega\right) & ={\tilde{d}'^{2}}\left[S_{{E_x}}\left(\omega\right)+S_{{ E}_y}\left(\omega\right)\right],\label{Gammadprime} \\
\Gamma_{d_{\parallel}}\left(\omega\right)&={\tilde{d}_{\parallel}^{2}}S_{{ E_z}}\left(\omega\right), \label{Gammadpara}
\end{align} 
with ${d}_{\parallel} =\tilde{d}_{\parallel}/2\pi$, ${d}_{\perp}=\tilde{d}_{\perp}/ 2\pi$, ${d}'=\tilde{d}'/2\pi$. Two rates are correspondingly associated with the magnetic noise spectral densities,
\begin{align}
\Gamma_{\gamma_{\perp}}\left(\omega\right)&={\tilde{\gamma}_{\perp}^{2}}\left[S_{{ B_x}}\left(\omega\right)+S_{{ B}_y}\left(\omega\right)\right], \label{gammaperp} \\ \Gamma_{\gamma_{\parallel}}\left(\omega\right)&={\tilde{\gamma}_{\parallel}^{2}}S_{{ B_z}}\left(\omega\right) \label{gammapara},  
\end{align} 
with $\gamma_{\perp}=\tilde{\gamma}_{\perp}/2\pi$ and $\gamma_{\parallel}=\tilde{\gamma}_{\parallel}/2\pi$.
Finally, all the considerations above yield the Lindblad dynamical equation
\begin{equation}
\frac{d\langle\hat{\rho}_{\rm{I}}\left(t\right)\rangle}{dt} =  \sum_{k=1}^{8}\left[ {L}_{k,\rm{I}}\langle\hat{\rho}_{\rm{I}}\left(t\right)\rangle{L}_{k,\rm{I}}^{\dagger}-\frac{1}{2}\left\{{L}_{k,\rm{I}}^{\dagger}{L}_{k,\rm{I}},\langle\hat{\rho}_{\rm{I}}\left(t\right)\rangle\right\}\right], \label{lindbladeq}
\end{equation}
with Lindblad operators in the interaction picture, $L_{k,\rm{I}}$, given by 
\begin{align}
L_{1,\rm{I}} & =\sqrt{\Gamma_{d_{\parallel}}\left(0\right)}\left(S_{z}^{2}-2/3\right), \label{L1}\\
L_{2,\rm{I}} & =\frac{1}{2}\sqrt{\Gamma_{d_{\perp}}\left(\omega_{+-}\right)}S_{+}^2,  \label{L2}\\
L_{3,\rm{I}} & =\frac{1}{2}\sqrt{\Gamma_{d_{\perp}}\left(\omega_{+-}\right)} S_{-}^2, \label{L3}  \\
L_{4,\rm{I}} &= \frac{1}{2}\sqrt{{\Gamma_{d'}\left(\omega_{+0}\right)+\Gamma_{\gamma_{\perp}}\left(\omega_{+0}\right)}}\thinspace {S_z S_+}, \label{L4} \\
L_{5,\rm{I}} &= \frac{1}{2}\sqrt{{\Gamma_{d'}\left(\omega_{+0}\right)+\Gamma_{\gamma_{\perp}}\left(\omega_{+0}\right)}}\thinspace {S_- S_z} ,\label{L5} \\
L_{6,\rm{I}} &= \frac{1}{2}\sqrt{{\Gamma_{d'}\left(\omega_{-0}\right)+\Gamma_{\gamma_{\perp}}\left(\omega_{-0}\right)}}\thinspace {S_+ S_z}, \label{L6} \\
L_{7,\rm{I}} &= \frac{1}{2}\sqrt{{\Gamma_{d'}\left(\omega_{-0}\right)+\Gamma_{\gamma_{\perp}}\left(\omega_{-0}\right)}}\thinspace {S_z S_-}, \label{L7} \\
L_{8,\rm{I}} &= \sqrt{\Gamma_{\gamma_{\parallel}}\left(0\right)} \thinspace S_z. \label{L8}
\end{align}
Here the operator $S_{z}S_+$ ($S_{-}S_z$) represents the raising (lowering) operator within the subspace spanned by $\{ \ket{T_+},\ket{T_0} \}$, while $S_{+}S_z$ ($S_{z}S_-$) represents the raising (lowering) operator within the subspace spanned by $\{\ket{T_0},\ket{T_-}\}$. Additionally, the operator $S_{+}^2$ ($S_{-}^{2}$) is the raising (lowering) operator within the subspaced spanned by $\{ \ket{T_+}, \ket{T_-}\}$.

Using the Lindblad operators [Eqs.~(\ref{L1})--(\ref{L8})] within the Lindblad equation [Eq.~(\ref{lindbladeq})], we can also obtain the following differential equation that governs the dynamics of the density matrix $\langle\hat{\rho}_{\rm{I}}\left(t\right)\rangle _{\mu \nu}= \rho_{\mu \nu}(t)$, namely,
\begin{equation}
\frac{d}{dt}\left[\begin{array}{c}
\vdots\\
\rho_{\mu\nu}(t)\\
\vdots
\end{array}\right]_{1\times9}={\cal L}_{9\times9}\left[\begin{array}{c}
\vdots\\
\rho_{\mu\nu}(t)\\
\vdots
\end{array}\right]_{1\times9},
\end{equation}
with $\mu,\nu = \{0,\pm \}$ and the corresponding Lindbladian or Liouvillian matrix, ${\cal L}_{9\times 9}$. For our case, ${\cal L}_{9\times 9}$ is composed of a $3\times 3$ block diagonal matrix that governs the relaxation process of our quantum states, and a diagonal $6\times6$ matrix governing the dephasings between different subspaces. Both processes will be investigated in the next two subsections.

\subsubsection{Spin center relaxation }

The part of the Lindbladian governing the relaxation process is described by the evolution of the diagonal elements of $\langle\hat{\rho}_{\rm{I}}\left(t\right)\rangle$, namely
\begin{widetext}
\begin{equation}
    \frac{d}{dt}\left[\begin{array}{c}
\rho_{++}(t)\\
\rho_{00}(t)\\
\rho_{--}(t)
\end{array}\right]=\left[\begin{array}{ccc}
-\frac{1}{2}\Gamma_{\gamma d'}\left(\omega_{+0}\right)-\Gamma_{d_{\perp}}\left(\omega_{+-}\right) & \frac{1}{2}\Gamma_{\gamma d'}\left(\omega_{+0}\right) & \Gamma_{d_{\perp}}\left(\omega_{+-}\right)\\
\frac{1}{2}\Gamma_{\gamma d'}\left(\omega_{+0}\right) & -\frac{1}{2}\Gamma_{\gamma d'}\left(\omega_{+0}\right)-\frac{1}{2}\Gamma_{\gamma d'}\left(\omega_{-0}\right) & \frac{1}{2}\Gamma_{\gamma d'}\left(\omega_{-0}\right)\\
\Gamma_{d_{\perp}}\left(\omega_{+-}\right) & \frac{1}{2}\Gamma_{\gamma d'}\left(\omega_{-0}\right) & -\frac{1}{2}\Gamma_{\gamma d'}\left(\omega_{-0}\right)-\Gamma_{d_{\perp}}\left(\omega_{+-}\right)
\end{array}\right]\left[\begin{array}{c}
\rho_{++}(t)\\
\rho_{00}(t)\\
\rho_{--}(t)
\end{array}\right], \label{matrix-T1}
\end{equation}
\end{widetext}
where we define $\Gamma_{\gamma d'}(\omega)=\Gamma_{d'}(\omega)+\Gamma_{\gamma_{\perp}}(\omega)$. The solution of this equation is obtained through the ansatz 
\begin{equation}
    \left[\begin{array}{c}
\rho{}_{++}\left(t\right)\\
\rho{}_{00}\left(t\right)\\
\rho{}_{--}\left(t\right)
\end{array}\right]=\sum_{i=1}^{3}c_{i}\left[\begin{array}{c}
a_{i}^{+}\\
a_{i}^{0}\\
a_{i}^{-}
\end{array}\right]e^{\lambda_i t}.
\end{equation}
Here, $\lambda_i$ are the three eigenvalues of the $3\times 3$ matrix within Eq.~(\ref{matrix-T1}), and $\left[\begin{array}{ccc}
a_{i}^{+} & a_{i}^{0} & a_{i}^{-}\end{array}\right]^{T}$ are the corresponding eigenvectors. In principle, the three eigenvalues ($\lambda_i$) define three different relaxation rates, namely, $T_{1}^i=-\lambda_i$, which are
\begin{align}
1/T_{1}^+& =\gamma+\Omega_{-}+\Omega_{+}\label{t1p}\\
&+\sqrt{\gamma^{2}+\gamma\left(\Omega_{+}+\Omega_{-}\right)-\Omega_{+}\Omega_{-}+\Omega_{+}^{2}+\Omega_{-}^{2}},\nonumber\\
1/T_{1}^- & =\gamma+\Omega_{-}+\Omega_{+}\label{t1m} \\
&-\sqrt{\gamma^{2}-\gamma\left(\Omega_{+}+\Omega_{-}\right)-\Omega_{+}\Omega_{-}+\Omega_{+}^{2}+\Omega_{-}^{2}},\nonumber\\
1/T_{1}^0 & =0,
\end{align}
with $\Omega_{\pm}=({1}/{2})\Gamma_{\gamma d'}\left(\omega_{\pm0}\right)$
and $\gamma=\Gamma_{d_{\perp}}\left(\omega_{+-}\right)$, defined similarly to Refs.~\cite{PhysRevB.87.235436,magneticnoise6}. Accordingly, the general solution for the evolution of the diagonal density matrix elements is
\begin{widetext}
\begin{equation}
\left[\begin{array}{c}
\rho{}_{++} (t)\\
\rho{}_{00} (t)\\
\rho{}_{--} (t)
\end{array}\right]=c_1\left[\begin{array}{c}
1\\
1\\
1
\end{array}\right]+c_{2}\left[\begin{array}{c}
\frac{\gamma+2\Omega_{-}-1/T_{1}^{+}}{\gamma-\Omega_{-}}\\
\frac{2\Omega_{+}+\Omega_{-}-1/T_{1}^{-}}{\gamma-\Omega_{-}}\\
1
\end{array}\right]e^{-t/T_{1}^{+}}+c_{3}\left[\begin{array}{c}
\frac{\gamma+2\Omega_{-}-1/T_{1}^{-}}{\gamma-\Omega_{-}}\\
\frac{2\Omega_{+}+\Omega_{-}-1/T_{1}^{+}}{\gamma-\Omega_{-}}\\
1
\end{array}\right]e^{-t/T_{1}^{-}}.
\end{equation}
\end{widetext}
Using ${\rm Tr}[\langle\hat{\rho}_{\rm{I}}\left(t\right)\rangle]=1$ we obtain $c_1 = 1/3$, and the remaining coefficients $c_{2,3}$ are determined by the initial condition of the density matrix. For the initial condition ${\rho_{--}(t=0)=1}$, 
\begin{equation}
c_2=\frac{\gamma+\Omega_{-}-2/3T_{1}^{-}}{1/T_{1}^{+}-1/T_{1}^{-}}
\end{equation}
 and 
 \begin{equation}
 c_3=-\frac{\gamma+\Omega_{-}-2/3T_{1}^{+}}{1/T_{1}^{+}-1/T_{1}^{-}}.
 \end{equation}
  Interestingly, and differently from the spin-$1/2$ case, here we obtain a bi-exponential relaxation. This bi-exponential behavior was already discussed and presented in the literature for spin center dephasing~\cite{candido-pin} and relaxation processes~\cite{PhysRevB.87.235436,magneticnoise6}.

Now we analyze some cases of particular interest. The first one corresponds to $\omega_{+0}\approx\omega_{-0}\equiv \omega_{\pm0}$, which produces $\Gamma_{\gamma d'}(\omega_{+0})\approx\Gamma_{\gamma d'}(\omega_{-0})$. Accordingly, this leads to $\Omega_+=\Omega_-\equiv \Omega$, which was already discussed in the literature \cite{magneticnoise6} and produces the relaxation times
\begin{align}
\frac{1}{{T}_{1,eq}^+} &=2\gamma + \Omega = 2\Gamma_{d_{\perp}}\left(\omega_{+-}\right) + \frac{1}{2}\left[\Gamma_{\gamma_{\perp}}\left(\omega_{\pm0}\right) +\Gamma_{d'}\left(\omega_{\pm0}\right) \right],\\ \frac{1}{T_{1,eq}^{-}}&=3\Omega = \frac{3}{2}\left[\Gamma_{\gamma_{\perp}}\left(\omega_{\pm0}\right) +\Gamma_{d'}\left(\omega_{\pm0}\right) \right].
\end{align}
Importantly and contrary to \textcite{PhysRevB.87.235436,magneticnoise6}, we can see that when we consider all symmetry-allowed terms within our Hamiltonian Eq.~(\ref{H-matrix}), the charge noise also contributes to the $\Omega$ rate. Therefore, this rate does have an exclusive magnetic noise origin. Associating $\Omega$ to the magnetic noise is only accurate for $\Gamma_{\gamma_{\perp}}\left(\omega_{\pm0}\right)\gg \Gamma_{d'}\left(\omega_{\pm0}\right) $. The identification of a significant role for electric field noise in this rates can only be obtained due to the  inclusion of the $d'$ dipole terms allowed within the Hamiltonian for spin centers with $C_{3v}$ symmetries. 

We stress that the assumption, $\Gamma_{\gamma d'}(\omega_{+0})\approx\Gamma_{\gamma d'}(\omega_{-0})$\cite{PhysRevB.87.235436,magneticnoise6}, which results in $\Omega_-=\Omega_+$, depends strongly on the charge and magnetic noise spectral densities. It can only hold if the corresponding spectral noise densities are nearly ``flat'' (in frequency) between $\omega = \omega_{-0}$ and $\omega = \omega_{+0}$. More specifically, if we assume $S_{E_i}(\omega)=\left\langle \delta E^2\right\rangle \tau_{e,c}/(1+\omega^2\tau_{e,c}^2)$ [$S_{B_i}(\omega)=\left\langle \delta B^2\right\rangle \tau_{b,c}/(1+\omega^2\tau_{b,c}^2)$], $\Omega_-\approx\Omega_+$ holds typically for $\omega \ll 1/\tau_{e,c}$ [$\omega \ll 1/\tau_{b,c}$] as $S_{E_i}(\omega)$ [$S_{B_i}(\omega)$] becomes frequency independent. On the other hand, for $\omega\gg 1/\tau_{e,c}$ [$\omega\gg1/\tau_{b,c}$], we have $S_{E_i}(\omega)\propto 1/\omega^2$ [$S_{B_i}(\omega)\propto 1/\omega^2$], resulting in a very sensitive spectral noise density with respect to the frequency, and violation of $\Omega_-\approx\Omega_+$.

Experimental comparison of the measured rates, $\Omega_+$ and $\Omega_-$, indicates whether the noise spectral density is constant within the $\omega_{-0}<\omega <\omega_{+0}$ region. As a consequence, the difference between the nominal values of $\Omega_+$ and $\Omega_-$ can be used to obtain  information about the flatness of any noise spectral density. Hence the spin-1 relaxation mechanisms of $C_{3v}$ spin-defects can also be used to probe the presence of flat regions of the spectral noise density.

\begin{figure*}
  \includegraphics[width=\textwidth]{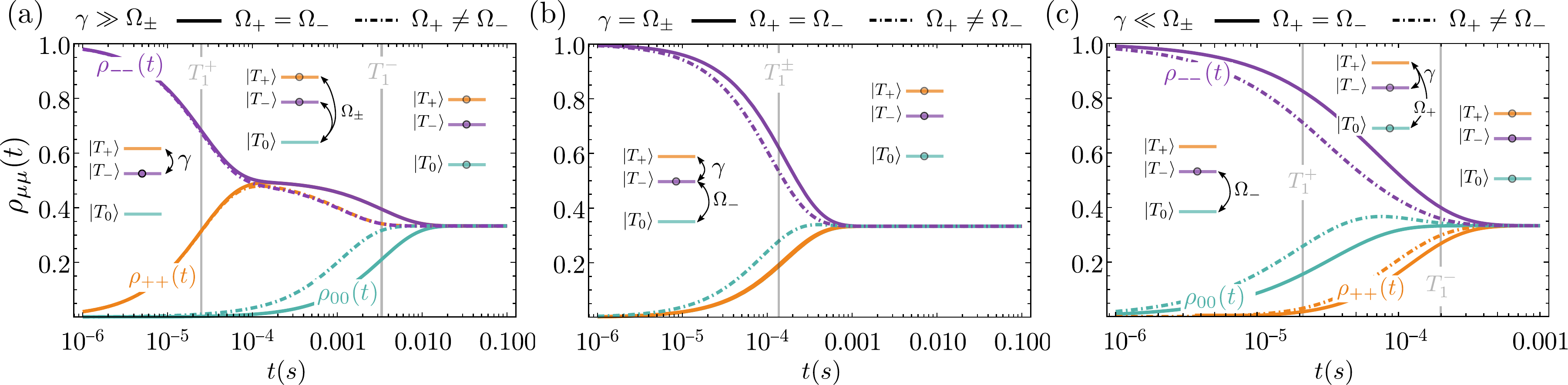}
  \caption{Population dynamics for the spin-defect energy levels as a function of time for $\gamma\gg \Omega_{\pm}$, $\gamma=  \Omega_{\pm}$ and $\gamma\ll \Omega_{\pm}$, respectively. (a) solid (dot-dashed) lines: $\gamma=20\times 10^3$~s$^{-1}$ with $\Omega_+=\Omega_- = 0.1\times 10^{3}$ ~s$^{-1}$ ($\Omega_+=\Omega_{-} /5= 0.1\times 10^{3}$ ~s$^{-1}$),  (b) solid (dot-dashed) lines: $\gamma=2\times 10^3$~s$^{-1}$ with $\Omega_+=\Omega_- = 2 \times 10^{3}$ ~s$^{-1}$ ($\Omega_+ =\Omega_{-}/2 = 2\times 10^{3}$ ~s$^{-1}$) and (c) solid (dot-dashed) lines: $\gamma=0.1\times 10^3$~s$^{-1}$ with $\Omega_+=\Omega_- = 10\times 10^{3}$ ~s$^{-1}$ ($\Omega_+ =\Omega_{-}/2 = 10\times 10^{3}$ ~s$^{-1}$).  }
  \label{fig2}
\end{figure*}

In Fig.~\ref{fig2}, we plot the three level population dynamics of our spin center as a function of time. We use three different regimes of parameters with the same initial condition, $\rho_{--}(t=0)=1$. In Fig.~\ref{fig2}(a), we have the corresponding dynamics for $\gamma \gg \Omega_{\pm}$. Although this regime is usually attributed to the dominance of the charge noise, the rate $\Omega_{\pm}$ also contains charge noise contributions via $\Gamma_{d'}(\omega_{\pm0})$. As a consequence, this  attribution is only accurate for $\Gamma_{d'}(\omega_{\pm0}) \ll \Gamma_{d_{\perp}}(\omega_{+-})$. Since different works suggest $d'\approx d_{\perp}$~\cite{PhysRevApplied.13.054068,doherty2012}, this translates to small values of magnetic fields satisfying $\omega_{+-}\ll\omega_{\pm 0}$ for $S_{E_i}(\omega)\propto 1/\omega^\alpha$. Nevertheless, in this case we see that a system initially prepared in the $\ket{T_{-}}$ state will start to increase the population of its $\ket{T_{+}}$ state due to the faster $\gamma$ relaxation rate. The corresponding relaxation is characterized by the $T_{1}^{+}$ timescale, Eq.~(\ref{t1p}), plotted as a vertical gray line. We see that in this process, the system first reaches an approximately equal population of $\ket{T_{-}}$ and $\ket{T_{+}}$. After this, the $\Omega_{\pm}$ rate starts playing a role, and we see an increase in the population of $\ket{T_{0}}$, with a characteristic timescale given by $T_{1}^-$, Eq.~(\ref{t1m}) (vertical gray line). Finally, for long times $t\gg T_{1}^{\pm}$, we obtain an equal population for all the levels. Similarly, the dot-dashed lines represent the same process but for $\Omega_{-}\neq\Omega_+$.

In Fig.~\ref{fig2}(b), the solid lines represent the plot for the population dynamics corresponding to $\gamma = \Omega_{\pm} = \Omega$. Here both the initially unpopulated levels, $\ket{T_{0}}$ and $\ket{T_{+}}$, experience a population increase with equal rate, and with a characteristic timescale defined by $T_{1}^{\pm}=(3\Omega)^{-1}$ [Eqs.~(\ref{t1p}) and (\ref{t1m})] (see vertical gray line). As a consequence, the populations of the states $\ket{T_{0}}$ and $\ket{T_{+}}$ increase equally. Similarly, the dot-dashed lines represent the same process but for $\gamma=\Omega_{+}<\Omega_-$. As a consequence of $\Omega_- > \gamma$, we see that the population of $\ket{T_{0}}$ increases faster than the population of $\ket{T_{+}}$.

Lastly, in Fig.~\ref{fig2}(c) we plot the population dynamics for $\gamma \ll \Omega_{\pm}$, usually associated with the dominance of  magnetic noise. However, as $\Omega_{\pm}=\Gamma_{\gamma d'}(\omega_{\pm 0})$ also contains a charge noise contribution (since $\Gamma_{\gamma d'}(\omega_{\pm 0})=\Gamma_{d'}(\omega_{\pm 0})+ \Gamma_{\gamma_{\perp}}(\omega_{\pm 0})$) we stress this identification of magnetic noise dominance will  only be accurate if $\Gamma_{d'}(\omega_{\pm 0})\ll \Gamma_{\gamma_{\perp}}(\omega_{\pm 0})$. We first see $\Omega_-$ as responsible for the increase of population of $\ket{T_{0}}$, with characteristic timescale $T_{1}^+$. At this point, as we have a finite population of $\ket{T_{0}}$, the rate $\Omega_+$ will begin to contribute, increasing the population of $\ket{T_{+}}$. Depending on the relative value of $\gamma$ and $\Omega_+$, $\gamma$ can also contribute to the increase of the $\ket{T_{+}}$ population.

\subsubsection{Spin center dephasing}
Dephasing within different subspaces is a consequence of the time dependence of the off-diagonal density matrix elements ${\rho_{\mu \nu}^{dep}(t)=\{ \rho_{+0}(t), \rho_{-0}(t), \rho_{0+}(t), \rho_{0-}(t), \rho_{+-}(t),\rho_{-+}(t)\}}$. Accordingly, our Lindblad operators [Eqs.~(\ref{L1})--(\ref{L8})] lead to three different dephasing rates within the three corresponding subspaces $\{\ket{T_0},\ket{T_+}\}$, $\{\ket{T_0},\ket{T_-}\}$, and $\{\ket{T_-},\ket{T_+}\}$, namely
\begin{equation}
\frac{d}{dt} \rho_{\mu \nu}^{dep}(t)   = - \frac{1}{T_{2}^{\mu \nu}} \rho_{\mu \nu}^{dep}(t),
\end{equation}
with corresponding dephasing times
\begin{align}
    \frac{1}{T_{2}^{0+}}&= \frac{1}{2}\left[\Gamma_{\gamma_{\parallel}}\left(0\right)+\Gamma_{\gamma d'}\left(\omega_{+0}\right)+\Gamma_{d_{\perp}}\left(\omega_{+-}\right)+\Gamma_{d_{\parallel}}\left(0\right)\right] \nonumber\\
    &+\frac{1}{4}\Gamma_{\gamma d'}\left(\omega_{-0}\right),\\
    \frac{1}{T_{2}^{0-}}&=\frac{1}{2}\left[\Gamma_{\gamma_{\parallel}}\left(0\right)+\Gamma_{\gamma d'}\left(\omega_{-0}\right)+\Gamma_{d_{\perp}}\left(\omega_{+-}\right)+\Gamma_{d_{\parallel}}\left(0\right)\right] \nonumber\\
    &+\frac{1}{4}\Gamma_{\gamma d'}\left(\omega_{+0}\right),\\
    \frac{1}{T_{2}^{-+}}&= 2\Gamma_{\gamma_{\parallel}}\left(0\right)+\Gamma_{d_{\perp}}\left(\omega_{+-}\right)+\frac{1}{4}\left[{\Gamma_{\gamma d'}\left(\omega_{-0}\right)+\Gamma_{\gamma d'}\left(\omega_{+0}\right)}\right].
\end{align}
We stress that this is the first work to report the accurate expression for the decoherence times including all the symmetry-allowed fluctuating terms. We see from these expressions that  dephasing within $\{\ket{T_\mu},\ket{T_\nu}\}$ is not solely originating from fluctuating terms within the same subspace. For example, even though $d' E_\pm (t)$ does not appear within the $\{\ket{T_-},\ket{T_+}\}$ subspace, the decoherence times $T_{2}^{0\pm}$ depend on $d'E_\pm (t)$. This shows that an indirect loss of coherence between coupled subspaces also happens. The same feature also occurs for  the $\{\ket{T_0},\ket{T_\pm}\}$ subspace, and was already discussed in Ref.~\cite{candido-pin}. In short, this shows the importance of taking into account fluctuators over the whole spin-defect manifold when calculating dephasing times.

\section{Theory of the fluctuating electric and magnetic fields}
\label{secIII}
Unintentional impurities within crystals can either donate electrons or accept electrons, leading to free electrons or holes in the crystal. As already discussed in Refs~\cite{magneticnoise6,candido-pin,candido2021theory}, these particles do not distribute uniformly and are also non-static, due to the thermal fluctuations of the electron and hole position, collisions between them, continual absorption and release by donors or acceptors, among other processes. Moreover, as the read-out and initialization of the spin center's state are performed with laser illumination, the measurement process additionally agitates the particles,  increasing these fluctuations. As different types of bulk noise were already discussed in Ref.~\cite{candido-pin,electricnoise3}, here we calculate the fluctuating surface fields arising from all the possible types of charge~\cite{electric-magnetic1,electric-magnetic2,electric-magnetic3,electric-magnetic4,electricnoise1,electricnoise2,electricnoise3,magneticnoise6} and magnetic noise~\cite{magneticnoise1,magneticnoise2,magneticnoise3,PhysRevB.87.235436,magneticnoise4,magneticnoise5,magneticnoise6}. We also provide an analysis for the competition between surface and bulk contributions. We emphasize here that  surface charge noise, $\delta \textbf{E}(t)$, can occur due to: (1)  electrons or holes trapped at the crystal surface, giving rise to a fluctuating dipole electric field; (2)  confined hole or electron gases produced by  band bending near the surface~\cite{PhysRevB.68.041304,surfacediamondreview,sussmann2009cvd,https://doi.org/10.1002/admi.201801449}, which produces a point-like fluctuating electric field; and (3)  electrons that are excited to the conduction band and therefore also contribute to a point-like fluctuating electric field. 
We  calculate the magnetic noise arising from  fluctuating magnetic moments at the surface, in addition to the magnetic noise produced by a random movement of charged particles (Biot-Savart law). For both charge and magnetic noise, we analyze the competition between their bulk and surface counterparts. 

The temporal difference of both electric and magnetic fields with respect to their averaged value, $\delta \textbf{E}(t)= \textbf{E}(t)-\left\langle  \textbf{E} \right\rangle$ and $\delta \textbf{B}(t)=\textbf{B}(t)-\left\langle  \textbf{B} \right\rangle$, respectively,  causes  decoherence of a prepared state~\cite{electric-magnetic1,electric-magnetic2,electric-magnetic3,electric-magnetic4,electricnoise1,electricnoise2,electricnoise3,magneticnoise1,magneticnoise2,magneticnoise3,PhysRevB.87.235436,magneticnoise4,magneticnoise5,magneticnoise6,magneticnoise7}, and also the increase of the photoluminescence linewidth of the defect emission~\cite{tamarat2006stark,anderson2019electrical,de2017stark,candido-pin}. This relevant quantities for the external fields are the magnetic and electric field correlations, $\left\langle B_\mu(t)B_\mu(0) \right\rangle $ and, $\left\langle E_\mu(t)E_\mu(0) \right\rangle$, respectively, presented in Eqs.~(\ref{specE}) and (\ref{specB}).


\subsection{Coordinate system}

The $z$ axis used in both Eqs.~(\ref{hgs}) and (\ref{H-matrix}) is defined with respect to the main symmetry axis of our defect spin center. This axis, however, can assume different directions with respect to the surface normal vector $\hat{n}$. Examples include diamond with surfaces perpendicular to either [111] or [001] crystalographic axis~\cite{chou2017nitrogen}, and for the different orientation of divacancies in SiC~\cite{candido-pin}. For general results we keep an arbitrary direction of the spin center main axis with respect to the surface normal, defined by $\hat{n}=\hat{z}'$. Accordingly, for the spin center and the surface, we have axis $\{\hat{x},\hat{y},\hat{z}\}$, and $\{\hat{x}',\hat{y}',\hat{z}'\}$, respectively. They are related to each other via a rotation of $\theta$ around the $x$ axis, $R_{\hat{x}}\left(\theta\right)$,
\begin{equation}
    \left(\begin{array}{c}
\hat{x}\\
\hat{y}\\
\hat{z}
\end{array}\right)=\left(\begin{array}{ccc}
1 & 0 & 0\\
0 & \cos\theta & -\sin\theta\\
0 & \sin\theta & \cos\theta
\end{array}\right)\left(\begin{array}{c}
\hat{x}'\\
\hat{y}'\\
\hat{z}'
\end{array}\right).
\end{equation}
This distinction is  important as the fluctuations of both charges  and magnetic moments at the surface produce a large fluctuating field along $\hat{z}$. Accordingly, we define our surface as ${\mathcal{S}= \left\{ (x',y',z'), \thinspace -L/2 \leq x',y' \leq L/2,\thinspace \thinspace z'=0 \right\}}$, and our defect position as $\textbf{r}_{\rm{def}}=(0,0,-z_{\rm{def}})=-z_{\rm{def}}\hat{z}'$.

\subsection{Analytical calculation for the surface fluctuating point-like electric field}

We now calculate the electric field correlation function $\left\langle E_{p,\mu}\left(t\right)E_{p,\mu}\left(0\right)\right\rangle$ arising from  fluctuating surface point-like charges~\cite{electric-magnetic1,electric-magnetic2,electric-magnetic3,magneticnoise6}. First, for the point-like charges we assume the electric field at  $\textbf{r}=\textbf{r}_{\rm{def}}$ produced by the $i$'th point-like charge is 
\begin{align}
\textbf{E}_{i}^{p} \left( \textbf{R}_{i}^p \right) & =\frac{Q_i}{4\pi\epsilon }\frac{\textbf{R}_{i}^p}{({R}_{i}^{p})^3},
\label{dipole2}
\end{align}
where $Q_i$ are both positive and negative trapped surface charges localized at $\textbf{r}_{i}^{p}\approx\left(x_{i}'^{p},y_{i}'^{p},0\right)$ with  $\textbf{R}_{i}^{p}=\textbf{r}_{i}^{p}-\textbf{r}_{\rm{def}}$ and $R_{i}^p=|\textbf{R}_{i}^p|$. Assuming now a total number of point-like charges given by $N_p$, the total electric field experienced by the defect is $\textbf{E}_{} =  \sum_{i =1}^{N_{p}} \textbf{E}_{i}^{p}\left( \textbf{R}_{i}^p \right)$, and the correlation is
\begin{equation}
    \left\langle E_{\mu}^{p}(t)E_{\mu}^{p}(0) \right\rangle= \sum_{i=1}^{N_p} \left \langle {E}_{i,\mu}^p\left(t \right) {E}_{i,\mu}^p\left( 0 \right)\right \rangle,
\end{equation}
where we assume there is no correlation between electric fields produced by different point-like charges. The inclusion of the non-null correlation between different particles was extensively analyzed in Ref.~\cite{candido2021theory}. We can evaluate this expression by assuming a continuous probability distribution for the positions $\textbf{r}_i$, $p_{\mathcal{S}}=p_{\mathcal{S}}(\textbf{r}')$, yielding 
\begin{equation}
    \left\langle \sum_{i=1}^{N} f\left(\textbf{r}_i\right)\right\rangle  \rightarrow \int_{\cal S} dS \thinspace n_{\cal S}\left(\textbf{r}'\right)f\left(\textbf{r}'\right) \label{conti-trans}
\end{equation}
where ${\cal S}$ is the surface containing the fluctuating charges, and $n_{\cal S}\left(\textbf{r}'\right)=N_p p_{\cal S}\left(\textbf{r}'\right)$ is the surface density of the particles $i$. Using an uniform probability distribution for the positions of the point-like charges, we obtain
\begin{equation}
   \left\langle E_{\mu}^{p}\left(t\right)E_{\mu}^{p}\left(0\right)\right\rangle=  \left(\frac{e}{4\pi\varepsilon}\right)^{2}\int dS\thinspace n\left(\textbf{r}'\right)\frac{R_{\mu}^{2}}{\left[R_{x}^{2}+R_{y}^{2}+R_{z}^{2}\right]^{3}}
\end{equation}
with 
with $\mu=x,y,z$, $R^2=R_{x}^2+R_{y}^2+R_{z}^2$ and 
\begin{align}
R_{x} & =x', \label{Rx}\\
R_{y} & =y'\cos\theta-z_{\rm{def}}\sin\theta,  \label{Ry}\\
R_{z} & =y'\sin\theta+z_{\rm{def}}\cos\theta,  \label{Rz}
\end{align}
yielding
\begin{align}
    |\delta E_{x}^p|^2 = \left\langle E_{x}^{p}\left(t\right)E_{x}^{p}\left(0\right)\right\rangle &= \left(\frac{e}{4\pi\varepsilon}\right)^{2}\frac{\pi n_{\cal S}}{4z_{\rm{def}}^{2}}, \label{point-like1}\\
        |\delta E_{y}^p|^2=\left\langle E_{y}^{p}\left(t\right)E_{y}^{p}\left(0\right)\right\rangle &= \left(\frac{e}{4\pi\varepsilon}\right)^{2}\frac{\pi n_{\cal S}}{8z_{\rm{def}}^{2}}(3-\cos2\theta), \\
      |\delta E_{z}^p|^2=\left\langle E_{z}^{p}\left(t\right)E_{z}^{p}\left(0\right)\right\rangle & =\left(\frac{e}{4\pi\varepsilon}\right)^{2}\frac{\pi n_{\cal S}}{8z_{\rm{def}}^{2}}(3+\cos2\theta). \label{point-like3}
\end{align}
These equations  were obtained assuming the maximum value for the correlation of the fluctuating charge positions.
For this type of noise, the frequency dependence of spectral noise density can be obtained from many different processess, e.g., absorption-release of electrons and holes by different traps and diffusion of electron and holes within our crystal. For those, the corresponding spectral noise density can be obtained similarly to Ref.~\cite{candido-pin}, yielding
\begin{equation}
    S_{{\rm E_{i}^p}} (\omega)= |\delta E_{i}^p|^2 \frac{2\tau_{p}}{1+\omega^2 \tau_{p}^2}.
    \label{Sep}
    \end{equation}

We now assess an important characteristic of our system, namely, how does the surface charge noise compete with the bulk noise arising from the fluctuating charges within the bulk of the sample~\cite{candido-pin}. This becomes important when spin centers are used for sensing, which requires maximizing the signal-to-noise ratio. The depth at which both contributions are nearly equal sets the optimal defect depth, $z_{\rm{opt}}$. To perform this analysis, we assume that the bulk noise comes predominantly from the near noise contribution of Ref.~\cite{candido-pin}, namely
\begin{equation}
 |\delta \textbf{E}_{\rm{b}}| = \frac{e}{\sqrt{2}\pi \epsilon} n_{\rm{\mathcal{V}}}^{2/3},
 \label{fluc-bulk}
\end{equation}
where $n_{\rm{\mathcal{V}}}^{2/3}$ is the volume density of fluctuating charges and ${\left|\delta\textbf{E}_b\right|=\sqrt{| \delta E_{b}^{x}|^2+| \delta E_{b}^{y}|^2+| \delta E_{b}^{z}|^2}}$ with $| \delta E_{b}^{x}|=| \delta E_{b}^{y}|=| \delta E_{b}^{z}|$. Assuming that the surface noise arises from point-like fluctuating charges,
Eqs.~(\ref{point-like1})--(\ref{point-like3}), we obtain the optimal defect depth, 
\begin{equation}
    z_{\rm{opt}} = \sqrt{2\pi} \frac{n_{\cal S}^{1/2}}{n_{\cal V}^{2/3}}.
\end{equation}

\begin{figure}[t!]
\begin{center}
\includegraphics[clip=true,width=1.0\columnwidth]{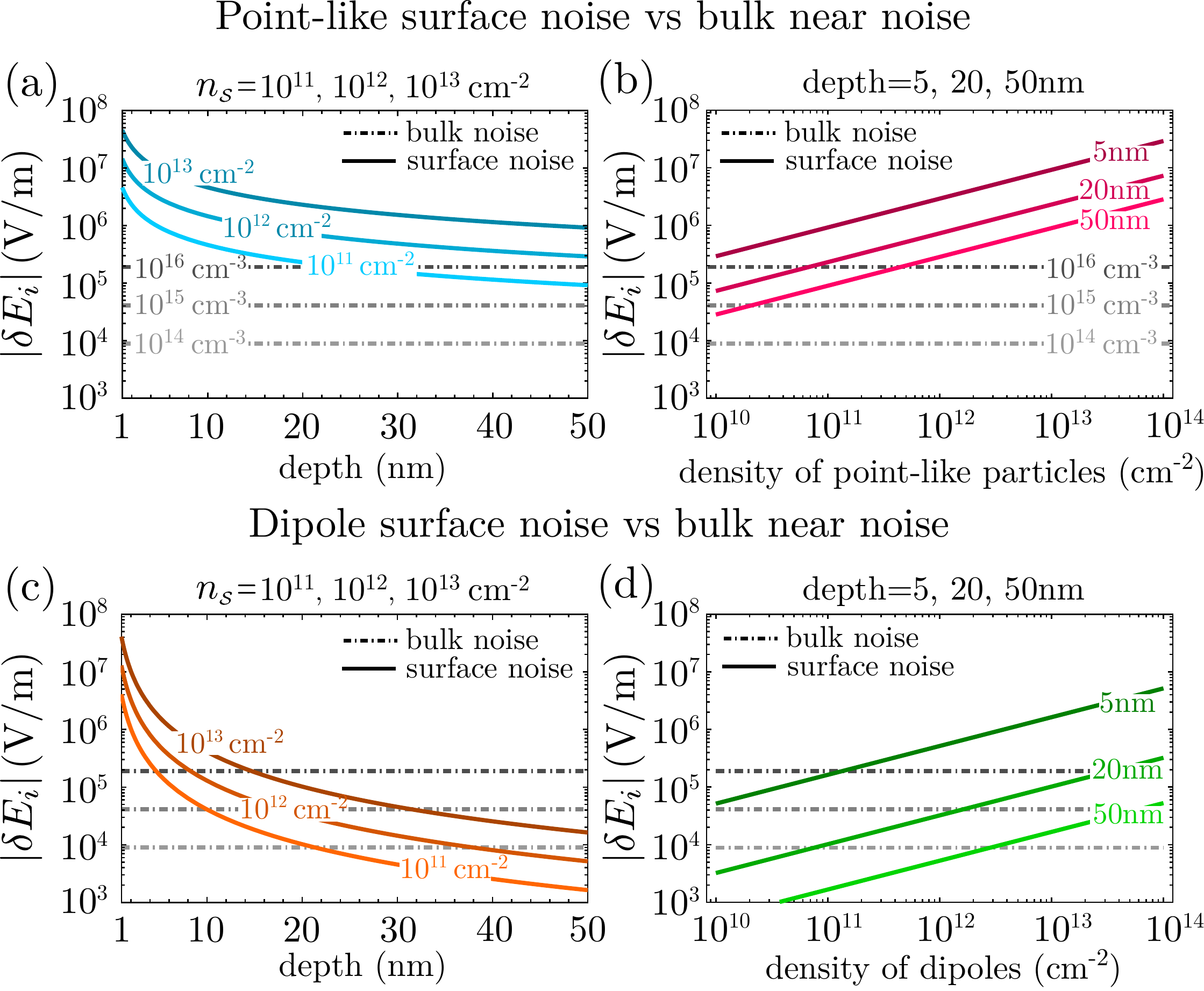}
\caption{Competition between fluctuating electric field for point-like surface noise [Eqs.~(\ref{point-like1})-(\ref{point-like3})] and bulk near noise [Eq.~(\ref{fluc-bulk})] as a function of depth for different surface densities (a), and as a function of surface density for different defect depths (b). Similar to (a) and (b), but representing the competition between fluctuating electric field for dipole surface noise [Eqs.~(\ref{dip-fluc1})-(\ref{dip-fluc3})] and bulk near noise [Eq.~(\ref{fluc-bulk})].
 In all the graphs, the fluctuating electric field due to the bulk near noise Eq.~(\ref{fluc-bulk}) is plotted for $n_{\cal V}=10^{14}$, $10^{15}$ and $10^{16}$~cm$^{-3}$} 
\label{fig3}
\end{center}
\end{figure}

In Figs.~\ref{fig3}(a) and (b) we plot the quantity $|\delta E_i|=\sqrt{\left\langle E_{i}^{p}\left(t\right)E_{i}^{p}\left(0\right)\right\rangle }$ for the surface charge noise arising from fluctuations of point-like charges at the surface, Eqs.~(\ref{point-like1})-(\ref{point-like3}), and the bulk near noise $|\delta \textbf{E}_b|$ arising from the fluctuation of point-like charges in bulk, Eq.~(\ref{fluc-bulk}). Here, the solid lines represent the surface charge noise and the dot-dashed ones represent the bulk near noise. In Fig.~\ref{fig3}(a) we plot both contributions for typical surface densities $n_{\cal S}=10^{11},10^{12}$ and $10^{13}$~cm$^{-2}$ together with bulk densities $n_{\cal V}=10^{14},10^{15}$ and $10^{16}$~cm$^{-3}$. While the surface noise contribution shows a $1/z_{def}^2$ depth dependence, the bulk one shows no dependence. Assuming an approximately bulk density of fluctuators of $10^{15}$~cm$^{-3}$, the surface noise always dominates, thus showing its critical importance for shallow defect implantation. Additionally, to have a better understanding of this competition, in Fig.~\ref{fig3}(b) we also plot both contributions as a function of the surface density of point-like fluctuators for different defect depths, $5,20$ and $50$~nm. We observe that for very shallow defects with $z_{def}\approx5$~nm, the surface noise will dominate even for low surface densities $n_{\cal S}\approx10^{10}$~cm$^{-2}$. On the other hand, for $z_{def}=50$~nm and assuming  $n_{\cal V}=10^{15}$~cm$^{-3}$, we obtain a dominance of the surface noise only for $n_{\cal S}>10^{12}$~cm$^{-2}$.

\subsection{Analytical calculation for the surface fluctuating dipole electric field}

Here we calculate the electric field correlation function $\left\langle E_\mu(t)E_\mu(0) \right\rangle$ arising from fluctuating surface dipole charges~\cite{electricnoise2,magneticnoise6}. Hence, the electric field arising from the $i$'th dipole, corresponding to the displacement of charges $Q_i$ and $-Q_i$ located at  $\textbf{r}_{i}^{d}\approx\left(x_{i}'^{d},y_{i}'^{d},0\right)$ 
and separated by the dipole distance, $\textbf{d}_{i}$, is written as
\begin{align}
\textbf{E}_{i}^{d}  & =\frac{Q_i}{4\pi\epsilon (R_{i}^d)^{5}}\left[3\left(\textbf{d}_{i}\cdot\textbf{R}_{i}^d\right)\textbf{R}_{i}^d-\textbf{d}_{i} (R_{i}^d)^{2}\right],
\label{dipole3}
\end{align}
with $\textbf{R}_{i}^{d}=\textbf{r}_{i}^{d}-\textbf{r}_{\rm{def}}$ and $R_{i}^d=|\textbf{R}_{i}^d|$. Assuming a total number of dipoles given by $N_d$, the total electric field experienced by our defect is $\textbf{E}  = \sum_{i=1}^{N_d} \textbf{E}_{i}^d\left( \textbf{R}_{d}^i \right)$, and we find
\begin{equation}
    \left\langle E_{\mu}^{d}(t)E_{\mu}^{d}(0) \right\rangle= \sum_{i,j=1}^{N_d} \left \langle {E}_{i,\mu}^{d}\left(t \right) {E}_{j,\mu}^{d}\left( 0 \right)\right \rangle.
\end{equation}
We first assume that the electric field produced by two different dipoles are not correlated, i.e., $\left \langle  {E}_{i,\mu}^{d}\left(t \right) {E}_{j,\mu}^d\left( 0 \right) \right \rangle  \approx \delta_{ij}  \left \langle  {E}_{i,\mu}^d\left(t \right) {E}_{i,\mu}^d\left( 0 \right) \right \rangle $. Secondly, we assume that the charge dipole displacements $\textbf{d}_{i}$ are randomly distributed along $\hat{x}'$, $\hat{y}'$ and $\hat{z}'$ with equal probability~\cite{candido-pin} and amplitude, yielding $\left\langle d_i (t)d_i (0)\right\rangle=\left\langle d(t)d(0)\right\rangle $ for $i=x,y,z$. For large number of dipoles, the above expression can be calculated by assuming a continuous density for the discrete dipole charge positions $\textbf{r}_i$, $n_{\mathcal{S}}=n_{\mathcal{S}}(\textbf{r})$, 
\begin{equation}
    \sum_{i=1}^{N_d} \left\langle \cdots \right\rangle\rightarrow \int_{{\cal S}} dS\thinspace n_{\cal S}(\textbf{r}')  \left\langle \cdots \right\rangle 
\end{equation}
Finally, we can write 
\begin{align}
\left\langle E_{\mu}^{d}\left(t\right)E_{\mu}^{d}\left(0\right)\right\rangle  & =\nonumber \\ 
\left(\frac{e}{4\pi\varepsilon}\right)^{2} & \int_{\mathcal{S}}dS\frac{n_{\cal S}(\textbf{r}')}{3}\left\langle d\left(t\right)d\left(0\right)\right\rangle \frac{3R_{\mu}^{2}+R^{2}}{R^{8}}.
\end{align}
For a position independent areal density $n_{\mathcal{S}}(\textbf{r})=n_{\mathcal{S}}^{d}$ and considering both surface lengths much larger than $|\textbf{r}_{\rm{def}}|=z_{\rm{def}}$, we obtain 
\begin{align}
    \left\langle E_{x}^{d}\left(t\right)E_{x}^{d}\left(0\right)\right\rangle &= \left(\frac{e}{4\pi\varepsilon}\right)^{2}\frac{\pi n_{\cal S}^{d}\left\langle d\left(t\right)d\left(0\right)\right\rangle }{4z_{\rm{def}}^{4}},\label{dip-fluc1}\\
   \left\langle E_{y}^{d}\left(t\right)E_{y}^{d}\left(0\right)\right\rangle & =\left(\frac{e}{4\pi\varepsilon}\right)^{2}\frac{\pi n_{\cal S}^{d}\left\langle d\left(t\right)d\left(0\right)\right\rangle }{8z_{\rm{def}}^{4}}\left(3-\cos2\theta\right),\\
   \left\langle E_{z}^{d}\left(t\right)E_{z}^{d}\left(0\right)\right\rangle & =\left(\frac{e}{4\pi\varepsilon}\right)^{2}\frac{\pi n_{\cal S}^{d}\left\langle d\left(t\right)d\left(0\right)\right\rangle }{8z_{\rm{def}}^{4}}\left(3+\cos2\theta\right).\label{dip-fluc3}
\end{align}

From Eqs.~(\ref{dip-fluc1})--(\ref{dip-fluc3}) and Eqs.~(\ref{point-like1})--(\ref{point-like3}), we see that the point-like and dipole field contributions yield different dependences on the defect depth $z_{\rm{def}}$. Whereas for point-like charge fluctuations we obtain a dependence $z_{\rm{def}}^{-2}$, the dipole one gives a weaker dependence $z_{\rm{def}}^{-4}$. This different defect depth dependence can be verified experimentally, and has already been discussed in the literature~\cite{magneticnoise4,magneticnoise6}. In the last section of this paper we will discuss how different surface treatments result in different types of charge noise (point-like or dipole).


Here we also obtain the optimal defect depth that defines the depth in which the bulk near noise becomes comparable to the surface noise, namely
\begin{equation}
    z_{\rm{opt}} = \left[\frac{\pi^{1/2}\bar{d}_{\cal S}\thinspace (n_{\mathcal{S}}^{d})^{1/2}}{2^{3/2}_{\textcolor{white}{V}}n_{\mathcal{V}}^{2/3}}\right]^{1/2}.
\end{equation}
with $\bar{d}_{\cal S}=\sqrt{\left\langle d\left(t\right)d\left(0\right)\right\rangle}\approx \sqrt{\left \langle d^2\right \rangle}$. Assuming a typical value $\bar{d}_{\cal S}=0.5$~nm, we plot in Fig.~\ref{fig3}(c) and (d) the competition between the surface charge noise arising from the fluctuation of dipole charges, Eqs.~(\ref{dip-fluc1})-(\ref{dip-fluc3}), and the bulk near noise Eq.~(\ref{fluc-bulk}). Due to the weaker character of the surface dipole charge noise, we can see that for a bulk density of fluctuators between $10^{14}$ and $10^{16}$~cm$^{-3}$ and $n_{\cal S}^{d}=10^{11},10^{12}$ and $10^{13}$~cm$^{-2}$, the optimal depth is always smaller than $20$~nm, showing the large contribution of the bulk noise to the dephasing and relaxation of shallow spin centers.

In Figs.~\ref{fig3}(c) and (d) the calculation of surface dipole charge noise contribution was done using $\bar{d}_{\cal S}=0.5$~nm. This, however, ignored the temporal dependence of the fluctuating dipoles. A more precise and accurate analysis includes the power spectral density of the dipole, yielding
\begin{align}
   S_{\rm{E_{x}^{d}}}\left(\omega\right)&=\left(\frac{e}{4\pi\varepsilon}\right)^{2}\frac{\pi n_{\cal S}^{d}}{4z_{\rm{def}}^{4}}S_{d}\left(\omega\right), \label{Sex-dip}\\
   S_{\rm{E_{y}^{d}}}\left(\omega\right)&=\left(\frac{e}{4\pi\varepsilon}\right)^{2}\frac{\pi n_{\cal S}^{d}\left(3-\cos2\theta\right)}{8z_{\rm{def}}^{4}}S_{d}\left(\omega\right),\label{Sey-dip}\\
   S_{\rm{E_{z}^{d}}}\left(\omega\right)&=\left(\frac{e}{4\pi\varepsilon}\right)^{2}\frac{\pi n_{\cal S}^{d}\left(3+\cos2\theta\right)}{8z_{\rm{def}}^{4}}S_{d}\left(\omega\right), \label{Sez-dip}
\end{align}
with $\left\langle X\left(t\right)X\left(0\right)\right\rangle=\int_{-\infty}^{\infty} \frac{d\omega}{2\pi} S_{X}(\omega)e^{i \omega t} $. 
Although it is still necessary to know and characterize $S_{d}(\omega)$, different studies have already investigated that in a similar context~\cite{PhysRevA.84.023412}.  Here, we first assume that the electric dipole will be represented by a trapped charge at the surface, which behaves as a harmonic oscillator with frequency $\omega_d$ and mass $m^*$. Using the equipartition theorem, i.e., $\frac{3}{2}k_{B}T \approx\frac{1}{2}m^* \omega_{d}^{2}\left\langle d^{2}\right\rangle$,  we obtain 
\begin{align}
\left\langle d^{2}\right\rangle  =\frac{3k_{B}T}{m\omega_{d}^{2}}.
\end{align}
The frequency $\omega_d$ is commonly associated with the energy difference between two  traps ($E_{0}$ and $E_1$), in which the charge can be imprisoned or trapped, yielding $\omega_d=(E_1-E_0)/\hbar$~\cite{PhysRevA.84.023412,PhysRevB.79.094520}. Single dipole fluctuators are usually described by a Lorenztian spectral noise density, namely 
\begin{equation}
    S_{d}(\omega)=\left\langle d^{2}\right\rangle \frac{ \Gamma_d}{(\omega-\omega_d)^2+(\Gamma_{d}/2)^2},
    \label{Sdip}
\end{equation}
where $\Gamma_{d}$ is the rate associated with the transition between energy levels $E_0$ and $E_1$. If we assume this rate is dominated by thermal activation, the rate will be given by $\Gamma_d \propto e^{-(E_1 -E_0)/ k_B T}$. When the temperature is not high enough to thermally activate the high-energy levels, $\Gamma_d$ is attributed to trapped electrons that diffuse at the surface through quantum tunneling over an energy barrier, $E_b$, with a corresponding width $b$. For the case of a double-well potential described by a parabolic one-dimensional potential,~\cite{PhysRevLett.95.195901,PhysRevA.87.023421}
\begin{equation}
\Gamma_0  = \frac{2\omega_d}{\pi^{3/2}} \sqrt{\frac{2E_b}{\hbar \omega_d}} e^{-{2E_b}/{\hbar \omega_d}},
\end{equation}
with $\omega_d=\sqrt{2E_b/m^* b^2}$ assuming hopping between two 1D parabolic double wells with an effective electronic mass $m^*$.

\subsection{Competition between point-like and dipole surface charge noise}

Here we analyze the competition between the charge noise due to point-like and dipole fluctuations, Eqs.~(\ref{point-like1})--(\ref{point-like3}) with Eq.~(\ref{Sep}), and Eqs.~(\ref{dip-fluc1})--(\ref{dip-fluc3}) with Eq.~(\ref{Sdip}), respectively. The ratio
\begin{equation}
    \frac{S_{\rm{E_{i}^{p}}}\left(\omega\right) }{S_{\rm{E_{i}^{d}}}\left(\omega\right)}=\frac{n_{\cal S}}{n_{\cal S}^d} \frac{ z_{def}^{2}}{ \bar{d}_{\cal S}^{\thinspace 2}}\frac{2\tau_p}{\Gamma_d}\frac{(\omega-\omega_d)^2+(\Gamma_d /2)^2}{1+\omega^2 \tau_{p}^2}, \label{comp-E}
\end{equation}
defines the condition driving this competition. The characteristic lengths (dipole length and defect depth) and densities of these different mechanisms play a role in this competition, however the ratio of the correlation time of the fluctuations $\tau_p$ and $\Gamma_d^{-1}$ also has a strong effect. If we assume that both fluctuations originate from the same physical mechanisms, we can assume $\tau_p \approx \Gamma_d^{-1}$. For this condition, the right-hand side of the equation above becomes ${n_{\cal S}}{ z_{def}^{2}} / {n_{\cal S}^d}{\bar{d}_{\cal S}^{\thinspace 2}}$. Hence, if the number of fluctuating point-like charges is approximately the same as the number of dipole fluctuators, i.e., $n_{\cal S} 
\approx n_{\cal S}^d$, the decoherence due to point-like fluctuations will dominate for $z_{def}\gtrsim \bar{d}_{\cal S}$, which is usually the case as $z_{def}\gtrsim 5$~nm and $\bar{d}_{\cal S}\lesssim 1$~nm. Although this would imply that the fluctuation of the point-like charges are always the dominant ones, this result relies exclusively on $\tau_p \approx 1/\Gamma_d$ and $n_{\cal S} \approx n_{\cal S}^d$. If we think of the fluctuating dipole charges as being described by trapped electrons (holes) due to surface acceptors (donors), and the point-like fluctuating charges as being produced by the surface hole (electron) gas produced by the electrons (holes) imprisonment, this would yield  $n_{\cal S} \approx n_{\cal S}^d$. However, as different experiments have already shown the dominant dipole character of the surface noise,  the trapped charges at the surface must be obtained either by the treatment of the surface crystal or its contact with the atmosphere~\cite{electric-magnetic3}.

\subsection{Frequency dependence of the electric spectral noise density}

In both previous two sections, we have used the fact that the frequency dependence of the spectral noise density is given by the Lorenztian
\begin{equation}
    S_{E_i}^{0}(\omega)\propto\frac{\tau }{1+\omega^2 \tau^2}.
    \label{S-part}
\end{equation}
Generally speaking, this follows when the dynamics of the fluctuation is characterized by only one characteristic time given by $\tau$. As a consequence, we have a correlation function for the electric field produced by the  fluctuators scaling as $e^{-t/\tau}$, whose Fourier transform yields Eq.~(\ref{S-part}). Although this is not incorrect, this form relies on a key assumption:  that either dipole or point-like charges have the same timescale associated to their fluctuations ($\tau$). This is not typically the case, and as a consequence, the spectral noise density deviates from $ S_{\rm E_{i}}^{0}(\omega)$ in realistic physical situations. For instance, recent experiments on shallow NV centers suggest a $ S_{\rm E_{i}}(\omega) \propto \omega^{-1}$ dependence~\cite{electricnoise1,magneticnoise6,electric-magnetic3}. There are many studies~\cite{RevModPhys.53.497,kogan2008electronic,RevModPhys.86.361} explaining the origin of the $\omega^{-1}$ spectral noise dependence. Here, we will use a similar approach, and apply it directly to our case. We first assume that $\tau$ has an origin in activation processes, either due to the continual trapping and release of electrons and holes, or due to the diffusion of them. We  then assume $\tau= \tau_0 e^{E/k_B T}$ where $E$ is the energy associated with either the tunneling between different trap centers or the activation energy. Due to roughness of the surface crystal and the range of trap centers, we cannot assume only one particular value for $E$ but rather a distribution for it, and here we take $E_1<E<E_2$. Taking this into account, our spectral density noise becomes 
\begin{equation}
    S_{E_i}(\omega) = \int_{E_1}^{E_2}dE g(E) P(E) S_{E_i}^{0}(\omega,E),
\end{equation}
with $g(E)$  the density of states of $\tau$ with respect to $E$, and $P(E)$  the weight associated to $S_{E_i}^{0}(\omega,E)$. By requiring $\int_{0}^{\infty}dEg\left(E\right)P\left(E\right)=1$ we obtain ${P\left(E\right)=\frac{k_{B}T}{E_{2}-E_{1}}\tau_{0}^{-1}e^{-{E}/{k_{B}T}}}$. The integration produces
\begin{widetext}
\begin{equation}
   S_{E_i}(\omega)= \frac{2k_{B}T}{E_{2}-E_{1}}\left[\frac{\tan^{-1}\left(\omega\tau_{0}e^{E_{2}/k_{B}T}\right)-\tan^{-1}\left(\omega\tau_{0}e^{E_{1}/k_{B}T}\right)}{\omega}\right],
\end{equation}
which yields the following three different frequency dependences
\begin{equation}
    S_{E_{i}}\left(\omega\right)=\begin{cases}
\frac{2k_{B}T}{E_{2}-E_{1}} \tau_{0}\left(e^{E_{2}/k_{B}T}-e^{E_{1}/k_{B}T}\right) & \omega\tau_{0}\ll e^{-E_{2}/k_{B}T},\\
\frac{k_{B}T}{E_{2}-E_{1}}\frac{\pi}{\omega} & e^{-E_{2}/k_{B}T}\ll\omega\tau_{0}\ll e^{-E_{1}/k_{B}T},\\
\frac{2k_{B}T}{E_{2}-E_{1}}\frac{\tau_{0}}{\omega^{2}}\left(e^{-E_{1}/k_{B}T}-e^{-E_{2}/k_{B}T}\right) & \omega\tau_{0}\gg e^{-E_{1}/k_{B}T},
\end{cases}
\end{equation}
\end{widetext}
which not only captures the $\omega^{-1}$ dependence but also shows a $\omega^{-2}$ dependence at higher frequencies.

\subsection{Charges and dipoles at the interface}

Within the last sections, we have assumed that all of our particles were always within the crystal region, thus experiencing a dielectric constant $\epsilon$. Conversely, if these charges and dipoles are now placed right at the interface between our material and the external medium, we need to perform the following change 
\begin{equation}
    \frac{1}{4\pi \epsilon} \rightarrow \frac{1}{2\pi ({\epsilon+\epsilon_{ext}})}
\end{equation}
in the equations above, where $\epsilon_{ext}$ is the dielectric constant of the external environment. Hence, for defect spins embedded within a high dielectric constant material a reduction of the charge noise is expected, which was already verified through an enhancement of the spin coherence time~\cite{electricnoise1}. 

\subsection{Results for charge noise}

\begin{figure*}
\includegraphics[width=\textwidth]{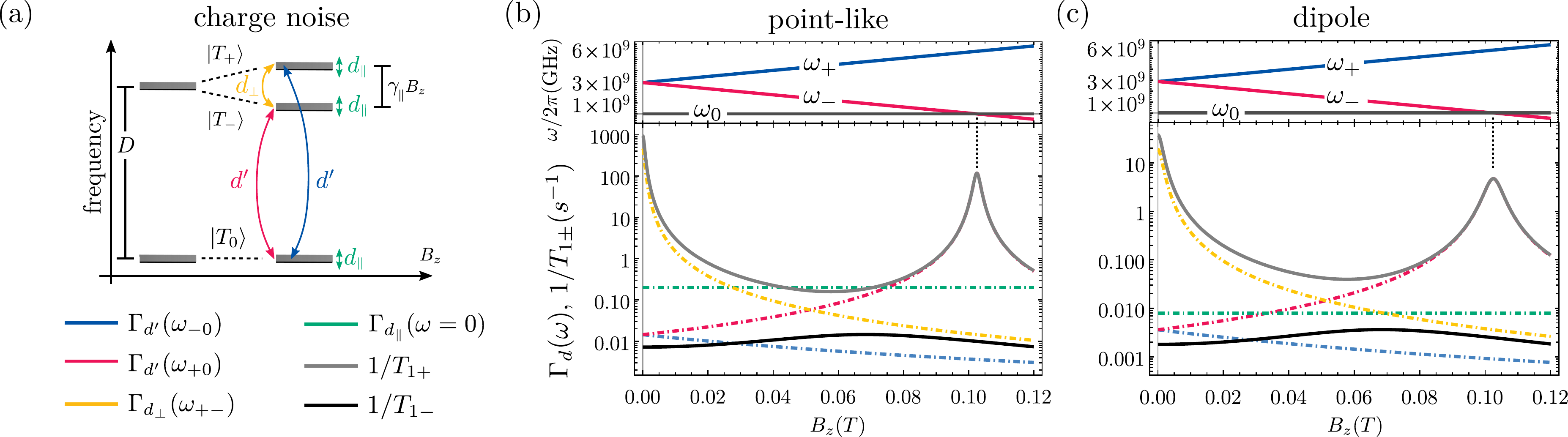}
  \caption{(a) Charge noise and the corresponding terms causing relaxation ($d_{\perp}$ and $d'$) and dephasing ($d_{\perp}$, $d'$ and $d_{\parallel}$). (b) $1/T_{1\pm}$ [Eqs.~(\ref{t1p}) and (\ref{t1m})] and the rates associated with charge noise arising from point-like charge fluctuators [$\Gamma_{d_{\perp}}(\omega_{+-})$, $\Gamma_{d'}(\omega_{\pm 0})$ and $\Gamma_{d_{\parallel}}(\omega=0)$, Eqs.~(\ref{Gammadperp})--(\ref{Gammadpara}), calculated through Eq.~(\ref{Sep}) for $n_{\cal S}=10^{11}$~cm$^{-2}$, $z_{def}=5$~nm and $\tau_p=5$~ns] as a function of magnetic field $B_z$. The magnetic field controls the frequency separation between the spin-defect levels, which can be seen in the upper part of the graph. (c) Same as (b) but for dipole charge fluctuators, i.e., calculated through Eqs.~(\ref{dip-fluc1})--(\ref{dip-fluc3}) with Eq.~(\ref{Sdip}) using $n_{\cal S}^{d}=10^{12}$~cm$^{-2}$, $z_{def}=5$~nm, $\bar{d}_{\cal S}=0.5$~nm, $\omega_d=0$ and $\Gamma_d=10^{9}$~s. For all the graphs, we have used the NV-center parameters $\gamma_{\parallel}=\gamma_{\perp}=28$~GHz/T, $d_{\perp}=17\times 10^{-2}$~Hz.m/V, $d_{\parallel}=0.35\times 10^{-2}$~Hz.m/V and $d'=d_{\perp}/2$.}
  \label{fig4}
\end{figure*}

In Fig.~\ref{fig4} we study the response of the rates (and associated relaxation times) generated by the charge noise as a function of the magnetic field ($B_z$) controlling the separation between the spin-defect energy levels [See upper plot of Figs.~\ref{fig4}(b) and (c)]. In Fig.~\ref{fig4}(b), we plot (dot-dashed lines) the rates $\Gamma_{d_{\perp}}(\omega_{+-})$, $\Gamma_{d'}(\omega_{\pm 0})$, and $\Gamma_{d_{\parallel}}(\omega=0)$ [Eqs.~(\ref{Gammadperp})--(\ref{Gammadpara})], assuming the charge noise is dominated by fluctuations of point-like charges [Eq.~(\ref{Sep})]. First, we observe that the rate $\Gamma_{d_{\perp}}(\omega_{+-})$ (yellow curve) decreases as a function of $B_z$. This happens because the larger the $B_z$, the larger the frequency separation between $\ket{T_-}$ and $\ket{T_+}$ ($\omega_{+-}$), thus suppressing the relaxation rate between these states. Similarly, we see the same behavior for $\Gamma_{d'}(\omega_{+0})$, represented by the blue dot-dashed line. Conversely, the rate $\Gamma_{d'}(\omega_{-0})$, indicated by the red plot, increases as a function of $B_z$. This happens because the frequency separation between $\ket{T_+}$ and $\ket{T_0}$ decreases when we increase the magnetic field, until it reaches the degeneracy point defined by $B_c \approx 0.105$~T, where it starts to increase again. When the frequencies $\omega_-$ and $\omega_0$ becomes degenerate, $\Gamma_{d'}(\omega_{-0})$ reaches its maximum. In addition, we also plot the relaxation times $1/T_{1\pm}$ [Eqs.~(\ref{t1p}) and (\ref{t1m}), here gray and black solid lines, respectively], with $1/T_{1+}$ resembling a sum of all the rates previously discussed. Interestingly, the response of our levels to the magnetic field causes a non-monotonic behavior of $1/T_{1+}$, which was already discussed in the literature for different systems~\cite{PhysRevB.89.115427}. Finally, we see that the only rate that does not change as a function of the magnetic field is $\Gamma_{d_{\parallel}}(\omega=0)$. This rate does not give rise to an extra relaxation process, as it does not couple two different levels. Therefore, it will not depend on the frequency difference between the energy levels, and will only cause decoherence. 

Similarly, in Fig.~\ref{fig4}(c) we plot the same rates as a function of $B_z$, but assuming the charge noise is dominated by a fluctuation of dipole charges. These are now calculated with Eqs.~(\ref{dip-fluc1})--(\ref{dip-fluc3}) with a corresponding frequency dependent spectral noise density given by Eq.~(\ref{Sdip}). For the chosen parameters, the charge noise arising from the fluctuation of point-like charges is stronger, thus generating faster relaxation rates. 

Most importantly, this whole analysis shows that to achieve the maximum suppression of the charge noise in NV center spin defects, we must choose magnetic fields values $B_z \approx 0.06$~T as this represents the maximum value for $T_{1\pm}$. This value depends on the ratio of the dipole moments $d_{\perp}/d'$ of our spin-defect, in addition to the assumption of a Lorentzian spectral noise density [Eq.~(\ref{Sdip})]. 

\subsection{Analytical calculation for  surface fluctuating magnetic fields}

As we saw before, the fluctuation of the charged particles (electrons) creates the surface charge noise. However, in addition to the charge, electrons also contain spin, and hence a fluctuation of the orientation of the particle's intrinsic magnetic moment is also expected~\cite{magneticnoise1,magneticnoise2,magneticnoise3,magneticnoise4,magneticnoise5,PhysRevB.87.235436,electricnoise3}. Similarly, the magnetic field experienced by the defect at $\textbf{r}_{\rm{def}}$ due to a magnetic dipole moment $\boldsymbol{\mu}_i$ located at $\textbf{r}_{i}^{m}$ is given by
\begin{equation}
    \textbf{B}_{i}^{m}\left(\textbf{R}_{i}^{m} \right)=\frac{\mu_{0}}{4\pi\left|\textbf{R}_{i}^{m}\right|^{5}}\left[3\left(\boldsymbol{\mu}_{i}\cdot\textbf{R}_{i}^{m}\right)\textbf{R}_{i}^{m}-\boldsymbol{\mu}_{i}\left|\textbf{R}_{i}^{m}\right|^{2}\right], \label{mag-mu}
\end{equation}
with $\textbf{R}_{i}^{m}=\textbf{r}_{i}^{m}-\textbf{r}_{\rm{def}}$ and $|\boldsymbol{\mu}_i|=h \gamma/2=g\mu_B/2$ for spin-1/2. Assuming that the density of charged particles is equal to the density of particles with magnetic moment, $n_S$, we can also determine the temporal correlation of the magnetic field with Eq.~(\ref{specB}), namely
\begin{align}
    \left\langle B_{x}^m\left(t\right)B_{x}^m\left(0\right)\right\rangle &= \left(\frac{\mu_{0}}{4\pi}\right)^{2}\frac{\pi n_{\cal S}^{d}\left\langle \mu\left(t\right)\mu\left(0\right)\right\rangle }{4z_{\rm{def}}^{4}},\\
   \left\langle B_{y}^m\left(t\right)B_{y}^m\left(0\right)\right\rangle & =\left(\frac{\mu_{0}}{4\pi}\right)^{2}\frac{\pi n_{\cal S}^{d}\left\langle \mu\left(t\right)\mu\left(0\right)\right\rangle }{8z_{\rm{def}}^{4}}\left(3-\cos2\theta\right),\\
   \left\langle B_{z}^m\left(t\right)B_{z}^m\left(0\right)\right\rangle & =\left(\frac{\mu_{0}}{4\pi}\right)^{2}\frac{\pi n_{\cal S}^{d}\left\langle \mu\left(t\right)\mu\left(0\right)\right\rangle }{8z_{\rm{def}}^{4}}\left(3+\cos2\theta\right),
\end{align}
where we have assumed the $\boldsymbol{\mu}_i$'s are randomly distributed along $\hat{x}'$, $\hat{y}'$ and $\hat{z}'$ with equal probability and amplitude, yielding $\left\langle \mu_i (t)\mu_i (0)\right\rangle=\left\langle \mu(t) \mu(0)\right\rangle $ for $i=x,y,z$. Using the definition of power spectral density, we obtain
\begin{align}
   S_{\rm{B_{x}^{m}}}\left(\omega\right)&=\left(\frac{\mu_{0}}{4\pi}\right)^{2}\frac{\pi n_{\cal S}^{d}}{4z_{\rm{def}}^{4}}S_{\mu}\left(\omega\right), \label{Sbx}\\
   S_{\rm{B_{y}^{m}}}\left(\omega\right)&=\left(\frac{\mu_{0}}{4\pi}\right)^{2}\frac{\pi n_{\cal S}^{d}\left(3-\cos2\theta\right)}{8z_{\rm{def}}^{4}}S_{\mu}\left(\omega\right),\label{Sby}\\
   S_{\rm{B_{z}^{m}}}\left(\omega\right)&=\left(\frac{\mu_{0}}{4\pi}\right)^{2}\frac{\pi n_{\cal S}^{d}\left(3+\cos2\theta\right)}{8z_{\rm{def}}^{4}}S_{\mu}\left(\omega\right) \label{Sbz}.
\end{align}
Similarly to the result obtained in Eqs.~(\ref{Sex-dip})--(\ref{Sez-dip}), here we also obtain a $z_{\rm{def}}^{-4}$ dependence on the defect depth due to the dipole character of Eq.~(\ref{mag-mu}). Moreover, if we treat the magnetic moments as being predominantly due to particles with spin-1/2, the characteristic frequency of the system will be set by $\Delta \omega_\mu=\omega_{\mu}^+ -\omega_{\mu}^- $ with $\hbar \omega_{\mu}^{\pm} =\pm(1/2)g \mu_B B = \pm (1/2)\gamma h B$, and hence we can assume a power spectral density corresponding to a Lorenztian peaked at $\omega=\Delta\omega_\mu$,  
\begin{equation}
    S_{\mu}\left(\omega\right) = \frac{2(\hbar \gamma/2)^2\tau}{1+(\omega-\Delta\omega_\mu)^2\tau^2}.
    \label{Smu}
\end{equation}
 
 Within the derivation above, the random character of the magnetic moment orientation was assumed. We emphasize this is only consistent at ``high'' temperatures defined by $k_B T \gg g \mu_B B$, which gives $N_{\uparrow}/N_{\downarrow} =e^{-\frac{2g\mu_B B}{k_B T}}\approx 1$, where $N_{\uparrow},N_{\downarrow}$ are the populations of spin polarization. Hence our results are valid for temperatures ${T\gg g\mu_B B/k_B}$, which for $g\approx 2$ and $0\leq B \lesssim 0.1$~T yields $T\gg 130 $~mK. For $T \lesssim 130$~mK the spins starts to align and the magnetic noise starts to be suppressed.

\subsection{Fluctuating magnetic field due to the movement of charged particles}

In addition to the magnetic noise produced by the fluctuating magnetic moments, we also have a magnetic field noise produced by the movement of charged particles, e.g., the Johnson-Nyquist noise in metals~\cite{PhysRev.32.97,PhysRev.32.110,Kolkowitz1129}. Accordingly, this can be obtained through Biot-Savart law
\begin{equation}
\textbf{B}=\sum_{i=1}^{N_q} \textbf{B}_{i}^{q}=\sum_{i=1}^{N_{q}}\frac{\mu_{0}}{4\pi}\frac{Q_{i}\textbf{v}_{i}\times\textbf{R}_{i}}{\textbf{R}_{i}^{3}},
\end{equation}
where $Q_i$ is the charge of the $i$-th particle, $\textbf{v}_i$ is its velocity and $N_q$ is the total number of mobile charged particles. Since we are assuming an areal density of charges, our velocity components can be approximated to $\textbf{v}_i=v_{i,x'}\hat{x}'+v_{i,y'}\hat{y}'$, i.e., no velocity perpendicular to the surface.  Similarly to the previous section, we assume no correlation between different particles' position, and a continuous probability distribution for the particles' positions $n_{\cal S}(\textbf{r})$, yielding 
\begin{align}
&\left\langle B_{x}^q\left(t\right)B_{x}^q\left(0\right)\right\rangle =\left(\frac{\mu_{0}e}{4\pi}\right)^{2} \left\langle v_{x'}\left(t\right)v_{x'}\left(0\right)\right\rangle  \int dS {n}_{\cal S}(\textbf{r}') \frac{z_{\rm{def}}^2}{R^6}, \\
&\left\langle B_{y}^q\left(t\right)B_{y}^q\left(0\right)\right\rangle =\left(\frac{\mu_{0}e}{4\pi}\right)^{2} \left\langle v_{x'}\left(t\right)v_{x'}\left(0\right)\right\rangle  \times \nonumber\\
&\int dS {n}_{\cal S}(\textbf{r}') \left\{ \cos^{2}\theta\frac{z_{\rm{def}}^{2}}{{R}^{6}}+\sin^{2}\theta\frac{y'^{2}+x'^{2}}{{R}^{6}}-\sin2\theta\frac{z_{\rm{def}}y'}{{R}^{6}}\right\}, \\
&\left\langle B_{z}^q\left(t\right)B_{z}^q\left(0\right)\right\rangle =\left(\frac{\mu_{0}e}{4\pi}\right)^{2} \left\langle v_{x'}\left(t\right)v_{x'}\left(0\right)\right\rangle  \times \nonumber\\
&\int dS {n}_{\cal S}(\textbf{r}') \left\{ \sin^{2}\theta\frac{z_{\rm{def}}^{2}}{{R}^{6}}+\cos^{2}\theta\frac{y'^{2}+x'^{2}}{{R}^{6}}+\sin2\theta\frac{z_{\rm{def}}y'}{{R}^{6}}\right\},    
\end{align}
where we have assumed a density ${n}_{\cal S}$ of mobile charged particles behaving as a 2D Brownian-Drude model, and hence $\left\langle v_{\mu}\left(t\right)v_{\nu}\left(0\right)\right\rangle =\delta_{\mu\nu}\left\langle v_{\nu}^{2}\right\rangle e^{-t/\tau}$ for $\mu,\nu=x',y',z'$, where $\tau$ is the relaxation time~\cite{kogan2008electronic}. Additionally, for particles in thermal equilibrium we have $\left\langle v_{\nu}^{2}\right\rangle= k_B T/m^*$ (equipartition theorem), where $k_B$ is the Boltzmann constant, $T$ is the temperature and $m^*$ is the effective mass of the charged particles. Hence we obtain 
\begin{equation}
    S_{v_\nu}\left(\omega \right)=\frac{2 \frac{k_B T}{m^*}\tau}{1+\omega^2 \tau^2}.
\end{equation}
Using now the definition of power spectral density, we write
\begin{equation}
   S_{\rm{B_{\mu}^{q}}}\left(\omega\right)=\frac{\mu_{0}^{2}k_{B}T\sigma_{2D}\left(\omega\right)}{16\pi}\frac{1}{z_{\rm{def}}^{2}}, \label{Sbmu-chargedvel}
  \end{equation}
for $\mu=x,y,z$, where $\sigma_{2D}(\omega)=\frac{{n}_{\cal S} e^2}{m^*}\frac{\tau}{1+\omega^2 \tau^2}$ is the 2D conductivity. These results show a stronger dependence on the defect depth  compared to the ones obtained from the fluctuation of magnetic moments, Eqs.~(\ref{Sbx})--(\ref{Sbz}), and cannot be neglected. This type of noise was already studied in the literature in the context of NV-sensing the electrons within a metallic  medium~\cite{Kolkowitz1129,ariyaratne2018nanoscale}.

\subsection{Competition between dipole magnetic noise and magnetic noise produced by charged particles}

Here we compare the spectral noise density for  the point-like and dipole surface charge noise, Eqs.~(\ref{Sbx})--(\ref{Sbz}) and Eq.~(\ref{Sbmu-chargedvel}). The ratio
\begin{equation}
    \frac{S_{{\rm B}_{\mu}^q}(\omega)}{S_{{\rm B}_{\mu}^{m}}(\omega)}=\frac{1}{n_{\cal S}^d}\frac{k_{B}T\sigma_{2D}\left(\omega\right)}{S_{\mu}\left(\omega\right)}z_{def}^2,
\end{equation}
defines the condition driving this competition. This expression is very similar to  Eq.~(\ref{comp-E}). 

\begin{figure*}
  \includegraphics[width=\textwidth]{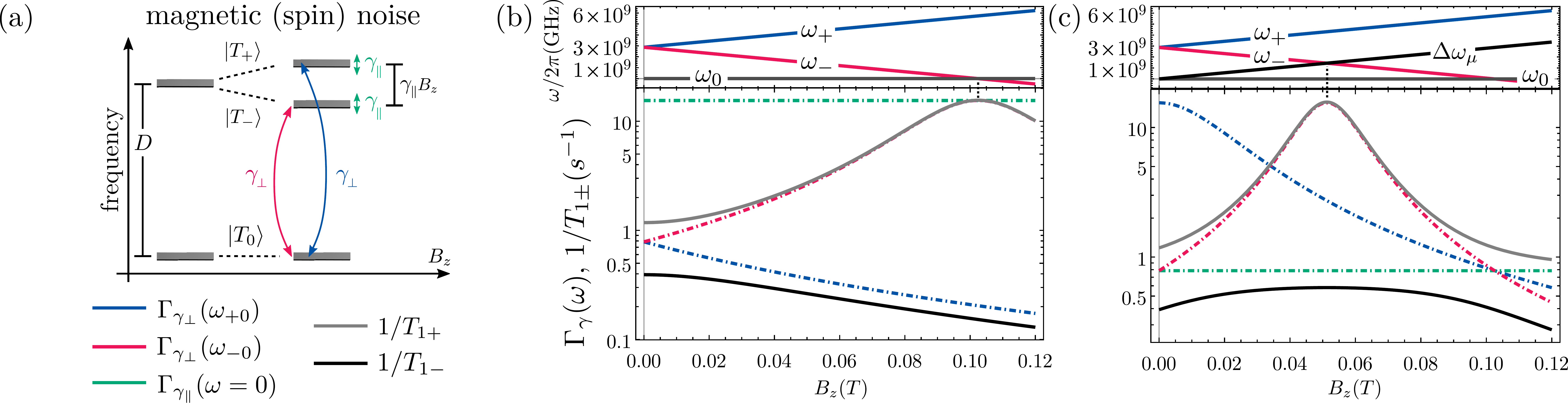}
  \caption{(a) Magnetic (spin) noise and the corresponding terms causing relaxation ($\gamma_{\perp}$) and dephasing ($\gamma_{\perp}$ and $\gamma_{\parallel}$). (b) $1/T_{1\pm}$ [Eqs.~(\ref{t1p}) and (\ref{t1m})] and the rates associated to the dipole magnetic noise [$\Gamma_{\gamma_{\perp}}(\omega_{\pm 0})$ and $\Gamma_{\gamma_{\parallel}}(\omega=0)$, Eqs.~(\ref{gammaperp}) and (\ref{gammapara}), calculated through Eqs.~(\ref{Sbx})--(\ref{Sbz}) and (\ref{Smu}) for $n_{\cal S}=5\times 10^{12}$~cm$^{-2}$, $z_{def}=5$~nm, $\Delta\omega_\mu=0$ and $\tau=0.24$~ns] as a function of magnetic field $B_z$. The magnetic field controls the frequency separation between the spin-defect levels, plotted in the upper part of the graph. (c) Same as (b) but assuming magnetic noise due to fluctuation of a two-level system with frequency splitting $\Delta\omega_\mu=\gamma h B_z/\hbar$.}
  \label{fig5}
\end{figure*}

\subsection{Results for  magnetic noise}

Here we study and investigate the dependence of the rates (and associated relaxation times) generated by the magnetic noise as a function of the magnetic field ($B_z$) [See Fig.~\ref{fig5}(a)]. In Fig.~\ref{fig5}(b), we plot (dot-dashed lines) the rates $\Gamma_{\gamma_{\perp}}(\omega_{\pm 0})$, and $\Gamma_{\gamma_{\parallel}}(\omega=0)$ [Eqs.~(\ref{gammaperp}) and (\ref{gammapara})], assuming the magnetic noise is dominated by fluctuations of magnetic moments [Eqs. (\ref{Sbx})--(\ref{Sbz}) and (\ref{Smu})] with $\Delta\omega_\mu =0$. First, we observe that the rate $\Gamma_{\gamma_{\perp}}(\omega_{+0})$ (blue curve) decreases as a function of $B_z$. This happens because the larger the $B_z$, the larger the frequency separation between $\ket{T_+}$ and $\ket{T_0}$ ($\omega_{+0}$), thus suppressing the relaxation rate between these states. On the other hand, the rate $\Gamma_{\gamma_{\perp}}(\omega_{-0})$ (red curve) increases as a function of $B_z$. This happens because the frequency separation between $\ket{T_-}$ and $\ket{T_0}$ decreases when we increase the magnetic field. Similarly to the case of charge noise, we see that the only rate that does not change as a function of the magnetic field is $\Gamma_{\gamma_{\parallel}}(\omega=0)$. This rate does not give rise to an extra relaxation process, as it does not couple two different levels; thus it only causes decoherence. 

In Fig.~\ref{fig4}(c) we plot the same rates as a function of $B_z$ assuming $\Delta\omega_\mu=\gamma h B_z$. For this case, the maximum of the spectral noise density is achieved when the spin-defect frequency transition $\omega_{-0}$ matches the energy splitting associated to the magnetic moments causing the noise, i.e., $\omega_{-0}=\Delta\omega_\mu$. This can be understood from the point of view of energy conservation, implying that the maximum relaxation rate happens when we have a relaxation of the spin-defect states ($\ket{T_-}\rightarrow \ket{T_0}$), followed by the excitation of the two-level systems defined by the spin-$1/2$ magnetic moments. This behavior was already seen experimentally, with magnetic spin noise given in terms of spin-$1/2$ nitrogen impurities~\cite{PhysRevLett.108.197601}. 

For both cases discussed above, we  plot the relaxation times $1/T_{1\pm}$ [Eqs.~(\ref{t1p}) and (\ref{t1m}), gray and black solid lines, respectively]. Similarly to the charge noise case, we  observe a non-monotonic behavior of $1/T_{1+}$. Here, however, the maximum of the rates occurs for $B_z$ such that $\omega_{+0}=\Delta \omega_\mu$.  Finally, we also stress that the experimental $B_z$-dependence of $1/T_1$ shows the same trend as the results developed in Ref.~\cite{PhysRevLett.108.197601}.

\section{Comparison with Experimental data}
\label{seciv}

In this section, we analyze and interpret the depth dependence of the decoherence times of NV-centers measured within the experimental work of Ref.~\cite{electric-magnetic3}. Using our theory developed in the previous sections, we analyze these experimental results and provide possible explanations for the different depth dependence originating from the treatment of different samples. 

\begin{figure*}
  \includegraphics[width=\textwidth]{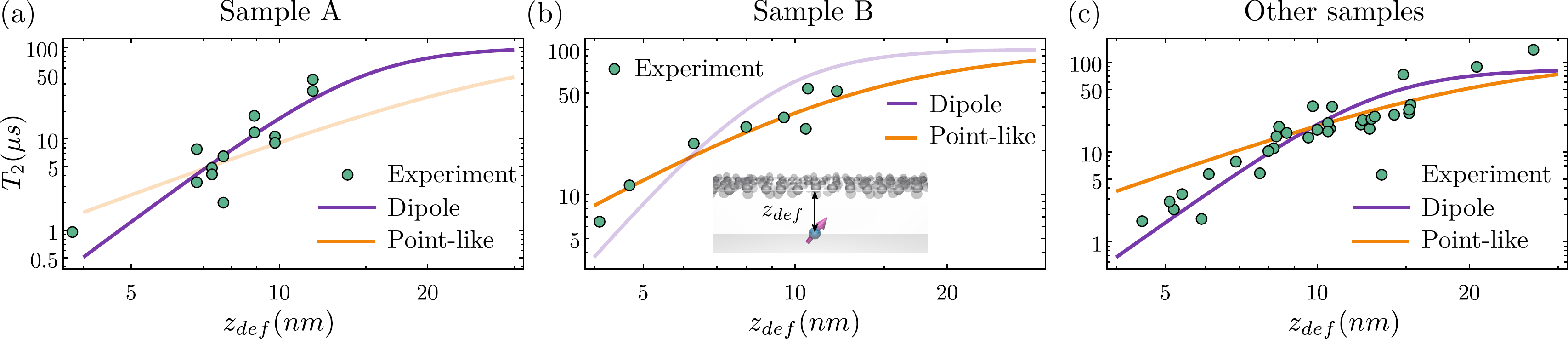}
  \caption{Comparison between theoretical and experimental $T_{2}$ decoherence time of NV-centers as a function of the depth for different samples of Ref.~\cite{electric-magnetic3}. (a) Sample A presents a $T_2$ depth-dependence that is better explained with dipole fluctuations (magnetic or electric), i.e., $1/T_2 \propto 1/z_{def}^{4}$ [Eqs.~(\ref{dip-fluc1})--(\ref{dip-fluc3}) and (\ref{Sbx})--(\ref{Sbz})]. (b) Sample B, which has been treated with high-temperature annealing and oxygen annealing, shows $T_2$ follows a depth-dependence that better explained with point-like fluctuations, i.e., $1/T_2 \propto 1/z_{def}^{2}$ [Eqs.~(\ref{point-like1})--(\ref{point-like3}) and (\ref{Sbmu-chargedvel})] (c) Other samples that were not annealed present $T_2$ depth-dependence better explained by the dipole fluctuations.}
  \label{fig6}
\end{figure*}

The experimental data of the NV-center decoherence time as a function of depth ($z_{def}$) is plotted in Fig.~\ref{fig6}. Figs.~\ref{fig6}(a) and (b), contain the decoherence time for samples A and B, respectively, while Fig.~\ref{fig6}(c) shows the decoherence for other samples. Many samples of Ref.~\cite{electric-magnetic3}, including sample A, present a depth dependence that is consistent with dipole fluctuations (e.g., fluctuating dipole charges or magnetic moments) [See Fig.~\ref{fig6}(a)]. This yields a $T_2 \propto z_{def}^{4}$ dependence as can be seen from Eqs.~(\ref{dip-fluc1})--(\ref{dip-fluc3}) and (\ref{Sbx})--(\ref{Sbz}). On the other hand, we see that for sample B, the decoherence time follows better the depth dependence related to fluctuation of point-like charges, described by Eqs.~(\ref{point-like1})--(\ref{point-like3}) and (\ref{Sbmu-chargedvel}), and $T_2 \propto z_{def}^{2}$. Furthermore, we also see that sample B is the one with the longest coherence time for shallow NV-centers. Unlike sample A [Fig.~\ref{fig6}(a)] and the other samples of Fig.~\ref{fig6}(c), sample B is the only one that was subjected to high temperature annealing followed by oxygen annealing (even though it was was part of the same initial crystal as sample A). Accordingly, through the comparison between data of sample A and sample B, we understand that the roughness of the crystal surface tends to produce fluctuating dipole-like fields, which are the dominant source of noise. As sample B had its surface treated, this type of noise was suppressed, leaving only the contribution of point-like fluctuating noise due to either surface-confined electron or hole gases, or electrons at the conduction band excited during the laser readout or initialization.

\section{Conclusion}

In this work, we first present a complete theory for the decoherence and relaxation of qutrit (spin-1) spin centers with $C_{3v}$ point-group symmetry. Accordingly, we obtain all the Lindblad operators arising from both charge and magnetic noise, and from that, we calculate the associated decoherence and relaxation times. We further present the relaxation dynamics for both charge and magnetic noise dominances. In the second part, we develop a microscopic theory for the charge noise arising from both point-like and dipole fluctuating charges, as well as for the magnetic noise arising from fluctuation of the magnetic moment and from randomness of the movement of charged particles. Using this quantitative theory, we study the evolution of the rates associated with both charge and magnetic noises as a function of the energy separation between the defect energy levels, which is produced by a finite magnetic field along the defect main symmetry axis. Finally, we use the theory developed in this work to analyze the depth dependence of decoherence times for samples with different treatments.


\begin{acknowledgments}
We thank  C. P. Anderson, D. D. Awschalom,  A. Bourassa,  P. E. Faria  G. D. Fuchs, M. Fukami, N. P. de Leon,  P. Mauer, L. V. Rodgers, S. Sangtawesin, and U. Zvi for useful discussions. This research was supported as part of the Center for Molecular Quantum Transduction (CMQT), an Energy Frontier Research Center funded by the U.S. Department of Energy, Office of Science, Basic Energy Sciences under Award No. DE-SC0021314. 
\end{acknowledgments}

\appendix

\nocite{*}

\bibliography{apssamp}

\providecommand{\noopsort}[1]{}\providecommand{\singleletter}[1]{#1}%
\begin{thebibliography}{130}%
\makeatletter
\providecommand \@ifxundefined [1]{%
 \@ifx{#1\undefined}
}%
\providecommand \@ifnum [1]{%
 \ifnum #1\expandafter \@firstoftwo
 \else \expandafter \@secondoftwo
 \fi
}%
\providecommand \@ifx [1]{%
 \ifx #1\expandafter \@firstoftwo
 \else \expandafter \@secondoftwo
 \fi
}%
\providecommand \natexlab [1]{#1}%
\providecommand \enquote  [1]{``#1''}%
\providecommand \bibnamefont  [1]{#1}%
\providecommand \bibfnamefont [1]{#1}%
\providecommand \citenamefont [1]{#1}%
\providecommand \href@noop [0]{\@secondoftwo}%
\providecommand \href [0]{\begingroup \@sanitize@url \@href}%
\providecommand \@href[1]{\@@startlink{#1}\@@href}%
\providecommand \@@href[1]{\endgroup#1\@@endlink}%
\providecommand \@sanitize@url [0]{\catcode `\\12\catcode `\$12\catcode
  `\&12\catcode `\#12\catcode `\^12\catcode `\_12\catcode `\%12\relax}%
\providecommand \@@startlink[1]{}%
\providecommand \@@endlink[0]{}%
\providecommand \url  [0]{\begingroup\@sanitize@url \@url }%
\providecommand \@url [1]{\endgroup\@href {#1}{\urlprefix }}%
\providecommand \urlprefix  [0]{URL }%
\providecommand \Eprint [0]{\href }%
\providecommand \doibase [0]{https://doi.org/}%
\providecommand \selectlanguage [0]{\@gobble}%
\providecommand \bibinfo  [0]{\@secondoftwo}%
\providecommand \bibfield  [0]{\@secondoftwo}%
\providecommand \translation [1]{[#1]}%
\providecommand \BibitemOpen [0]{}%
\providecommand \bibitemStop [0]{}%
\providecommand \bibitemNoStop [0]{.\EOS\space}%
\providecommand \EOS [0]{\spacefactor3000\relax}%
\providecommand \BibitemShut  [1]{\csname bibitem#1\endcsname}%
\let\auto@bib@innerbib\@empty
\bibitem [{\citenamefont {Taylor}\ \emph {et~al.}(2008)\citenamefont {Taylor},
  \citenamefont {Cappellaro}, \citenamefont {Childress}, \citenamefont {Jiang},
  \citenamefont {Budker}, \citenamefont {Hemmer}, \citenamefont {Yacoby},
  \citenamefont {Walsworth},\ and\ \citenamefont {Lukin}}]{taylor2008high}%
  \BibitemOpen
  \bibfield  {author} {\bibinfo {author} {\bibfnamefont {J.~M.}\ \bibnamefont
  {Taylor}}, \bibinfo {author} {\bibfnamefont {P.}~\bibnamefont {Cappellaro}},
  \bibinfo {author} {\bibfnamefont {L.}~\bibnamefont {Childress}}, \bibinfo
  {author} {\bibfnamefont {L.}~\bibnamefont {Jiang}}, \bibinfo {author}
  {\bibfnamefont {D.}~\bibnamefont {Budker}}, \bibinfo {author} {\bibfnamefont
  {P.~R.}\ \bibnamefont {Hemmer}}, \bibinfo {author} {\bibfnamefont
  {A.}~\bibnamefont {Yacoby}}, \bibinfo {author} {\bibfnamefont
  {R.}~\bibnamefont {Walsworth}},\ and\ \bibinfo {author} {\bibfnamefont
  {M.~D.}\ \bibnamefont {Lukin}},\ }\bibfield  {title} {\bibinfo {title}
  {High-sensitivity diamond magnetometer with nanoscale resolution},\
  }\href@noop {} {\bibfield  {journal} {\bibinfo  {journal} {Nat. Phys.}\
  }\textbf {\bibinfo {volume} {4}},\ \bibinfo {pages} {810} (\bibinfo {year}
  {2008})}\BibitemShut {NoStop}%
\bibitem [{\citenamefont {Dolde}\ \emph {et~al.}(2011)\citenamefont {Dolde}
  \emph {et~al.}}]{electric-magnetic1}%
  \BibitemOpen
  \bibfield  {author} {\bibinfo {author} {\bibfnamefont {F.}~\bibnamefont
  {Dolde}} \emph {et~al.},\ }\bibfield  {title} {\bibinfo {title}
  {Electric-field sensing using single diamond spins},\ }\href@noop {}
  {\bibfield  {journal} {\bibinfo  {journal} {Nature Physics}\ }\textbf
  {\bibinfo {volume} {7}},\ \bibinfo {pages} {459} (\bibinfo {year}
  {2011})}\BibitemShut {NoStop}%
\bibitem [{\citenamefont {Dolde}\ \emph
  {et~al.}(2014{\natexlab{a}})\citenamefont {Dolde}, \citenamefont {Doherty},
  \citenamefont {Michl}, \citenamefont {Jakobi}, \citenamefont {Naydenov},
  \citenamefont {Pezzagna}, \citenamefont {Meijer}, \citenamefont {Neumann},
  \citenamefont {Jelezko}, \citenamefont {Manson},\ and\ \citenamefont
  {Wrachtrup}}]{dolde2014}%
  \BibitemOpen
  \bibfield  {author} {\bibinfo {author} {\bibfnamefont {F.}~\bibnamefont
  {Dolde}}, \bibinfo {author} {\bibfnamefont {M.~W.}\ \bibnamefont {Doherty}},
  \bibinfo {author} {\bibfnamefont {J.}~\bibnamefont {Michl}}, \bibinfo
  {author} {\bibfnamefont {I.}~\bibnamefont {Jakobi}}, \bibinfo {author}
  {\bibfnamefont {B.}~\bibnamefont {Naydenov}}, \bibinfo {author}
  {\bibfnamefont {S.}~\bibnamefont {Pezzagna}}, \bibinfo {author}
  {\bibfnamefont {J.}~\bibnamefont {Meijer}}, \bibinfo {author} {\bibfnamefont
  {P.}~\bibnamefont {Neumann}}, \bibinfo {author} {\bibfnamefont
  {F.}~\bibnamefont {Jelezko}}, \bibinfo {author} {\bibfnamefont {N.~B.}\
  \bibnamefont {Manson}},\ and\ \bibinfo {author} {\bibfnamefont
  {J.}~\bibnamefont {Wrachtrup}},\ }\bibfield  {title} {\bibinfo {title}
  {Nanoscale detection of a single fundamental charge in ambient conditions
  using the $\mathrm{NV}{}^{\ensuremath{-}}$ center in diamond},\ }\href@noop
  {} {\bibfield  {journal} {\bibinfo  {journal} {Phys. Rev. Lett.}\ }\textbf
  {\bibinfo {volume} {112}},\ \bibinfo {pages} {097603} (\bibinfo {year}
  {2014}{\natexlab{a}})}\BibitemShut {NoStop}%
\bibitem [{\citenamefont {Schirhagl}\ \emph {et~al.}(2014)\citenamefont
  {Schirhagl}, \citenamefont {Chang}, \citenamefont {Loretz},\ and\
  \citenamefont {Degen}}]{schirhagl2014nitrogen}%
  \BibitemOpen
  \bibfield  {author} {\bibinfo {author} {\bibfnamefont {R.}~\bibnamefont
  {Schirhagl}}, \bibinfo {author} {\bibfnamefont {K.}~\bibnamefont {Chang}},
  \bibinfo {author} {\bibfnamefont {M.}~\bibnamefont {Loretz}},\ and\ \bibinfo
  {author} {\bibfnamefont {C.~L.}\ \bibnamefont {Degen}},\ }\bibfield  {title}
  {\bibinfo {title} {Nitrogen-vacancy centers in diamond: Nanoscale sensors for
  physics and biology},\ }\href
  {https://doi.org/10.1146/annurev-physchem-040513-103659} {\bibfield
  {journal} {\bibinfo  {journal} {Annual Review of Physical Chemistry}\
  }\textbf {\bibinfo {volume} {65}},\ \bibinfo {pages} {83} (\bibinfo {year}
  {2014})}\BibitemShut {NoStop}%
\bibitem [{\citenamefont {van~der Sar}\ \emph {et~al.}(2015)\citenamefont
  {van~der Sar}, \citenamefont {Casola}, \citenamefont {Walsworth},\ and\
  \citenamefont {Yacoby}}]{van2015nanometre}%
  \BibitemOpen
  \bibfield  {author} {\bibinfo {author} {\bibfnamefont {T.}~\bibnamefont
  {van~der Sar}}, \bibinfo {author} {\bibfnamefont {F.}~\bibnamefont {Casola}},
  \bibinfo {author} {\bibfnamefont {R.}~\bibnamefont {Walsworth}},\ and\
  \bibinfo {author} {\bibfnamefont {A.}~\bibnamefont {Yacoby}},\ }\bibfield
  {title} {\bibinfo {title} {Nanometre-scale probing of spin waves using single
  electron spins},\ }\href@noop {} {\bibfield  {journal} {\bibinfo  {journal}
  {Nat. Commun.}\ }\textbf {\bibinfo {volume} {6}},\ \bibinfo {pages} {7886}
  (\bibinfo {year} {2015})}\BibitemShut {NoStop}%
\bibitem [{\citenamefont {Degen}\ \emph
  {et~al.}(2017{\natexlab{a}})\citenamefont {Degen}, \citenamefont {Reinhard},\
  and\ \citenamefont {Cappellaro}}]{degen2017quantum}%
  \BibitemOpen
  \bibfield  {author} {\bibinfo {author} {\bibfnamefont {C.~L.}\ \bibnamefont
  {Degen}}, \bibinfo {author} {\bibfnamefont {F.}~\bibnamefont {Reinhard}},\
  and\ \bibinfo {author} {\bibfnamefont {P.}~\bibnamefont {Cappellaro}},\
  }\bibfield  {title} {\bibinfo {title} {Quantum sensing},\ }\href
  {https://doi.org/10.1103/RevModPhys.89.035002} {\bibfield  {journal}
  {\bibinfo  {journal} {Rev. Mod. Phys.}\ }\textbf {\bibinfo {volume} {89}},\
  \bibinfo {pages} {035002} (\bibinfo {year} {2017}{\natexlab{a}})}\BibitemShut
  {NoStop}%
\bibitem [{\citenamefont {Flebus}\ and\ \citenamefont
  {Tserkovnyak}(2018)}]{flebus2018quantum}%
  \BibitemOpen
  \bibfield  {author} {\bibinfo {author} {\bibfnamefont {B.}~\bibnamefont
  {Flebus}}\ and\ \bibinfo {author} {\bibfnamefont {Y.}~\bibnamefont
  {Tserkovnyak}},\ }\bibfield  {title} {\bibinfo {title} {Quantum-impurity
  relaxometry of magnetization dynamics},\ }\href
  {https://doi.org/10.1103/PhysRevLett.121.187204} {\bibfield  {journal}
  {\bibinfo  {journal} {Phys. Rev. Lett.}\ }\textbf {\bibinfo {volume} {121}},\
  \bibinfo {pages} {187204} (\bibinfo {year} {2018})}\BibitemShut {NoStop}%
\bibitem [{\citenamefont {Casola}\ \emph {et~al.}(2018)\citenamefont {Casola},
  \citenamefont {van~der Sar},\ and\ \citenamefont
  {Yacoby}}]{casola2018probing}%
  \BibitemOpen
  \bibfield  {author} {\bibinfo {author} {\bibfnamefont {F.}~\bibnamefont
  {Casola}}, \bibinfo {author} {\bibfnamefont {T.}~\bibnamefont {van~der
  Sar}},\ and\ \bibinfo {author} {\bibfnamefont {A.}~\bibnamefont {Yacoby}},\
  }\bibfield  {title} {\bibinfo {title} {Probing condensed matter physics with
  magnetometry based on nitrogen-vacancy centres in diamond},\ }\href@noop {}
  {\bibfield  {journal} {\bibinfo  {journal} {Nat. Rev. Mater.}\ }\textbf
  {\bibinfo {volume} {3}},\ \bibinfo {pages} {17088} (\bibinfo {year}
  {2018})}\BibitemShut {NoStop}%
\bibitem [{\citenamefont {Mittiga}\ \emph
  {et~al.}(2018{\natexlab{a}})\citenamefont {Mittiga}, \citenamefont {Hsieh},
  \citenamefont {Zu}, \citenamefont {Kobrin}, \citenamefont {Machado},
  \citenamefont {Bhattacharyya}, \citenamefont {Rui}, \citenamefont {Jarmola},
  \citenamefont {Choi}, \citenamefont {Budker} \emph
  {et~al.}}]{mittiga2018imaging}%
  \BibitemOpen
  \bibfield  {author} {\bibinfo {author} {\bibfnamefont {T.}~\bibnamefont
  {Mittiga}}, \bibinfo {author} {\bibfnamefont {S.}~\bibnamefont {Hsieh}},
  \bibinfo {author} {\bibfnamefont {C.}~\bibnamefont {Zu}}, \bibinfo {author}
  {\bibfnamefont {B.}~\bibnamefont {Kobrin}}, \bibinfo {author} {\bibfnamefont
  {F.}~\bibnamefont {Machado}}, \bibinfo {author} {\bibfnamefont
  {P.}~\bibnamefont {Bhattacharyya}}, \bibinfo {author} {\bibfnamefont {N.~Z.}\
  \bibnamefont {Rui}}, \bibinfo {author} {\bibfnamefont {A.}~\bibnamefont
  {Jarmola}}, \bibinfo {author} {\bibfnamefont {S.}~\bibnamefont {Choi}},
  \bibinfo {author} {\bibfnamefont {D.}~\bibnamefont {Budker}}, \emph
  {et~al.},\ }\bibfield  {title} {\bibinfo {title} {Imaging the local charge
  environment of nitrogen-vacancy centers in diamond},\ }\href
  {https://doi.org/10.1103/PhysRevLett.121.246402} {\bibfield  {journal}
  {\bibinfo  {journal} {Phys. Rev. Lett.}\ }\textbf {\bibinfo {volume} {121}},\
  \bibinfo {pages} {246402} (\bibinfo {year} {2018}{\natexlab{a}})}\BibitemShut
  {NoStop}%
\bibitem [{\citenamefont {Zhou}\ \emph {et~al.}(2020)\citenamefont {Zhou},
  \citenamefont {Jerger}, \citenamefont {Lee}, \citenamefont {Fukami},
  \citenamefont {Mujid}, \citenamefont {Park},\ and\ \citenamefont
  {Awschalom}}]{PhysRevX.10.011003}%
  \BibitemOpen
  \bibfield  {author} {\bibinfo {author} {\bibfnamefont {B.~B.}\ \bibnamefont
  {Zhou}}, \bibinfo {author} {\bibfnamefont {P.~C.}\ \bibnamefont {Jerger}},
  \bibinfo {author} {\bibfnamefont {K.-H.}\ \bibnamefont {Lee}}, \bibinfo
  {author} {\bibfnamefont {M.}~\bibnamefont {Fukami}}, \bibinfo {author}
  {\bibfnamefont {F.}~\bibnamefont {Mujid}}, \bibinfo {author} {\bibfnamefont
  {J.}~\bibnamefont {Park}},\ and\ \bibinfo {author} {\bibfnamefont {D.~D.}\
  \bibnamefont {Awschalom}},\ }\bibfield  {title} {\bibinfo {title}
  {Spatiotemporal mapping of a photocurrent vortex in monolayer
  $\mathrm{MoS}_{2}$ using diamond quantum sensors},\ }\href@noop {} {\bibfield
   {journal} {\bibinfo  {journal} {Phys. Rev. X}\ }\textbf {\bibinfo {volume}
  {10}},\ \bibinfo {pages} {011003} (\bibinfo {year} {2020})}\BibitemShut
  {NoStop}%
\bibitem [{\citenamefont {Lee-Wong}\ \emph {et~al.}(2020)\citenamefont
  {Lee-Wong}, \citenamefont {Xue}, \citenamefont {Ye}, \citenamefont {Kreisel},
  \citenamefont {van~der Sar}, \citenamefont {Yacoby},\ and\ \citenamefont
  {Du}}]{lee2020nanoscale}%
  \BibitemOpen
  \bibfield  {author} {\bibinfo {author} {\bibfnamefont {E.}~\bibnamefont
  {Lee-Wong}}, \bibinfo {author} {\bibfnamefont {R.}~\bibnamefont {Xue}},
  \bibinfo {author} {\bibfnamefont {F.}~\bibnamefont {Ye}}, \bibinfo {author}
  {\bibfnamefont {A.}~\bibnamefont {Kreisel}}, \bibinfo {author} {\bibfnamefont
  {T.}~\bibnamefont {van~der Sar}}, \bibinfo {author} {\bibfnamefont
  {A.}~\bibnamefont {Yacoby}},\ and\ \bibinfo {author} {\bibfnamefont {C.~R.}\
  \bibnamefont {Du}},\ }\bibfield  {title} {\bibinfo {title} {Nanoscale
  detection of magnon excitations with variable wavevectors through a quantum
  spin sensor},\ }\href@noop {} {\bibfield  {journal} {\bibinfo  {journal}
  {Nano Letters}\ }\textbf {\bibinfo {volume} {20}},\ \bibinfo {pages} {3284}
  (\bibinfo {year} {2020})}\BibitemShut {NoStop}%
\bibitem [{\citenamefont {Rustagi}\ \emph {et~al.}(2020)\citenamefont
  {Rustagi}, \citenamefont {Bertelli}, \citenamefont {van~der Sar},\ and\
  \citenamefont {Upadhyaya}}]{rustagi2020}%
  \BibitemOpen
  \bibfield  {author} {\bibinfo {author} {\bibfnamefont {A.}~\bibnamefont
  {Rustagi}}, \bibinfo {author} {\bibfnamefont {I.}~\bibnamefont {Bertelli}},
  \bibinfo {author} {\bibfnamefont {T.}~\bibnamefont {van~der Sar}},\ and\
  \bibinfo {author} {\bibfnamefont {P.}~\bibnamefont {Upadhyaya}},\ }\bibfield
  {title} {\bibinfo {title} {Sensing chiral magnetic noise via quantum impurity
  relaxometry},\ }\href {https://doi.org/10.1103/PhysRevB.102.220403}
  {\bibfield  {journal} {\bibinfo  {journal} {Phys. Rev. B}\ }\textbf {\bibinfo
  {volume} {102}},\ \bibinfo {pages} {220403(R)} (\bibinfo {year}
  {2020})}\BibitemShut {NoStop}%
\bibitem [{\citenamefont {Candido}\ and\ \citenamefont
  {Flatt{\'e}}(2021)}]{candido2021theory}%
  \BibitemOpen
  \bibfield  {author} {\bibinfo {author} {\bibfnamefont {D.~R.}\ \bibnamefont
  {Candido}}\ and\ \bibinfo {author} {\bibfnamefont {M.~E.}\ \bibnamefont
  {Flatt{\'e}}},\ }\bibfield  {title} {\bibinfo {title} {Theory of spin center
  sensing of diffusion},\ }\href@noop {} {\bibfield  {journal} {\bibinfo
  {journal} {arXiv preprint arXiv:2112.15581}\ } (\bibinfo {year}
  {2021})}\BibitemShut {NoStop}%
\bibitem [{\citenamefont {Tamarat}\ \emph {et~al.}(2006)\citenamefont
  {Tamarat}, \citenamefont {Gaebel}, \citenamefont {Rabeau}, \citenamefont
  {Khan}, \citenamefont {Greentree}, \citenamefont {Wilson}, \citenamefont
  {Hollenberg}, \citenamefont {Prawer}, \citenamefont {Hemmer}, \citenamefont
  {Jelezko},\ and\ \citenamefont {Wrachtrup}}]{tamarat2006stark}%
  \BibitemOpen
  \bibfield  {author} {\bibinfo {author} {\bibfnamefont {P.}~\bibnamefont
  {Tamarat}}, \bibinfo {author} {\bibfnamefont {T.}~\bibnamefont {Gaebel}},
  \bibinfo {author} {\bibfnamefont {J.~R.}\ \bibnamefont {Rabeau}}, \bibinfo
  {author} {\bibfnamefont {M.}~\bibnamefont {Khan}}, \bibinfo {author}
  {\bibfnamefont {A.~D.}\ \bibnamefont {Greentree}}, \bibinfo {author}
  {\bibfnamefont {H.}~\bibnamefont {Wilson}}, \bibinfo {author} {\bibfnamefont
  {L.~C.~L.}\ \bibnamefont {Hollenberg}}, \bibinfo {author} {\bibfnamefont
  {S.}~\bibnamefont {Prawer}}, \bibinfo {author} {\bibfnamefont
  {P.}~\bibnamefont {Hemmer}}, \bibinfo {author} {\bibfnamefont
  {F.}~\bibnamefont {Jelezko}},\ and\ \bibinfo {author} {\bibfnamefont
  {J.}~\bibnamefont {Wrachtrup}},\ }\bibfield  {title} {\bibinfo {title} {Stark
  shift control of single optical centers in diamond},\ }\href@noop {}
  {\bibfield  {journal} {\bibinfo  {journal} {Phys. Rev. Lett.}\ }\textbf
  {\bibinfo {volume} {97}},\ \bibinfo {pages} {083002} (\bibinfo {year}
  {2006})}\BibitemShut {NoStop}%
\bibitem [{\citenamefont {Anderson}\ \emph {et~al.}(2019)\citenamefont
  {Anderson}, \citenamefont {Bourassa}, \citenamefont {Miao}, \citenamefont
  {Wolfowicz}, \citenamefont {Mintun}, \citenamefont {Crook}, \citenamefont
  {Abe}, \citenamefont {Ul~Hassan}, \citenamefont {Son}, \citenamefont
  {Ohshima},\ and\ \citenamefont {Awschalom}}]{anderson2019electrical}%
  \BibitemOpen
  \bibfield  {author} {\bibinfo {author} {\bibfnamefont {C.~P.}\ \bibnamefont
  {Anderson}}, \bibinfo {author} {\bibfnamefont {A.}~\bibnamefont {Bourassa}},
  \bibinfo {author} {\bibfnamefont {K.~C.}\ \bibnamefont {Miao}}, \bibinfo
  {author} {\bibfnamefont {G.}~\bibnamefont {Wolfowicz}}, \bibinfo {author}
  {\bibfnamefont {P.~J.}\ \bibnamefont {Mintun}}, \bibinfo {author}
  {\bibfnamefont {A.~L.}\ \bibnamefont {Crook}}, \bibinfo {author}
  {\bibfnamefont {H.}~\bibnamefont {Abe}}, \bibinfo {author} {\bibfnamefont
  {J.}~\bibnamefont {Ul~Hassan}}, \bibinfo {author} {\bibfnamefont {N.~T.}\
  \bibnamefont {Son}}, \bibinfo {author} {\bibfnamefont {T.}~\bibnamefont
  {Ohshima}},\ and\ \bibinfo {author} {\bibfnamefont {D.~D.}\ \bibnamefont
  {Awschalom}},\ }\bibfield  {title} {\bibinfo {title} {Electrical and optical
  control of single spins integrated in scalable semiconductor devices},\
  }\href@noop {} {\bibfield  {journal} {\bibinfo  {journal} {Science}\ }\textbf
  {\bibinfo {volume} {366}},\ \bibinfo {pages} {1225} (\bibinfo {year}
  {2019})}\BibitemShut {NoStop}%
\bibitem [{\citenamefont {de~las Casas}\ \emph {et~al.}(2017)\citenamefont
  {de~las Casas}, \citenamefont {Christle}, \citenamefont {Ul~Hassan},
  \citenamefont {Ohshima}, \citenamefont {Son},\ and\ \citenamefont
  {Awschalom}}]{de2017stark}%
  \BibitemOpen
  \bibfield  {author} {\bibinfo {author} {\bibfnamefont {C.~F.}\ \bibnamefont
  {de~las Casas}}, \bibinfo {author} {\bibfnamefont {D.~J.}\ \bibnamefont
  {Christle}}, \bibinfo {author} {\bibfnamefont {J.}~\bibnamefont {Ul~Hassan}},
  \bibinfo {author} {\bibfnamefont {T.}~\bibnamefont {Ohshima}}, \bibinfo
  {author} {\bibfnamefont {N.~T.}\ \bibnamefont {Son}},\ and\ \bibinfo {author}
  {\bibfnamefont {D.~D.}\ \bibnamefont {Awschalom}},\ }\bibfield  {title}
  {\bibinfo {title} {Stark tuning and electrical charge state control of single
  divacancies in silicon carbide},\ }\href@noop {} {\bibfield  {journal}
  {\bibinfo  {journal} {Applied Physics Letters}\ }\textbf {\bibinfo {volume}
  {111}},\ \bibinfo {pages} {262403} (\bibinfo {year} {2017})}\BibitemShut
  {NoStop}%
\bibitem [{\citenamefont {Trifunovic}\ \emph {et~al.}(2013)\citenamefont
  {Trifunovic}, \citenamefont {Pedrocchi},\ and\ \citenamefont
  {Loss}}]{lukaprx}%
  \BibitemOpen
  \bibfield  {author} {\bibinfo {author} {\bibfnamefont {L.}~\bibnamefont
  {Trifunovic}}, \bibinfo {author} {\bibfnamefont {F.~L.}\ \bibnamefont
  {Pedrocchi}},\ and\ \bibinfo {author} {\bibfnamefont {D.}~\bibnamefont
  {Loss}},\ }\bibfield  {title} {\bibinfo {title} {Long-distance entanglement
  of spin qubits via ferromagnet},\ }\href
  {https://doi.org/10.1103/PhysRevX.3.041023} {\bibfield  {journal} {\bibinfo
  {journal} {Phys. Rev. X}\ }\textbf {\bibinfo {volume} {3}},\ \bibinfo {pages}
  {041023} (\bibinfo {year} {2013})}\BibitemShut {NoStop}%
\bibitem [{\citenamefont {Li}\ \emph {et~al.}(2015)\citenamefont {Li},
  \citenamefont {Liu}, \citenamefont {Gao}, \citenamefont {Xiang},
  \citenamefont {Rabl}, \citenamefont {Xiao},\ and\ \citenamefont
  {Li}}]{li2015hybrid}%
  \BibitemOpen
  \bibfield  {author} {\bibinfo {author} {\bibfnamefont {P.-B.}\ \bibnamefont
  {Li}}, \bibinfo {author} {\bibfnamefont {Y.-C.}\ \bibnamefont {Liu}},
  \bibinfo {author} {\bibfnamefont {S.-Y.}\ \bibnamefont {Gao}}, \bibinfo
  {author} {\bibfnamefont {Z.-L.}\ \bibnamefont {Xiang}}, \bibinfo {author}
  {\bibfnamefont {P.}~\bibnamefont {Rabl}}, \bibinfo {author} {\bibfnamefont
  {Y.-F.}\ \bibnamefont {Xiao}},\ and\ \bibinfo {author} {\bibfnamefont
  {F.-L.}\ \bibnamefont {Li}},\ }\bibfield  {title} {\bibinfo {title} {Hybrid
  quantum device based on {NV} centers in diamond nanomechanical resonators
  plus superconducting waveguide cavities},\ }\href
  {https://doi.org/10.1103/PhysRevApplied.4.044003} {\bibfield  {journal}
  {\bibinfo  {journal} {Phys. Rev. Applied}\ }\textbf {\bibinfo {volume} {4}},\
  \bibinfo {pages} {044003} (\bibinfo {year} {2015})}\BibitemShut {NoStop}%
\bibitem [{\citenamefont {Li}\ \emph {et~al.}(2016)\citenamefont {Li},
  \citenamefont {Xiang}, \citenamefont {Rabl},\ and\ \citenamefont
  {Nori}}]{li2016hybrid}%
  \BibitemOpen
  \bibfield  {author} {\bibinfo {author} {\bibfnamefont {P.-B.}\ \bibnamefont
  {Li}}, \bibinfo {author} {\bibfnamefont {Z.-L.}\ \bibnamefont {Xiang}},
  \bibinfo {author} {\bibfnamefont {P.}~\bibnamefont {Rabl}},\ and\ \bibinfo
  {author} {\bibfnamefont {F.}~\bibnamefont {Nori}},\ }\bibfield  {title}
  {\bibinfo {title} {Hybrid quantum device with nitrogen-vacancy centers in
  diamond coupled to carbon nanotubes},\ }\href
  {https://doi.org/10.1103/PhysRevLett.117.015502} {\bibfield  {journal}
  {\bibinfo  {journal} {Phys. Rev. Lett.}\ }\textbf {\bibinfo {volume} {117}},\
  \bibinfo {pages} {015502} (\bibinfo {year} {2016})}\BibitemShut {NoStop}%
\bibitem [{\citenamefont {Andrich}\ \emph {et~al.}(2017)\citenamefont
  {Andrich}, \citenamefont {de~las Casas}, \citenamefont {Liu}, \citenamefont
  {Bretscher}, \citenamefont {Berman}, \citenamefont {Heremans}, \citenamefont
  {Nealey},\ and\ \citenamefont {Awschalom}}]{andrich2017long}%
  \BibitemOpen
  \bibfield  {author} {\bibinfo {author} {\bibfnamefont {P.}~\bibnamefont
  {Andrich}}, \bibinfo {author} {\bibfnamefont {C.~F.}\ \bibnamefont {de~las
  Casas}}, \bibinfo {author} {\bibfnamefont {X.}~\bibnamefont {Liu}}, \bibinfo
  {author} {\bibfnamefont {H.~L.}\ \bibnamefont {Bretscher}}, \bibinfo {author}
  {\bibfnamefont {J.~R.}\ \bibnamefont {Berman}}, \bibinfo {author}
  {\bibfnamefont {F.~J.}\ \bibnamefont {Heremans}}, \bibinfo {author}
  {\bibfnamefont {P.~F.}\ \bibnamefont {Nealey}},\ and\ \bibinfo {author}
  {\bibfnamefont {D.~D.}\ \bibnamefont {Awschalom}},\ }\bibfield  {title}
  {\bibinfo {title} {Long-range spin wave mediated control of defect qubits in
  nanodiamonds},\ }\href@noop {} {\bibfield  {journal} {\bibinfo  {journal}
  {npj Quantum Inf}\ }\textbf {\bibinfo {volume} {3}},\ \bibinfo {pages} {28}
  (\bibinfo {year} {2017})}\BibitemShut {NoStop}%
\bibitem [{\citenamefont {Lemonde}\ \emph {et~al.}(2018)\citenamefont
  {Lemonde}, \citenamefont {Meesala}, \citenamefont {Sipahigil}, \citenamefont
  {Schuetz}, \citenamefont {Lukin}, \citenamefont {Loncar},\ and\ \citenamefont
  {Rabl}}]{lemonde2018phonon}%
  \BibitemOpen
  \bibfield  {author} {\bibinfo {author} {\bibfnamefont {M.-A.}\ \bibnamefont
  {Lemonde}}, \bibinfo {author} {\bibfnamefont {S.}~\bibnamefont {Meesala}},
  \bibinfo {author} {\bibfnamefont {A.}~\bibnamefont {Sipahigil}}, \bibinfo
  {author} {\bibfnamefont {M.~J.~A.}\ \bibnamefont {Schuetz}}, \bibinfo
  {author} {\bibfnamefont {M.~D.}\ \bibnamefont {Lukin}}, \bibinfo {author}
  {\bibfnamefont {M.}~\bibnamefont {Loncar}},\ and\ \bibinfo {author}
  {\bibfnamefont {P.}~\bibnamefont {Rabl}},\ }\bibfield  {title} {\bibinfo
  {title} {Phonon networks with silicon-vacancy centers in diamond
  waveguides},\ }\href {https://doi.org/10.1103/PhysRevLett.120.213603}
  {\bibfield  {journal} {\bibinfo  {journal} {Phys. Rev. Lett.}\ }\textbf
  {\bibinfo {volume} {120}},\ \bibinfo {pages} {213603} (\bibinfo {year}
  {2018})}\BibitemShut {NoStop}%
\bibitem [{\citenamefont {Flebus}\ and\ \citenamefont
  {Tserkovnyak}(2019)}]{flebus2019entangling}%
  \BibitemOpen
  \bibfield  {author} {\bibinfo {author} {\bibfnamefont {B.}~\bibnamefont
  {Flebus}}\ and\ \bibinfo {author} {\bibfnamefont {Y.}~\bibnamefont
  {Tserkovnyak}},\ }\bibfield  {title} {\bibinfo {title} {Entangling distant
  spin qubits via a magnetic domain wall},\ }\href
  {https://doi.org/10.1103/PhysRevB.99.140403} {\bibfield  {journal} {\bibinfo
  {journal} {Phys. Rev. B}\ }\textbf {\bibinfo {volume} {99}},\ \bibinfo
  {pages} {140403} (\bibinfo {year} {2019})}\BibitemShut {NoStop}%
\bibitem [{\citenamefont {M\"uhlherr}\ \emph
  {et~al.}(2019{\natexlab{a}})\citenamefont {M\"uhlherr}, \citenamefont
  {Shkolnikov},\ and\ \citenamefont {Burkard}}]{muhlherr2019magnetic}%
  \BibitemOpen
  \bibfield  {author} {\bibinfo {author} {\bibfnamefont {C.}~\bibnamefont
  {M\"uhlherr}}, \bibinfo {author} {\bibfnamefont {V.~O.}\ \bibnamefont
  {Shkolnikov}},\ and\ \bibinfo {author} {\bibfnamefont {G.}~\bibnamefont
  {Burkard}},\ }\bibfield  {title} {\bibinfo {title} {Magnetic resonance in
  defect spins mediated by spin waves},\ }\href
  {https://doi.org/10.1103/PhysRevB.99.195413} {\bibfield  {journal} {\bibinfo
  {journal} {Phys. Rev. B}\ }\textbf {\bibinfo {volume} {99}},\ \bibinfo
  {pages} {195413} (\bibinfo {year} {2019}{\natexlab{a}})}\BibitemShut
  {NoStop}%
\bibitem [{\citenamefont {Zou}\ \emph {et~al.}(2020)\citenamefont {Zou},
  \citenamefont {Kim},\ and\ \citenamefont {Tserkovnyak}}]{zou2020tuning}%
  \BibitemOpen
  \bibfield  {author} {\bibinfo {author} {\bibfnamefont {J.}~\bibnamefont
  {Zou}}, \bibinfo {author} {\bibfnamefont {S.~K.}\ \bibnamefont {Kim}},\ and\
  \bibinfo {author} {\bibfnamefont {Y.}~\bibnamefont {Tserkovnyak}},\
  }\bibfield  {title} {\bibinfo {title} {Tuning entanglement by squeezing
  magnons in anisotropic magnets},\ }\href
  {https://doi.org/10.1103/PhysRevB.101.014416} {\bibfield  {journal} {\bibinfo
   {journal} {Phys. Rev. B}\ }\textbf {\bibinfo {volume} {101}},\ \bibinfo
  {pages} {014416} (\bibinfo {year} {2020})}\BibitemShut {NoStop}%
\bibitem [{\citenamefont {Candido}\ \emph {et~al.}(2021)\citenamefont
  {Candido}, \citenamefont {Fuchs}, \citenamefont {Johnston-Halperin},\ and\
  \citenamefont {Flatt{\'{e}}}}]{candido2020predicted}%
  \BibitemOpen
  \bibfield  {author} {\bibinfo {author} {\bibfnamefont {D.~R.}\ \bibnamefont
  {Candido}}, \bibinfo {author} {\bibfnamefont {G.~D.}\ \bibnamefont {Fuchs}},
  \bibinfo {author} {\bibfnamefont {E.}~\bibnamefont {Johnston-Halperin}},\
  and\ \bibinfo {author} {\bibfnamefont {M.~E.}\ \bibnamefont {Flatt{\'{e}}}},\
  }\bibfield  {title} {\bibinfo {title} {Predicted strong coupling of
  solid-state spins via a single magnon mode},\ }\href
  {https://doi.org/10.1088/2633-4356/ab9a55} {\bibfield  {journal} {\bibinfo
  {journal} {Mat. Quantum Technol.}\ }\textbf {\bibinfo {volume} {1}},\
  \bibinfo {pages} {011001} (\bibinfo {year} {2021})}\BibitemShut {NoStop}%
\bibitem [{\citenamefont {Neuman}\ \emph {et~al.}(2020)\citenamefont {Neuman},
  \citenamefont {Wang},\ and\ \citenamefont {Narang}}]{neumanprl2020}%
  \BibitemOpen
  \bibfield  {author} {\bibinfo {author} {\bibfnamefont {T.}~\bibnamefont
  {Neuman}}, \bibinfo {author} {\bibfnamefont {D.~S.}\ \bibnamefont {Wang}},\
  and\ \bibinfo {author} {\bibfnamefont {P.}~\bibnamefont {Narang}},\
  }\bibfield  {title} {\bibinfo {title} {Nanomagnonic cavities for strong
  spin-magnon coupling and magnon-mediated spin-spin interactions},\ }\href
  {https://doi.org/10.1103/PhysRevLett.125.247702} {\bibfield  {journal}
  {\bibinfo  {journal} {Phys. Rev. Lett.}\ }\textbf {\bibinfo {volume} {125}},\
  \bibinfo {pages} {247702} (\bibinfo {year} {2020})}\BibitemShut {NoStop}%
\bibitem [{\citenamefont {Wang}\ \emph {et~al.}(2021)\citenamefont {Wang},
  \citenamefont {Neuman},\ and\ \citenamefont
  {Narang}}]{doi:10.1021/acs.jpcc.0c11536}%
  \BibitemOpen
  \bibfield  {author} {\bibinfo {author} {\bibfnamefont {D.~S.}\ \bibnamefont
  {Wang}}, \bibinfo {author} {\bibfnamefont {T.}~\bibnamefont {Neuman}},\ and\
  \bibinfo {author} {\bibfnamefont {P.}~\bibnamefont {Narang}},\ }\bibfield
  {title} {\bibinfo {title} {Spin emitters beyond the point dipole
  approximation in nanomagnonic cavities},\ }\href
  {https://doi.org/10.1021/acs.jpcc.0c11536} {\bibfield  {journal} {\bibinfo
  {journal} {The Journal of Physical Chemistry C}\ }\textbf {\bibinfo {volume}
  {125}},\ \bibinfo {pages} {6222} (\bibinfo {year} {2021})}\BibitemShut
  {NoStop}%
\bibitem [{\citenamefont {Solanki}\ \emph {et~al.}(2020)\citenamefont
  {Solanki}, \citenamefont {Bogdanov}, \citenamefont {Rustagi}, \citenamefont
  {Dilley}, \citenamefont {Shen}, \citenamefont {Rahman}, \citenamefont {Tong},
  \citenamefont {Debashis}, \citenamefont {Chen}, \citenamefont {Appenzeller}
  \emph {et~al.}}]{solanki}%
  \BibitemOpen
  \bibfield  {author} {\bibinfo {author} {\bibfnamefont {A.~B.}\ \bibnamefont
  {Solanki}}, \bibinfo {author} {\bibfnamefont {S.~I.}\ \bibnamefont
  {Bogdanov}}, \bibinfo {author} {\bibfnamefont {A.}~\bibnamefont {Rustagi}},
  \bibinfo {author} {\bibfnamefont {N.~R.}\ \bibnamefont {Dilley}}, \bibinfo
  {author} {\bibfnamefont {T.}~\bibnamefont {Shen}}, \bibinfo {author}
  {\bibfnamefont {M.~M.}\ \bibnamefont {Rahman}}, \bibinfo {author}
  {\bibfnamefont {W.}~\bibnamefont {Tong}}, \bibinfo {author} {\bibfnamefont
  {P.}~\bibnamefont {Debashis}}, \bibinfo {author} {\bibfnamefont
  {Z.}~\bibnamefont {Chen}}, \bibinfo {author} {\bibfnamefont {J.}~\bibnamefont
  {Appenzeller}}, \emph {et~al.},\ }\bibfield  {title} {\bibinfo {title}
  {Electric field control of interaction between magnons and quantum spin
  defects},\ }\href@noop {} {\bibfield  {journal} {\bibinfo  {journal}
  {arXiv:2012.01497}\ } (\bibinfo {year} {2020})}\BibitemShut {NoStop}%
\bibitem [{\citenamefont {Fukami}\ \emph {et~al.}(2021)\citenamefont {Fukami},
  \citenamefont {Candido}, \citenamefont {Awschalom},\ and\ \citenamefont
  {Flatt\'e}}]{fukami2021}%
  \BibitemOpen
  \bibfield  {author} {\bibinfo {author} {\bibfnamefont {M.}~\bibnamefont
  {Fukami}}, \bibinfo {author} {\bibfnamefont {D.~R.}\ \bibnamefont {Candido}},
  \bibinfo {author} {\bibfnamefont {D.~D.}\ \bibnamefont {Awschalom}},\ and\
  \bibinfo {author} {\bibfnamefont {M.~E.}\ \bibnamefont {Flatt\'e}},\
  }\bibfield  {title} {\bibinfo {title} {Opportunities for long-range
  magnon-mediated entanglement of spin qubits via on- and off-resonant
  coupling},\ }\href {https://doi.org/10.1103/PRXQuantum.2.040314} {\bibfield
  {journal} {\bibinfo  {journal} {PRX Quantum}\ }\textbf {\bibinfo {volume}
  {2}},\ \bibinfo {pages} {040314} (\bibinfo {year} {2021})}\BibitemShut
  {NoStop}%
\bibitem [{\citenamefont {Jamonneau}\ \emph {et~al.}(2016)\citenamefont
  {Jamonneau}, \citenamefont {Lesik}, \citenamefont {Tetienne}, \citenamefont
  {Alvizu}, \citenamefont {Mayer}, \citenamefont {Dr\'eau}, \citenamefont
  {Kosen}, \citenamefont {Roch}, \citenamefont {Pezzagna}, \citenamefont
  {Meijer}, \citenamefont {Teraji}, \citenamefont {Kubo}, \citenamefont
  {Bertet}, \citenamefont {Maze},\ and\ \citenamefont
  {Jacques}}]{electric-magnetic2}%
  \BibitemOpen
  \bibfield  {author} {\bibinfo {author} {\bibfnamefont {P.}~\bibnamefont
  {Jamonneau}}, \bibinfo {author} {\bibfnamefont {M.}~\bibnamefont {Lesik}},
  \bibinfo {author} {\bibfnamefont {J.~P.}\ \bibnamefont {Tetienne}}, \bibinfo
  {author} {\bibfnamefont {I.}~\bibnamefont {Alvizu}}, \bibinfo {author}
  {\bibfnamefont {L.}~\bibnamefont {Mayer}}, \bibinfo {author} {\bibfnamefont
  {A.}~\bibnamefont {Dr\'eau}}, \bibinfo {author} {\bibfnamefont
  {S.}~\bibnamefont {Kosen}}, \bibinfo {author} {\bibfnamefont {J.-F.}\
  \bibnamefont {Roch}}, \bibinfo {author} {\bibfnamefont {S.}~\bibnamefont
  {Pezzagna}}, \bibinfo {author} {\bibfnamefont {J.}~\bibnamefont {Meijer}},
  \bibinfo {author} {\bibfnamefont {T.}~\bibnamefont {Teraji}}, \bibinfo
  {author} {\bibfnamefont {Y.}~\bibnamefont {Kubo}}, \bibinfo {author}
  {\bibfnamefont {P.}~\bibnamefont {Bertet}}, \bibinfo {author} {\bibfnamefont
  {J.~R.}\ \bibnamefont {Maze}},\ and\ \bibinfo {author} {\bibfnamefont
  {V.}~\bibnamefont {Jacques}},\ }\bibfield  {title} {\bibinfo {title}
  {Competition between electric field and magnetic field noise in the
  decoherence of a single spin in diamond},\ }\href@noop {} {\bibfield
  {journal} {\bibinfo  {journal} {Phys. Rev. B}\ }\textbf {\bibinfo {volume}
  {93}},\ \bibinfo {pages} {024305} (\bibinfo {year} {2016})}\BibitemShut
  {NoStop}%
\bibitem [{\citenamefont {Sangtawesin}\ \emph {et~al.}(2019)\citenamefont
  {Sangtawesin}, \citenamefont {Dwyer}, \citenamefont {Srinivasan},
  \citenamefont {Allred}, \citenamefont {Rodgers}, \citenamefont {De~Greve},
  \citenamefont {Stacey}, \citenamefont {Dontschuk}, \citenamefont {O'Donnell},
  \citenamefont {Hu}, \citenamefont {Evans}, \citenamefont {Jaye},
  \citenamefont {Fischer}, \citenamefont {Markham}, \citenamefont {Twitchen},
  \citenamefont {Park}, \citenamefont {Lukin},\ and\ \citenamefont
  {de~Leon}}]{electric-magnetic3}%
  \BibitemOpen
  \bibfield  {author} {\bibinfo {author} {\bibfnamefont {S.}~\bibnamefont
  {Sangtawesin}}, \bibinfo {author} {\bibfnamefont {B.~L.}\ \bibnamefont
  {Dwyer}}, \bibinfo {author} {\bibfnamefont {S.}~\bibnamefont {Srinivasan}},
  \bibinfo {author} {\bibfnamefont {J.~J.}\ \bibnamefont {Allred}}, \bibinfo
  {author} {\bibfnamefont {L.~V.~H.}\ \bibnamefont {Rodgers}}, \bibinfo
  {author} {\bibfnamefont {K.}~\bibnamefont {De~Greve}}, \bibinfo {author}
  {\bibfnamefont {A.}~\bibnamefont {Stacey}}, \bibinfo {author} {\bibfnamefont
  {N.}~\bibnamefont {Dontschuk}}, \bibinfo {author} {\bibfnamefont {K.~M.}\
  \bibnamefont {O'Donnell}}, \bibinfo {author} {\bibfnamefont {D.}~\bibnamefont
  {Hu}}, \bibinfo {author} {\bibfnamefont {D.~A.}\ \bibnamefont {Evans}},
  \bibinfo {author} {\bibfnamefont {C.}~\bibnamefont {Jaye}}, \bibinfo {author}
  {\bibfnamefont {D.~A.}\ \bibnamefont {Fischer}}, \bibinfo {author}
  {\bibfnamefont {M.~L.}\ \bibnamefont {Markham}}, \bibinfo {author}
  {\bibfnamefont {D.~J.}\ \bibnamefont {Twitchen}}, \bibinfo {author}
  {\bibfnamefont {H.}~\bibnamefont {Park}}, \bibinfo {author} {\bibfnamefont
  {M.~D.}\ \bibnamefont {Lukin}},\ and\ \bibinfo {author} {\bibfnamefont
  {N.~P.}\ \bibnamefont {de~Leon}},\ }\bibfield  {title} {\bibinfo {title}
  {Origins of diamond surface noise probed by correlating single-spin
  measurements with surface spectroscopy},\ }\href@noop {} {\bibfield
  {journal} {\bibinfo  {journal} {Phys. Rev. X}\ }\textbf {\bibinfo {volume}
  {9}},\ \bibinfo {pages} {031052} (\bibinfo {year} {2019})}\BibitemShut
  {NoStop}%
\bibitem [{\citenamefont {Shin}\ \emph {et~al.}(2013)\citenamefont {Shin},
  \citenamefont {Avalos}, \citenamefont {Butler}, \citenamefont {Wang},
  \citenamefont {Seltzer}, \citenamefont {Liu}, \citenamefont {Pines},\ and\
  \citenamefont {Bajaj}}]{electric-magnetic4}%
  \BibitemOpen
  \bibfield  {author} {\bibinfo {author} {\bibfnamefont {C.~S.}\ \bibnamefont
  {Shin}}, \bibinfo {author} {\bibfnamefont {C.~E.}\ \bibnamefont {Avalos}},
  \bibinfo {author} {\bibfnamefont {M.~C.}\ \bibnamefont {Butler}}, \bibinfo
  {author} {\bibfnamefont {H.-J.}\ \bibnamefont {Wang}}, \bibinfo {author}
  {\bibfnamefont {S.~J.}\ \bibnamefont {Seltzer}}, \bibinfo {author}
  {\bibfnamefont {R.-B.}\ \bibnamefont {Liu}}, \bibinfo {author} {\bibfnamefont
  {A.}~\bibnamefont {Pines}},\ and\ \bibinfo {author} {\bibfnamefont {V.~S.}\
  \bibnamefont {Bajaj}},\ }\bibfield  {title} {\bibinfo {title} {Suppression of
  electron spin decoherence of the diamond {NV} center by a transverse magnetic
  field},\ }\href@noop {} {\bibfield  {journal} {\bibinfo  {journal} {Phys.
  Rev. B}\ }\textbf {\bibinfo {volume} {88}},\ \bibinfo {pages} {161412}
  (\bibinfo {year} {2013})}\BibitemShut {NoStop}%
\bibitem [{\citenamefont {Kim}\ \emph {et~al.}(2015)\citenamefont {Kim},
  \citenamefont {Mamin}, \citenamefont {Sherwood}, \citenamefont {Ohno},
  \citenamefont {Awschalom},\ and\ \citenamefont {Rugar}}]{electricnoise1}%
  \BibitemOpen
  \bibfield  {author} {\bibinfo {author} {\bibfnamefont {M.}~\bibnamefont
  {Kim}}, \bibinfo {author} {\bibfnamefont {H.~J.}\ \bibnamefont {Mamin}},
  \bibinfo {author} {\bibfnamefont {M.~H.}\ \bibnamefont {Sherwood}}, \bibinfo
  {author} {\bibfnamefont {K.}~\bibnamefont {Ohno}}, \bibinfo {author}
  {\bibfnamefont {D.~D.}\ \bibnamefont {Awschalom}},\ and\ \bibinfo {author}
  {\bibfnamefont {D.}~\bibnamefont {Rugar}},\ }\bibfield  {title} {\bibinfo
  {title} {Decoherence of near-surface nitrogen-vacancy centers due to electric
  field noise},\ }\href@noop {} {\bibfield  {journal} {\bibinfo  {journal}
  {Phys. Rev. Lett.}\ }\textbf {\bibinfo {volume} {115}},\ \bibinfo {pages}
  {087602} (\bibinfo {year} {2015})}\BibitemShut {NoStop}%
\bibitem [{\citenamefont {Chrostoski}\ \emph {et~al.}(2018)\citenamefont
  {Chrostoski}, \citenamefont {Sadeghpour},\ and\ \citenamefont
  {Santamore}}]{electricnoise2}%
  \BibitemOpen
  \bibfield  {author} {\bibinfo {author} {\bibfnamefont {P.}~\bibnamefont
  {Chrostoski}}, \bibinfo {author} {\bibfnamefont {H.~R.}\ \bibnamefont
  {Sadeghpour}},\ and\ \bibinfo {author} {\bibfnamefont {D.~H.}\ \bibnamefont
  {Santamore}},\ }\bibfield  {title} {\bibinfo {title} {Electric noise spectra
  of a near-surface nitrogen-vacancy center in diamond with a protective
  layer},\ }\href@noop {} {\bibfield  {journal} {\bibinfo  {journal} {Phys.
  Rev. Applied}\ }\textbf {\bibinfo {volume} {10}},\ \bibinfo {pages} {064056}
  (\bibinfo {year} {2018})}\BibitemShut {NoStop}%
\bibitem [{\citenamefont {Mittiga}\ \emph
  {et~al.}(2018{\natexlab{b}})\citenamefont {Mittiga}, \citenamefont {Hsieh},
  \citenamefont {Zu}, \citenamefont {Kobrin}, \citenamefont {Machado},
  \citenamefont {Bhattacharyya}, \citenamefont {Rui}, \citenamefont {Jarmola},
  \citenamefont {Choi}, \citenamefont {Budker},\ and\ \citenamefont
  {Yao}}]{electricnoise3}%
  \BibitemOpen
  \bibfield  {author} {\bibinfo {author} {\bibfnamefont {T.}~\bibnamefont
  {Mittiga}}, \bibinfo {author} {\bibfnamefont {S.}~\bibnamefont {Hsieh}},
  \bibinfo {author} {\bibfnamefont {C.}~\bibnamefont {Zu}}, \bibinfo {author}
  {\bibfnamefont {B.}~\bibnamefont {Kobrin}}, \bibinfo {author} {\bibfnamefont
  {F.}~\bibnamefont {Machado}}, \bibinfo {author} {\bibfnamefont
  {P.}~\bibnamefont {Bhattacharyya}}, \bibinfo {author} {\bibfnamefont {N.~Z.}\
  \bibnamefont {Rui}}, \bibinfo {author} {\bibfnamefont {A.}~\bibnamefont
  {Jarmola}}, \bibinfo {author} {\bibfnamefont {S.}~\bibnamefont {Choi}},
  \bibinfo {author} {\bibfnamefont {D.}~\bibnamefont {Budker}},\ and\ \bibinfo
  {author} {\bibfnamefont {N.~Y.}\ \bibnamefont {Yao}},\ }\bibfield  {title}
  {\bibinfo {title} {Imaging the local charge environment of nitrogen-vacancy
  centers in diamond},\ }\href@noop {} {\bibfield  {journal} {\bibinfo
  {journal} {Phys. Rev. Lett.}\ }\textbf {\bibinfo {volume} {121}},\ \bibinfo
  {pages} {246402} (\bibinfo {year} {2018}{\natexlab{b}})}\BibitemShut
  {NoStop}%
\bibitem [{\citenamefont {Balasubramanian}\ \emph {et~al.}(2009)\citenamefont
  {Balasubramanian}, \citenamefont {Neumann}, \citenamefont {Twitchen},
  \citenamefont {Markham}, \citenamefont {Kolesov}, \citenamefont {Mizuochi},
  \citenamefont {Isoya}, \citenamefont {Achard}, \citenamefont {Beck},
  \citenamefont {Tissler}, \citenamefont {Jacques}, \citenamefont {Hemmer},
  \citenamefont {Jelezko},\ and\ \citenamefont {Wrachtrup}}]{magneticnoise1}%
  \BibitemOpen
  \bibfield  {author} {\bibinfo {author} {\bibfnamefont {G.}~\bibnamefont
  {Balasubramanian}}, \bibinfo {author} {\bibfnamefont {P.}~\bibnamefont
  {Neumann}}, \bibinfo {author} {\bibfnamefont {D.}~\bibnamefont {Twitchen}},
  \bibinfo {author} {\bibfnamefont {M.}~\bibnamefont {Markham}}, \bibinfo
  {author} {\bibfnamefont {R.}~\bibnamefont {Kolesov}}, \bibinfo {author}
  {\bibfnamefont {N.}~\bibnamefont {Mizuochi}}, \bibinfo {author}
  {\bibfnamefont {J.}~\bibnamefont {Isoya}}, \bibinfo {author} {\bibfnamefont
  {J.}~\bibnamefont {Achard}}, \bibinfo {author} {\bibfnamefont
  {J.}~\bibnamefont {Beck}}, \bibinfo {author} {\bibfnamefont {J.}~\bibnamefont
  {Tissler}}, \bibinfo {author} {\bibfnamefont {V.}~\bibnamefont {Jacques}},
  \bibinfo {author} {\bibfnamefont {P.~R.}\ \bibnamefont {Hemmer}}, \bibinfo
  {author} {\bibfnamefont {F.}~\bibnamefont {Jelezko}},\ and\ \bibinfo {author}
  {\bibfnamefont {J.}~\bibnamefont {Wrachtrup}},\ }\bibfield  {title} {\bibinfo
  {title} {Ultralong spin coherence time in isotopically engineered diamond},\
  }\href@noop {} {\bibfield  {journal} {\bibinfo  {journal} {Nature Materials}\
  }\textbf {\bibinfo {volume} {8}},\ \bibinfo {pages} {383} (\bibinfo {year}
  {2009})}\BibitemShut {NoStop}%
\bibitem [{\citenamefont {Meriles}\ \emph {et~al.}(2010)\citenamefont
  {Meriles}, \citenamefont {Jiang}, \citenamefont {Goldstein}, \citenamefont
  {Hodges}, \citenamefont {Maze}, \citenamefont {Lukin},\ and\ \citenamefont
  {Cappellaro}}]{magneticnoise2}%
  \BibitemOpen
  \bibfield  {author} {\bibinfo {author} {\bibfnamefont {C.~A.}\ \bibnamefont
  {Meriles}}, \bibinfo {author} {\bibfnamefont {L.}~\bibnamefont {Jiang}},
  \bibinfo {author} {\bibfnamefont {G.}~\bibnamefont {Goldstein}}, \bibinfo
  {author} {\bibfnamefont {J.~S.}\ \bibnamefont {Hodges}}, \bibinfo {author}
  {\bibfnamefont {J.}~\bibnamefont {Maze}}, \bibinfo {author} {\bibfnamefont
  {M.~D.}\ \bibnamefont {Lukin}},\ and\ \bibinfo {author} {\bibfnamefont
  {P.}~\bibnamefont {Cappellaro}},\ }\bibfield  {title} {\bibinfo {title}
  {Imaging mesoscopic nuclear spin noise with a diamond magnetometer},\
  }\href@noop {} {\bibfield  {journal} {\bibinfo  {journal} {The Journal of
  Chemical Physics}\ }\textbf {\bibinfo {volume} {133}},\ \bibinfo {pages}
  {124105} (\bibinfo {year} {2010})}\BibitemShut {NoStop}%
\bibitem [{\citenamefont {Tetienne}\ \emph {et~al.}(2013)\citenamefont
  {Tetienne}, \citenamefont {Hingant}, \citenamefont {Rondin}, \citenamefont
  {Cavaill\`es}, \citenamefont {Mayer}, \citenamefont {Dantelle}, \citenamefont
  {Gacoin}, \citenamefont {Wrachtrup}, \citenamefont {Roch},\ and\
  \citenamefont {Jacques}}]{PhysRevB.87.235436}%
  \BibitemOpen
  \bibfield  {author} {\bibinfo {author} {\bibfnamefont {J.-P.}\ \bibnamefont
  {Tetienne}}, \bibinfo {author} {\bibfnamefont {T.}~\bibnamefont {Hingant}},
  \bibinfo {author} {\bibfnamefont {L.}~\bibnamefont {Rondin}}, \bibinfo
  {author} {\bibfnamefont {A.}~\bibnamefont {Cavaill\`es}}, \bibinfo {author}
  {\bibfnamefont {L.}~\bibnamefont {Mayer}}, \bibinfo {author} {\bibfnamefont
  {G.}~\bibnamefont {Dantelle}}, \bibinfo {author} {\bibfnamefont
  {T.}~\bibnamefont {Gacoin}}, \bibinfo {author} {\bibfnamefont
  {J.}~\bibnamefont {Wrachtrup}}, \bibinfo {author} {\bibfnamefont {J.-F.}\
  \bibnamefont {Roch}},\ and\ \bibinfo {author} {\bibfnamefont
  {V.}~\bibnamefont {Jacques}},\ }\bibfield  {title} {\bibinfo {title} {Spin
  relaxometry of single nitrogen-vacancy defects in diamond nanocrystals for
  magnetic noise sensing},\ }\href {https://doi.org/10.1103/PhysRevB.87.235436}
  {\bibfield  {journal} {\bibinfo  {journal} {Phys. Rev. B}\ }\textbf {\bibinfo
  {volume} {87}},\ \bibinfo {pages} {235436} (\bibinfo {year}
  {2013})}\BibitemShut {NoStop}%
\bibitem [{\citenamefont {Rosskopf}\ \emph {et~al.}(2014)\citenamefont
  {Rosskopf}, \citenamefont {Dussaux}, \citenamefont {Ohashi}, \citenamefont
  {Loretz}, \citenamefont {Schirhagl}, \citenamefont {Watanabe}, \citenamefont
  {Shikata}, \citenamefont {Itoh},\ and\ \citenamefont
  {Degen}}]{magneticnoise3}%
  \BibitemOpen
  \bibfield  {author} {\bibinfo {author} {\bibfnamefont {T.}~\bibnamefont
  {Rosskopf}}, \bibinfo {author} {\bibfnamefont {A.}~\bibnamefont {Dussaux}},
  \bibinfo {author} {\bibfnamefont {K.}~\bibnamefont {Ohashi}}, \bibinfo
  {author} {\bibfnamefont {M.}~\bibnamefont {Loretz}}, \bibinfo {author}
  {\bibfnamefont {R.}~\bibnamefont {Schirhagl}}, \bibinfo {author}
  {\bibfnamefont {H.}~\bibnamefont {Watanabe}}, \bibinfo {author}
  {\bibfnamefont {S.}~\bibnamefont {Shikata}}, \bibinfo {author} {\bibfnamefont
  {K.~M.}\ \bibnamefont {Itoh}},\ and\ \bibinfo {author} {\bibfnamefont
  {C.~L.}\ \bibnamefont {Degen}},\ }\bibfield  {title} {\bibinfo {title}
  {Investigation of surface magnetic noise by shallow spins in diamond},\
  }\href@noop {} {\bibfield  {journal} {\bibinfo  {journal} {Phys. Rev. Lett.}\
  }\textbf {\bibinfo {volume} {112}},\ \bibinfo {pages} {147602} (\bibinfo
  {year} {2014})}\BibitemShut {NoStop}%
\bibitem [{\citenamefont {Myers}\ \emph {et~al.}(2014)\citenamefont {Myers},
  \citenamefont {Das}, \citenamefont {Dartiailh}, \citenamefont {Ohno},
  \citenamefont {Awschalom},\ and\ \citenamefont
  {Bleszynski~Jayich}}]{magneticnoise4}%
  \BibitemOpen
  \bibfield  {author} {\bibinfo {author} {\bibfnamefont {B.~A.}\ \bibnamefont
  {Myers}}, \bibinfo {author} {\bibfnamefont {A.}~\bibnamefont {Das}}, \bibinfo
  {author} {\bibfnamefont {M.~C.}\ \bibnamefont {Dartiailh}}, \bibinfo {author}
  {\bibfnamefont {K.}~\bibnamefont {Ohno}}, \bibinfo {author} {\bibfnamefont
  {D.~D.}\ \bibnamefont {Awschalom}},\ and\ \bibinfo {author} {\bibfnamefont
  {A.~C.}\ \bibnamefont {Bleszynski~Jayich}},\ }\bibfield  {title} {\bibinfo
  {title} {Probing surface noise with depth-calibrated spins in diamond},\
  }\href@noop {} {\bibfield  {journal} {\bibinfo  {journal} {Phys. Rev. Lett.}\
  }\textbf {\bibinfo {volume} {113}},\ \bibinfo {pages} {027602} (\bibinfo
  {year} {2014})}\BibitemShut {NoStop}%
\bibitem [{\citenamefont {Romach}\ \emph {et~al.}(2015)\citenamefont {Romach},
  \citenamefont {M\"uller}, \citenamefont {Unden}, \citenamefont {Rogers},
  \citenamefont {Isoda}, \citenamefont {Itoh}, \citenamefont {Markham},
  \citenamefont {Stacey}, \citenamefont {Meijer}, \citenamefont {Pezzagna},
  \citenamefont {Naydenov}, \citenamefont {McGuinness}, \citenamefont
  {Bar-Gill},\ and\ \citenamefont {Jelezko}}]{magneticnoise5}%
  \BibitemOpen
  \bibfield  {author} {\bibinfo {author} {\bibfnamefont {Y.}~\bibnamefont
  {Romach}}, \bibinfo {author} {\bibfnamefont {C.}~\bibnamefont {M\"uller}},
  \bibinfo {author} {\bibfnamefont {T.}~\bibnamefont {Unden}}, \bibinfo
  {author} {\bibfnamefont {L.~J.}\ \bibnamefont {Rogers}}, \bibinfo {author}
  {\bibfnamefont {T.}~\bibnamefont {Isoda}}, \bibinfo {author} {\bibfnamefont
  {K.~M.}\ \bibnamefont {Itoh}}, \bibinfo {author} {\bibfnamefont
  {M.}~\bibnamefont {Markham}}, \bibinfo {author} {\bibfnamefont
  {A.}~\bibnamefont {Stacey}}, \bibinfo {author} {\bibfnamefont
  {J.}~\bibnamefont {Meijer}}, \bibinfo {author} {\bibfnamefont
  {S.}~\bibnamefont {Pezzagna}}, \bibinfo {author} {\bibfnamefont
  {B.}~\bibnamefont {Naydenov}}, \bibinfo {author} {\bibfnamefont {L.~P.}\
  \bibnamefont {McGuinness}}, \bibinfo {author} {\bibfnamefont
  {N.}~\bibnamefont {Bar-Gill}},\ and\ \bibinfo {author} {\bibfnamefont
  {F.}~\bibnamefont {Jelezko}},\ }\bibfield  {title} {\bibinfo {title}
  {Spectroscopy of surface-induced noise using shallow spins in diamond},\
  }\href@noop {} {\bibfield  {journal} {\bibinfo  {journal} {Phys. Rev. Lett.}\
  }\textbf {\bibinfo {volume} {114}},\ \bibinfo {pages} {017601} (\bibinfo
  {year} {2015})}\BibitemShut {NoStop}%
\bibitem [{\citenamefont {Myers}\ \emph {et~al.}(2017)\citenamefont {Myers},
  \citenamefont {Ariyaratne},\ and\ \citenamefont {Jayich}}]{magneticnoise6}%
  \BibitemOpen
  \bibfield  {author} {\bibinfo {author} {\bibfnamefont {B.~A.}\ \bibnamefont
  {Myers}}, \bibinfo {author} {\bibfnamefont {A.}~\bibnamefont {Ariyaratne}},\
  and\ \bibinfo {author} {\bibfnamefont {A.~C.~B.}\ \bibnamefont {Jayich}},\
  }\bibfield  {title} {\bibinfo {title} {Double-quantum spin-relaxation limits
  to coherence of near-surface nitrogen-vacancy centers},\ }\href@noop {}
  {\bibfield  {journal} {\bibinfo  {journal} {Phys. Rev. Lett.}\ }\textbf
  {\bibinfo {volume} {118}},\ \bibinfo {pages} {197201} (\bibinfo {year}
  {2017})}\BibitemShut {NoStop}%
\bibitem [{\citenamefont {Choi}\ \emph {et~al.}(2017)\citenamefont {Choi},
  \citenamefont {Choi}, \citenamefont {Kucsko}, \citenamefont {Maurer},
  \citenamefont {Shields}, \citenamefont {Sumiya}, \citenamefont {Onoda},
  \citenamefont {Isoya}, \citenamefont {Demler}, \citenamefont {Jelezko},
  \citenamefont {Yao},\ and\ \citenamefont {Lukin}}]{magneticnoise7}%
  \BibitemOpen
  \bibfield  {author} {\bibinfo {author} {\bibfnamefont {J.}~\bibnamefont
  {Choi}}, \bibinfo {author} {\bibfnamefont {S.}~\bibnamefont {Choi}}, \bibinfo
  {author} {\bibfnamefont {G.}~\bibnamefont {Kucsko}}, \bibinfo {author}
  {\bibfnamefont {P.~C.}\ \bibnamefont {Maurer}}, \bibinfo {author}
  {\bibfnamefont {B.~J.}\ \bibnamefont {Shields}}, \bibinfo {author}
  {\bibfnamefont {H.}~\bibnamefont {Sumiya}}, \bibinfo {author} {\bibfnamefont
  {S.}~\bibnamefont {Onoda}}, \bibinfo {author} {\bibfnamefont
  {J.}~\bibnamefont {Isoya}}, \bibinfo {author} {\bibfnamefont
  {E.}~\bibnamefont {Demler}}, \bibinfo {author} {\bibfnamefont
  {F.}~\bibnamefont {Jelezko}}, \bibinfo {author} {\bibfnamefont {N.~Y.}\
  \bibnamefont {Yao}},\ and\ \bibinfo {author} {\bibfnamefont {M.~D.}\
  \bibnamefont {Lukin}},\ }\bibfield  {title} {\bibinfo {title} {Depolarization
  dynamics in a strongly interacting solid-state spin ensemble},\ }\href@noop
  {} {\bibfield  {journal} {\bibinfo  {journal} {Phys. Rev. Lett.}\ }\textbf
  {\bibinfo {volume} {118}},\ \bibinfo {pages} {093601} (\bibinfo {year}
  {2017})}\BibitemShut {NoStop}%
\bibitem [{\citenamefont {Kolkowitz}\ \emph {et~al.}(2015)\citenamefont
  {Kolkowitz}, \citenamefont {Safira}, \citenamefont {High}, \citenamefont
  {Devlin}, \citenamefont {Choi}, \citenamefont {Unterreithmeier},
  \citenamefont {Patterson}, \citenamefont {Zibrov}, \citenamefont
  {Manucharyan}, \citenamefont {Park},\ and\ \citenamefont
  {Lukin}}]{Kolkowitz1129}%
  \BibitemOpen
  \bibfield  {author} {\bibinfo {author} {\bibfnamefont {S.}~\bibnamefont
  {Kolkowitz}}, \bibinfo {author} {\bibfnamefont {A.}~\bibnamefont {Safira}},
  \bibinfo {author} {\bibfnamefont {A.~A.}\ \bibnamefont {High}}, \bibinfo
  {author} {\bibfnamefont {R.~C.}\ \bibnamefont {Devlin}}, \bibinfo {author}
  {\bibfnamefont {S.}~\bibnamefont {Choi}}, \bibinfo {author} {\bibfnamefont
  {Q.~P.}\ \bibnamefont {Unterreithmeier}}, \bibinfo {author} {\bibfnamefont
  {D.}~\bibnamefont {Patterson}}, \bibinfo {author} {\bibfnamefont {A.~S.}\
  \bibnamefont {Zibrov}}, \bibinfo {author} {\bibfnamefont {V.~E.}\
  \bibnamefont {Manucharyan}}, \bibinfo {author} {\bibfnamefont
  {H.}~\bibnamefont {Park}},\ and\ \bibinfo {author} {\bibfnamefont {M.~D.}\
  \bibnamefont {Lukin}},\ }\bibfield  {title} {\bibinfo {title} {Probing
  johnson noise and ballistic transport in normal metals with a single-spin
  qubit},\ }\href {https://doi.org/10.1126/science.aaa4298} {\bibfield
  {journal} {\bibinfo  {journal} {Science}\ }\textbf {\bibinfo {volume}
  {347}},\ \bibinfo {pages} {1129} (\bibinfo {year} {2015})}\BibitemShut
  {NoStop}%
\bibitem [{\citenamefont {Candido}\ and\ \citenamefont
  {Flatt\'e}(2021)}]{candido-pin}%
  \BibitemOpen
  \bibfield  {author} {\bibinfo {author} {\bibfnamefont {D.~R.}\ \bibnamefont
  {Candido}}\ and\ \bibinfo {author} {\bibfnamefont {M.~E.}\ \bibnamefont
  {Flatt\'e}},\ }\bibfield  {title} {\bibinfo {title} {Suppression of the
  optical linewidth and spin decoherence of a quantum spin center in a $p$-$n$
  diode},\ }\href {https://doi.org/10.1103/PRXQuantum.2.040310} {\bibfield
  {journal} {\bibinfo  {journal} {PRX Quantum}\ }\textbf {\bibinfo {volume}
  {2}},\ \bibinfo {pages} {040310} (\bibinfo {year} {2021})}\BibitemShut
  {NoStop}%
\bibitem [{\citenamefont {Lindblad}(1976)}]{lindblad1976}%
  \BibitemOpen
  \bibfield  {author} {\bibinfo {author} {\bibfnamefont {G.}~\bibnamefont
  {Lindblad}},\ }\bibfield  {title} {\bibinfo {title} {On the generators of
  quantum dynamical semigroups},\ }\href {https://doi.org/10.1007/BF01608499}
  {\bibfield  {journal} {\bibinfo  {journal} {Commun. Math. Phys.}\ }\textbf
  {\bibinfo {volume} {48}},\ \bibinfo {pages} {119} (\bibinfo {year}
  {1976})}\BibitemShut {NoStop}%
\bibitem [{\citenamefont {Sque}\ \emph {et~al.}(2006)\citenamefont {Sque},
  \citenamefont {Jones},\ and\ \citenamefont {Briddon}}]{PhysRevB.73.085313}%
  \BibitemOpen
  \bibfield  {author} {\bibinfo {author} {\bibfnamefont {S.~J.}\ \bibnamefont
  {Sque}}, \bibinfo {author} {\bibfnamefont {R.}~\bibnamefont {Jones}},\ and\
  \bibinfo {author} {\bibfnamefont {P.~R.}\ \bibnamefont {Briddon}},\
  }\bibfield  {title} {\bibinfo {title} {Structure, electronics, and
  interaction of hydrogen and oxygen on diamond surfaces},\ }\href
  {https://doi.org/10.1103/PhysRevB.73.085313} {\bibfield  {journal} {\bibinfo
  {journal} {Phys. Rev. B}\ }\textbf {\bibinfo {volume} {73}},\ \bibinfo
  {pages} {085313} (\bibinfo {year} {2006})}\BibitemShut {NoStop}%
\bibitem [{\citenamefont {Sussmann}(2009)}]{sussmann2009cvd}%
  \BibitemOpen
  \bibfield  {author} {\bibinfo {author} {\bibfnamefont {R.~S.}\ \bibnamefont
  {Sussmann}},\ }\href@noop {} {\emph {\bibinfo {title} {CVD diamond for
  electronic devices and sensors}}},\ Vol.~\bibinfo {volume} {26}\ (\bibinfo
  {publisher} {John Wiley \& Sons},\ \bibinfo {year} {2009})\BibitemShut
  {NoStop}%
\bibitem [{\citenamefont {Kawarada}(1996)}]{KAWARADA1996205}%
  \BibitemOpen
  \bibfield  {author} {\bibinfo {author} {\bibfnamefont {H.}~\bibnamefont
  {Kawarada}},\ }\bibfield  {title} {\bibinfo {title} {Hydrogen-terminated
  diamond surfaces and interfaces},\ }\href
  {https://doi.org/https://doi.org/10.1016/S0167-5729(97)80002-7} {\bibfield
  {journal} {\bibinfo  {journal} {Surface Science Reports}\ }\textbf {\bibinfo
  {volume} {26}},\ \bibinfo {pages} {205} (\bibinfo {year} {1996})}\BibitemShut
  {NoStop}%
\bibitem [{\citenamefont {Maier}\ \emph {et~al.}(2000)\citenamefont {Maier},
  \citenamefont {Riedel}, \citenamefont {Mantel}, \citenamefont {Ristein},\
  and\ \citenamefont {Ley}}]{PhysRevLett.85.3472}%
  \BibitemOpen
  \bibfield  {author} {\bibinfo {author} {\bibfnamefont {F.}~\bibnamefont
  {Maier}}, \bibinfo {author} {\bibfnamefont {M.}~\bibnamefont {Riedel}},
  \bibinfo {author} {\bibfnamefont {B.}~\bibnamefont {Mantel}}, \bibinfo
  {author} {\bibfnamefont {J.}~\bibnamefont {Ristein}},\ and\ \bibinfo {author}
  {\bibfnamefont {L.}~\bibnamefont {Ley}},\ }\bibfield  {title} {\bibinfo
  {title} {Origin of surface conductivity in diamond},\ }\href
  {https://doi.org/10.1103/PhysRevLett.85.3472} {\bibfield  {journal} {\bibinfo
   {journal} {Phys. Rev. Lett.}\ }\textbf {\bibinfo {volume} {85}},\ \bibinfo
  {pages} {3472} (\bibinfo {year} {2000})}\BibitemShut {NoStop}%
\bibitem [{\citenamefont {Takeuchi}\ \emph {et~al.}(2003)\citenamefont
  {Takeuchi}, \citenamefont {Riedel}, \citenamefont {Ristein},\ and\
  \citenamefont {Ley}}]{PhysRevB.68.041304}%
  \BibitemOpen
  \bibfield  {author} {\bibinfo {author} {\bibfnamefont {D.}~\bibnamefont
  {Takeuchi}}, \bibinfo {author} {\bibfnamefont {M.}~\bibnamefont {Riedel}},
  \bibinfo {author} {\bibfnamefont {J.}~\bibnamefont {Ristein}},\ and\ \bibinfo
  {author} {\bibfnamefont {L.}~\bibnamefont {Ley}},\ }\bibfield  {title}
  {\bibinfo {title} {Surface band bending and surface conductivity of
  hydrogenated diamond},\ }\href {https://doi.org/10.1103/PhysRevB.68.041304}
  {\bibfield  {journal} {\bibinfo  {journal} {Phys. Rev. B}\ }\textbf {\bibinfo
  {volume} {68}},\ \bibinfo {pages} {041304} (\bibinfo {year}
  {2003})}\BibitemShut {NoStop}%
\bibitem [{\citenamefont {Crawford}\ \emph {et~al.}(2021)\citenamefont
  {Crawford}, \citenamefont {Maini}, \citenamefont {Macdonald},\ and\
  \citenamefont {Moran}}]{surfacediamondreview}%
  \BibitemOpen
  \bibfield  {author} {\bibinfo {author} {\bibfnamefont {K.~G.}\ \bibnamefont
  {Crawford}}, \bibinfo {author} {\bibfnamefont {I.}~\bibnamefont {Maini}},
  \bibinfo {author} {\bibfnamefont {D.~A.}\ \bibnamefont {Macdonald}},\ and\
  \bibinfo {author} {\bibfnamefont {D.~A.}\ \bibnamefont {Moran}},\ }\bibfield
  {title} {\bibinfo {title} {Surface transfer doping of diamond: A review},\
  }\href {https://doi.org/https://doi.org/10.1016/j.progsurf.2021.100613}
  {\bibfield  {journal} {\bibinfo  {journal} {Progress in Surface Science}\
  }\textbf {\bibinfo {volume} {96}},\ \bibinfo {pages} {100613} (\bibinfo
  {year} {2021})}\BibitemShut {NoStop}%
\bibitem [{\citenamefont {Stacey}\ \emph {et~al.}(2019)\citenamefont {Stacey},
  \citenamefont {Dontschuk}, \citenamefont {Chou}, \citenamefont {Broadway},
  \citenamefont {Schenk}, \citenamefont {Sear}, \citenamefont {Tetienne},
  \citenamefont {Hoffman}, \citenamefont {Prawer}, \citenamefont {Pakes},
  \citenamefont {Tadich}, \citenamefont {de~Leon}, \citenamefont {Gali},\ and\
  \citenamefont {Hollenberg}}]{https://doi.org/10.1002/admi.201801449}%
  \BibitemOpen
  \bibfield  {author} {\bibinfo {author} {\bibfnamefont {A.}~\bibnamefont
  {Stacey}}, \bibinfo {author} {\bibfnamefont {N.}~\bibnamefont {Dontschuk}},
  \bibinfo {author} {\bibfnamefont {J.-P.}\ \bibnamefont {Chou}}, \bibinfo
  {author} {\bibfnamefont {D.~A.}\ \bibnamefont {Broadway}}, \bibinfo {author}
  {\bibfnamefont {A.~K.}\ \bibnamefont {Schenk}}, \bibinfo {author}
  {\bibfnamefont {M.~J.}\ \bibnamefont {Sear}}, \bibinfo {author}
  {\bibfnamefont {J.-P.}\ \bibnamefont {Tetienne}}, \bibinfo {author}
  {\bibfnamefont {A.}~\bibnamefont {Hoffman}}, \bibinfo {author} {\bibfnamefont
  {S.}~\bibnamefont {Prawer}}, \bibinfo {author} {\bibfnamefont {C.~I.}\
  \bibnamefont {Pakes}}, \bibinfo {author} {\bibfnamefont {A.}~\bibnamefont
  {Tadich}}, \bibinfo {author} {\bibfnamefont {N.~P.}\ \bibnamefont {de~Leon}},
  \bibinfo {author} {\bibfnamefont {A.}~\bibnamefont {Gali}},\ and\ \bibinfo
  {author} {\bibfnamefont {L.~C.~L.}\ \bibnamefont {Hollenberg}},\ }\bibfield
  {title} {\bibinfo {title} {Evidence for primal sp2 defects at the diamond
  surface: Candidates for electron trapping and noise sources},\ }\href
  {https://doi.org/https://doi.org/10.1002/admi.201801449} {\bibfield
  {journal} {\bibinfo  {journal} {Advanced Materials Interfaces}\ }\textbf
  {\bibinfo {volume} {6}},\ \bibinfo {pages} {1801449} (\bibinfo {year}
  {2019})}\BibitemShut {NoStop}%
\bibitem [{\citenamefont {Reed}\ \emph {et~al.}(2022)\citenamefont {Reed},
  \citenamefont {Bathen}, \citenamefont {Ash}, \citenamefont {Meara},
  \citenamefont {Zakharov}, \citenamefont {Goss}, \citenamefont {Wells},
  \citenamefont {Evans},\ and\ \citenamefont {Cooil}}]{PhysRevB.105.205304}%
  \BibitemOpen
  \bibfield  {author} {\bibinfo {author} {\bibfnamefont {B.~P.}\ \bibnamefont
  {Reed}}, \bibinfo {author} {\bibfnamefont {M.~E.}\ \bibnamefont {Bathen}},
  \bibinfo {author} {\bibfnamefont {J.~W.~R.}\ \bibnamefont {Ash}}, \bibinfo
  {author} {\bibfnamefont {C.~J.}\ \bibnamefont {Meara}}, \bibinfo {author}
  {\bibfnamefont {A.~A.}\ \bibnamefont {Zakharov}}, \bibinfo {author}
  {\bibfnamefont {J.~P.}\ \bibnamefont {Goss}}, \bibinfo {author}
  {\bibfnamefont {J.~W.}\ \bibnamefont {Wells}}, \bibinfo {author}
  {\bibfnamefont {D.~A.}\ \bibnamefont {Evans}},\ and\ \bibinfo {author}
  {\bibfnamefont {S.~P.}\ \bibnamefont {Cooil}},\ }\bibfield  {title} {\bibinfo
  {title} {Diamond (111) surface reconstruction and epitaxial graphene
  interface},\ }\href {https://doi.org/10.1103/PhysRevB.105.205304} {\bibfield
  {journal} {\bibinfo  {journal} {Phys. Rev. B}\ }\textbf {\bibinfo {volume}
  {105}},\ \bibinfo {pages} {205304} (\bibinfo {year} {2022})}\BibitemShut
  {NoStop}%
\bibitem [{\citenamefont {Loubser$~$}\ and\ \citenamefont {van
  Wyk}(1978)}]{loubser1978electron}%
  \BibitemOpen
  \bibfield  {author} {\bibinfo {author} {\bibfnamefont {J.~H.~N.}\
  \bibnamefont {Loubser$~$}}\ and\ \bibinfo {author} {\bibfnamefont {J.~A.}\
  \bibnamefont {van Wyk}},\ }\bibfield  {title} {\bibinfo {title} {Electron
  spin resonance in the study of diamond},\ }\href@noop {} {\bibfield
  {journal} {\bibinfo  {journal} {Reports on Progress in Physics}\ }\textbf
  {\bibinfo {volume} {41}},\ \bibinfo {pages} {1201} (\bibinfo {year}
  {1978})}\BibitemShut {NoStop}%
\bibitem [{\citenamefont {Van Oort$~$}\ and\ \citenamefont
  {Glasbeek}(1990)}]{van1990electric}%
  \BibitemOpen
  \bibfield  {author} {\bibinfo {author} {\bibfnamefont {E.}~\bibnamefont {Van
  Oort$~$}}\ and\ \bibinfo {author} {\bibfnamefont {M.}~\bibnamefont
  {Glasbeek}},\ }\bibfield  {title} {\bibinfo {title} {Electric-field-induced
  modulation of spin echoes of $\mathrm{N-V}$ centers in diamond},\ }\href@noop
  {} {\bibfield  {journal} {\bibinfo  {journal} {Chemical Physics Letters}\
  }\textbf {\bibinfo {volume} {168}},\ \bibinfo {pages} {529 } (\bibinfo {year}
  {1990})}\BibitemShut {NoStop}%
\bibitem [{\citenamefont {Lenef}\ and\ \citenamefont
  {Rand}(1996)}]{PhysRevB.53.13441}%
  \BibitemOpen
  \bibfield  {author} {\bibinfo {author} {\bibfnamefont {A.}~\bibnamefont
  {Lenef}}\ and\ \bibinfo {author} {\bibfnamefont {S.~C.}\ \bibnamefont
  {Rand}},\ }\bibfield  {title} {\bibinfo {title} {Electronic structure of the
  n-v center in diamond: Theory},\ }\href
  {https://doi.org/10.1103/PhysRevB.53.13441} {\bibfield  {journal} {\bibinfo
  {journal} {Phys. Rev. B}\ }\textbf {\bibinfo {volume} {53}},\ \bibinfo
  {pages} {13441} (\bibinfo {year} {1996})}\BibitemShut {NoStop}%
\bibitem [{\citenamefont {Hossain}\ \emph {et~al.}(2008)\citenamefont
  {Hossain}, \citenamefont {Doherty}, \citenamefont {Wilson},\ and\
  \citenamefont {Hollenberg}}]{hossain2008ab}%
  \BibitemOpen
  \bibfield  {author} {\bibinfo {author} {\bibfnamefont {F.~M.}\ \bibnamefont
  {Hossain}}, \bibinfo {author} {\bibfnamefont {M.~W.}\ \bibnamefont
  {Doherty}}, \bibinfo {author} {\bibfnamefont {H.~F.}\ \bibnamefont
  {Wilson}},\ and\ \bibinfo {author} {\bibfnamefont {L.~C.~L.}\ \bibnamefont
  {Hollenberg}},\ }\bibfield  {title} {\bibinfo {title} {\textit{Ab Initio}
  electronic and optical properties of the
  ${N}\ensuremath{-}{V}^{\ensuremath{-}}$ center in diamond},\ }\href@noop {}
  {\bibfield  {journal} {\bibinfo  {journal} {Phys. Rev. Lett.}\ }\textbf
  {\bibinfo {volume} {101}},\ \bibinfo {pages} {226403} (\bibinfo {year}
  {2008})}\BibitemShut {NoStop}%
\bibitem [{\citenamefont {de~Lange}\ \emph {et~al.}(2010)\citenamefont
  {de~Lange}, \citenamefont {Wang}, \citenamefont {Rist{\`e}}, \citenamefont
  {Dobrovitski},\ and\ \citenamefont {Hanson}}]{de2010universal}%
  \BibitemOpen
  \bibfield  {author} {\bibinfo {author} {\bibfnamefont {G.}~\bibnamefont
  {de~Lange}}, \bibinfo {author} {\bibfnamefont {Z.~H.}\ \bibnamefont {Wang}},
  \bibinfo {author} {\bibfnamefont {D.}~\bibnamefont {Rist{\`e}}}, \bibinfo
  {author} {\bibfnamefont {V.~V.}\ \bibnamefont {Dobrovitski}},\ and\ \bibinfo
  {author} {\bibfnamefont {R.}~\bibnamefont {Hanson}},\ }\bibfield  {title}
  {\bibinfo {title} {Universal dynamical decoupling of a single solid-state
  spin from a spin bath},\ }\href@noop {} {\bibfield  {journal} {\bibinfo
  {journal} {Science}\ }\textbf {\bibinfo {volume} {330}},\ \bibinfo {pages}
  {60} (\bibinfo {year} {2010})}\BibitemShut {NoStop}%
\bibitem [{\citenamefont {Togan}\ \emph {et~al.}(2010)\citenamefont {Togan},
  \citenamefont {Chu}, \citenamefont {Trifonov}, \citenamefont {Jiang},
  \citenamefont {Maze}, \citenamefont {Childress}, \citenamefont {Dutt},
  \citenamefont {S{\o}rensen}, \citenamefont {Hemmer}, \citenamefont {Zibrov},\
  and\ \citenamefont {Lukin}}]{togan2010quantum}%
  \BibitemOpen
  \bibfield  {author} {\bibinfo {author} {\bibfnamefont {E.}~\bibnamefont
  {Togan}}, \bibinfo {author} {\bibfnamefont {Y.}~\bibnamefont {Chu}}, \bibinfo
  {author} {\bibfnamefont {A.~S.}\ \bibnamefont {Trifonov}}, \bibinfo {author}
  {\bibfnamefont {L.}~\bibnamefont {Jiang}}, \bibinfo {author} {\bibfnamefont
  {J.}~\bibnamefont {Maze}}, \bibinfo {author} {\bibfnamefont {L.}~\bibnamefont
  {Childress}}, \bibinfo {author} {\bibfnamefont {M.~V.~G.}\ \bibnamefont
  {Dutt}}, \bibinfo {author} {\bibfnamefont {A.~S.}\ \bibnamefont
  {S{\o}rensen}}, \bibinfo {author} {\bibfnamefont {P.~R.}\ \bibnamefont
  {Hemmer}}, \bibinfo {author} {\bibfnamefont {A.~S.}\ \bibnamefont {Zibrov}},\
  and\ \bibinfo {author} {\bibfnamefont {M.~D.}\ \bibnamefont {Lukin}},\
  }\bibfield  {title} {\bibinfo {title} {Quantum entanglement between an
  optical photon and a solid-state spin qubit},\ }\href@noop {} {\bibfield
  {journal} {\bibinfo  {journal} {Nature}\ }\textbf {\bibinfo {volume} {466}},\
  \bibinfo {pages} {730} (\bibinfo {year} {2010})}\BibitemShut {NoStop}%
\bibitem [{\citenamefont {Bassett}\ \emph {et~al.}(2011)\citenamefont
  {Bassett}, \citenamefont {Heremans}, \citenamefont {Yale}, \citenamefont
  {Buckley},\ and\ \citenamefont {Awschalom}}]{bassett2011electrical}%
  \BibitemOpen
  \bibfield  {author} {\bibinfo {author} {\bibfnamefont {L.~C.}\ \bibnamefont
  {Bassett}}, \bibinfo {author} {\bibfnamefont {F.~J.}\ \bibnamefont
  {Heremans}}, \bibinfo {author} {\bibfnamefont {C.~G.}\ \bibnamefont {Yale}},
  \bibinfo {author} {\bibfnamefont {B.~B.}\ \bibnamefont {Buckley}},\ and\
  \bibinfo {author} {\bibfnamefont {D.~D.}\ \bibnamefont {Awschalom}},\
  }\bibfield  {title} {\bibinfo {title} {Electrical tuning of single
  nitrogen-vacancy center optical transitions enhanced by photoinduced
  fields},\ }\href@noop {} {\bibfield  {journal} {\bibinfo  {journal} {Phys.
  Rev. Lett.}\ }\textbf {\bibinfo {volume} {107}},\ \bibinfo {pages} {266403}
  (\bibinfo {year} {2011})}\BibitemShut {NoStop}%
\bibitem [{\citenamefont {Maze}\ \emph {et~al.}(2011)\citenamefont {Maze},
  \citenamefont {Gali}, \citenamefont {Togan}, \citenamefont {Chu},
  \citenamefont {Trifonov}, \citenamefont {Kaxiras},\ and\ \citenamefont
  {Lukin}}]{maze2011}%
  \BibitemOpen
  \bibfield  {author} {\bibinfo {author} {\bibfnamefont {J.~R.}\ \bibnamefont
  {Maze}}, \bibinfo {author} {\bibfnamefont {A.}~\bibnamefont {Gali}}, \bibinfo
  {author} {\bibfnamefont {E.}~\bibnamefont {Togan}}, \bibinfo {author}
  {\bibfnamefont {Y.}~\bibnamefont {Chu}}, \bibinfo {author} {\bibfnamefont
  {A.}~\bibnamefont {Trifonov}}, \bibinfo {author} {\bibfnamefont
  {E.}~\bibnamefont {Kaxiras}},\ and\ \bibinfo {author} {\bibfnamefont {M.~D.}\
  \bibnamefont {Lukin}},\ }\bibfield  {title} {\bibinfo {title} {Properties of
  nitrogen-vacancy centers in diamond: the group theoretic approach},\
  }\href@noop {} {\bibfield  {journal} {\bibinfo  {journal} {New Journal of
  Physics}\ }\textbf {\bibinfo {volume} {13}},\ \bibinfo {pages} {025025}
  (\bibinfo {year} {2011})}\BibitemShut {NoStop}%
\bibitem [{\citenamefont {Doherty}\ \emph {et~al.}(2011)\citenamefont
  {Doherty}, \citenamefont {Manson}, \citenamefont {Delaney},\ and\
  \citenamefont {Hollenberg}}]{doherty2011negatively}%
  \BibitemOpen
  \bibfield  {author} {\bibinfo {author} {\bibfnamefont {M.~W.}\ \bibnamefont
  {Doherty}}, \bibinfo {author} {\bibfnamefont {N.~B.}\ \bibnamefont {Manson}},
  \bibinfo {author} {\bibfnamefont {P.}~\bibnamefont {Delaney}},\ and\ \bibinfo
  {author} {\bibfnamefont {L.~C.~L.}\ \bibnamefont {Hollenberg}},\ }\bibfield
  {title} {\bibinfo {title} {The negatively charged nitrogen-vacancy centre in
  diamond: the electronic solution},\ }\href@noop {} {\bibfield  {journal}
  {\bibinfo  {journal} {New Journal of Physics}\ }\textbf {\bibinfo {volume}
  {13}},\ \bibinfo {pages} {025019} (\bibinfo {year} {2011})}\BibitemShut
  {NoStop}%
\bibitem [{\citenamefont {Acosta}\ \emph {et~al.}(2012)\citenamefont {Acosta},
  \citenamefont {Santori}, \citenamefont {Faraon}, \citenamefont {Huang},
  \citenamefont {Fu}, \citenamefont {Stacey}, \citenamefont {Simpson},
  \citenamefont {Ganesan}, \citenamefont {Tomljenovic-Hanic}, \citenamefont
  {Greentree}, \citenamefont {Prawer},\ and\ \citenamefont
  {Beausoleil}}]{acosta2012dynamic}%
  \BibitemOpen
  \bibfield  {author} {\bibinfo {author} {\bibfnamefont {V.~M.}\ \bibnamefont
  {Acosta}}, \bibinfo {author} {\bibfnamefont {C.}~\bibnamefont {Santori}},
  \bibinfo {author} {\bibfnamefont {A.}~\bibnamefont {Faraon}}, \bibinfo
  {author} {\bibfnamefont {Z.}~\bibnamefont {Huang}}, \bibinfo {author}
  {\bibfnamefont {K.-M.~C.}\ \bibnamefont {Fu}}, \bibinfo {author}
  {\bibfnamefont {A.}~\bibnamefont {Stacey}}, \bibinfo {author} {\bibfnamefont
  {D.~A.}\ \bibnamefont {Simpson}}, \bibinfo {author} {\bibfnamefont
  {K.}~\bibnamefont {Ganesan}}, \bibinfo {author} {\bibfnamefont
  {S.}~\bibnamefont {Tomljenovic-Hanic}}, \bibinfo {author} {\bibfnamefont
  {A.~D.}\ \bibnamefont {Greentree}}, \bibinfo {author} {\bibfnamefont
  {S.}~\bibnamefont {Prawer}},\ and\ \bibinfo {author} {\bibfnamefont {R.~G.}\
  \bibnamefont {Beausoleil}},\ }\bibfield  {title} {\bibinfo {title} {Dynamic
  stabilization of the optical resonances of single nitrogen-vacancy centers in
  diamond},\ }\href@noop {} {\bibfield  {journal} {\bibinfo  {journal} {Phys.
  Rev. Lett.}\ }\textbf {\bibinfo {volume} {108}},\ \bibinfo {pages} {206401}
  (\bibinfo {year} {2012})}\BibitemShut {NoStop}%
\bibitem [{\citenamefont {Doherty}\ \emph {et~al.}(2012)\citenamefont
  {Doherty}, \citenamefont {Dolde}, \citenamefont {Fedder}, \citenamefont
  {Jelezko}, \citenamefont {Wrachtrup}, \citenamefont {Manson},\ and\
  \citenamefont {Hollenberg}}]{doherty2012}%
  \BibitemOpen
  \bibfield  {author} {\bibinfo {author} {\bibfnamefont {M.~W.}\ \bibnamefont
  {Doherty}}, \bibinfo {author} {\bibfnamefont {F.}~\bibnamefont {Dolde}},
  \bibinfo {author} {\bibfnamefont {H.}~\bibnamefont {Fedder}}, \bibinfo
  {author} {\bibfnamefont {F.}~\bibnamefont {Jelezko}}, \bibinfo {author}
  {\bibfnamefont {J.}~\bibnamefont {Wrachtrup}}, \bibinfo {author}
  {\bibfnamefont {N.~B.}\ \bibnamefont {Manson}},\ and\ \bibinfo {author}
  {\bibfnamefont {L.~C.~L.}\ \bibnamefont {Hollenberg}},\ }\bibfield  {title}
  {\bibinfo {title} {Theory of the ground-state spin of the
  $\mathrm{NV}{}^{\ensuremath{-}}$ center in diamond},\ }\href@noop {}
  {\bibfield  {journal} {\bibinfo  {journal} {Phys. Rev. B}\ }\textbf {\bibinfo
  {volume} {85}},\ \bibinfo {pages} {205203} (\bibinfo {year}
  {2012})}\BibitemShut {NoStop}%
\bibitem [{\citenamefont {Doherty}\ \emph {et~al.}(2013)\citenamefont
  {Doherty}, \citenamefont {Manson}, \citenamefont {Delaney}, \citenamefont
  {Jelezko}, \citenamefont {Wrachtrup},\ and\ \citenamefont
  {Hollenberg}}]{doherty2013nitrogen}%
  \BibitemOpen
  \bibfield  {author} {\bibinfo {author} {\bibfnamefont {M.~W.}\ \bibnamefont
  {Doherty}}, \bibinfo {author} {\bibfnamefont {N.~B.}\ \bibnamefont {Manson}},
  \bibinfo {author} {\bibfnamefont {P.}~\bibnamefont {Delaney}}, \bibinfo
  {author} {\bibfnamefont {F.}~\bibnamefont {Jelezko}}, \bibinfo {author}
  {\bibfnamefont {J.}~\bibnamefont {Wrachtrup}},\ and\ \bibinfo {author}
  {\bibfnamefont {L.~C.~L.}\ \bibnamefont {Hollenberg}},\ }\bibfield  {title}
  {\bibinfo {title} {The nitrogen-vacancy colour centre in diamond},\
  }\href@noop {} {\bibfield  {journal} {\bibinfo  {journal} {Physics Reports}\
  }\textbf {\bibinfo {volume} {528}},\ \bibinfo {pages} {1 } (\bibinfo {year}
  {2013})}\BibitemShut {NoStop}%
\bibitem [{\citenamefont {Dolde}\ \emph
  {et~al.}(2014{\natexlab{b}})\citenamefont {Dolde}, \citenamefont {Doherty},
  \citenamefont {Michl}, \citenamefont {Jakobi}, \citenamefont {Naydenov},
  \citenamefont {Pezzagna}, \citenamefont {Meijer}, \citenamefont {Neumann},
  \citenamefont {Jelezko}, \citenamefont {Manson},\ and\ \citenamefont
  {Wrachtrup}}]{dolde2014nanoscale}%
  \BibitemOpen
  \bibfield  {author} {\bibinfo {author} {\bibfnamefont {F.}~\bibnamefont
  {Dolde}}, \bibinfo {author} {\bibfnamefont {M.~W.}\ \bibnamefont {Doherty}},
  \bibinfo {author} {\bibfnamefont {J.}~\bibnamefont {Michl}}, \bibinfo
  {author} {\bibfnamefont {I.}~\bibnamefont {Jakobi}}, \bibinfo {author}
  {\bibfnamefont {B.}~\bibnamefont {Naydenov}}, \bibinfo {author}
  {\bibfnamefont {S.}~\bibnamefont {Pezzagna}}, \bibinfo {author}
  {\bibfnamefont {J.}~\bibnamefont {Meijer}}, \bibinfo {author} {\bibfnamefont
  {P.}~\bibnamefont {Neumann}}, \bibinfo {author} {\bibfnamefont
  {F.}~\bibnamefont {Jelezko}}, \bibinfo {author} {\bibfnamefont {N.~B.}\
  \bibnamefont {Manson}},\ and\ \bibinfo {author} {\bibfnamefont
  {J.}~\bibnamefont {Wrachtrup}},\ }\bibfield  {title} {\bibinfo {title}
  {Nanoscale detection of a single fundamental charge in ambient conditions
  using the $\mathrm{NV}{}^{\ensuremath{-}}$ center in diamond},\ }\href@noop
  {} {\bibfield  {journal} {\bibinfo  {journal} {Phys. Rev. Lett.}\ }\textbf
  {\bibinfo {volume} {112}},\ \bibinfo {pages} {097603} (\bibinfo {year}
  {2014}{\natexlab{b}})}\BibitemShut {NoStop}%
\bibitem [{\citenamefont {Rogers}\ \emph {et~al.}(2015)\citenamefont {Rogers},
  \citenamefont {Doherty}, \citenamefont {Barson}, \citenamefont {Onoda},
  \citenamefont {Ohshima},\ and\ \citenamefont {Manson}}]{rogers2015singlet}%
  \BibitemOpen
  \bibfield  {author} {\bibinfo {author} {\bibfnamefont {L.~J.}\ \bibnamefont
  {Rogers}}, \bibinfo {author} {\bibfnamefont {M.~W.}\ \bibnamefont {Doherty}},
  \bibinfo {author} {\bibfnamefont {M.~S.~J.}\ \bibnamefont {Barson}}, \bibinfo
  {author} {\bibfnamefont {S.}~\bibnamefont {Onoda}}, \bibinfo {author}
  {\bibfnamefont {T.}~\bibnamefont {Ohshima}},\ and\ \bibinfo {author}
  {\bibfnamefont {N.~B.}\ \bibnamefont {Manson}},\ }\bibfield  {title}
  {\bibinfo {title} {Singlet levels of the {NV}-centre in diamond},\
  }\href@noop {} {\bibfield  {journal} {\bibinfo  {journal} {New Journal of
  Physics}\ }\textbf {\bibinfo {volume} {17}},\ \bibinfo {pages} {013048}
  (\bibinfo {year} {2015})}\BibitemShut {NoStop}%
\bibitem [{\citenamefont {Iv\'ady}\ \emph {et~al.}(2015)\citenamefont
  {Iv\'ady}, \citenamefont {Sz\'asz}, \citenamefont {Falk}, \citenamefont
  {Klimov}, \citenamefont {Christle}, \citenamefont {Janz\'en}, \citenamefont
  {Abrikosov}, \citenamefont {Awschalom},\ and\ \citenamefont
  {Gali}}]{ivady2015theoretical}%
  \BibitemOpen
  \bibfield  {author} {\bibinfo {author} {\bibfnamefont {V.}~\bibnamefont
  {Iv\'ady}}, \bibinfo {author} {\bibfnamefont {K.}~\bibnamefont {Sz\'asz}},
  \bibinfo {author} {\bibfnamefont {A.~L.}\ \bibnamefont {Falk}}, \bibinfo
  {author} {\bibfnamefont {P.~V.}\ \bibnamefont {Klimov}}, \bibinfo {author}
  {\bibfnamefont {D.~J.}\ \bibnamefont {Christle}}, \bibinfo {author}
  {\bibfnamefont {E.}~\bibnamefont {Janz\'en}}, \bibinfo {author}
  {\bibfnamefont {I.~A.}\ \bibnamefont {Abrikosov}}, \bibinfo {author}
  {\bibfnamefont {D.~D.}\ \bibnamefont {Awschalom}},\ and\ \bibinfo {author}
  {\bibfnamefont {A.}~\bibnamefont {Gali}},\ }\bibfield  {title} {\bibinfo
  {title} {Theoretical model of dynamic spin polarization of nuclei coupled to
  paramagnetic point defects in diamond and silicon carbide},\ }\href@noop {}
  {\bibfield  {journal} {\bibinfo  {journal} {Phys. Rev. B}\ }\textbf {\bibinfo
  {volume} {92}},\ \bibinfo {pages} {115206} (\bibinfo {year}
  {2015})}\BibitemShut {NoStop}%
\bibitem [{\citenamefont {Seo}\ \emph {et~al.}(2016)\citenamefont {Seo},
  \citenamefont {Falk}, \citenamefont {Klimov}, \citenamefont {Miao},
  \citenamefont {Galli},\ and\ \citenamefont {Awschalom}}]{seo2016}%
  \BibitemOpen
  \bibfield  {author} {\bibinfo {author} {\bibfnamefont {H.}~\bibnamefont
  {Seo}}, \bibinfo {author} {\bibfnamefont {A.~L.}\ \bibnamefont {Falk}},
  \bibinfo {author} {\bibfnamefont {P.~V.}\ \bibnamefont {Klimov}}, \bibinfo
  {author} {\bibfnamefont {K.~C.}\ \bibnamefont {Miao}}, \bibinfo {author}
  {\bibfnamefont {G.}~\bibnamefont {Galli}},\ and\ \bibinfo {author}
  {\bibfnamefont {D.~D.}\ \bibnamefont {Awschalom}},\ }\bibfield  {title}
  {\bibinfo {title} {Quantum decoherence dynamics of divacancy spins in silicon
  carbide},\ }\href@noop {} {\bibfield  {journal} {\bibinfo  {journal} {Nature
  Communications}\ }\textbf {\bibinfo {volume} {7}},\ \bibinfo {pages} {12935}
  (\bibinfo {year} {2016})}\BibitemShut {NoStop}%
\bibitem [{\citenamefont {Klimov}\ \emph {et~al.}(2014)\citenamefont {Klimov},
  \citenamefont {Falk}, \citenamefont {Buckley},\ and\ \citenamefont
  {Awschalom}}]{klimov2014electrically}%
  \BibitemOpen
  \bibfield  {author} {\bibinfo {author} {\bibfnamefont {P.~V.}\ \bibnamefont
  {Klimov}}, \bibinfo {author} {\bibfnamefont {A.~L.}\ \bibnamefont {Falk}},
  \bibinfo {author} {\bibfnamefont {B.~B.}\ \bibnamefont {Buckley}},\ and\
  \bibinfo {author} {\bibfnamefont {D.~D.}\ \bibnamefont {Awschalom}},\
  }\bibfield  {title} {\bibinfo {title} {Electrically driven spin resonance in
  silicon carbide color centers},\ }\href@noop {} {\bibfield  {journal}
  {\bibinfo  {journal} {Phys. Rev. Lett.}\ }\textbf {\bibinfo {volume} {112}},\
  \bibinfo {pages} {087601} (\bibinfo {year} {2014})}\BibitemShut {NoStop}%
\bibitem [{\citenamefont {Christle}\ \emph {et~al.}(2017)\citenamefont
  {Christle}, \citenamefont {Klimov}, \citenamefont {de~las Casas},
  \citenamefont {Sz\'asz}, \citenamefont {Iv\'ady}, \citenamefont
  {Jokubavicius}, \citenamefont {Ul~Hassan}, \citenamefont {Syv\"aj\"arvi},
  \citenamefont {Koehl}, \citenamefont {Ohshima}, \citenamefont {Son},
  \citenamefont {Janz\'en}, \citenamefont {Gali},\ and\ \citenamefont
  {Awschalom}}]{christle2017}%
  \BibitemOpen
  \bibfield  {author} {\bibinfo {author} {\bibfnamefont {D.~J.}\ \bibnamefont
  {Christle}}, \bibinfo {author} {\bibfnamefont {P.~V.}\ \bibnamefont
  {Klimov}}, \bibinfo {author} {\bibfnamefont {C.~F.}\ \bibnamefont {de~las
  Casas}}, \bibinfo {author} {\bibfnamefont {K.}~\bibnamefont {Sz\'asz}},
  \bibinfo {author} {\bibfnamefont {V.}~\bibnamefont {Iv\'ady}}, \bibinfo
  {author} {\bibfnamefont {V.}~\bibnamefont {Jokubavicius}}, \bibinfo {author}
  {\bibfnamefont {J.}~\bibnamefont {Ul~Hassan}}, \bibinfo {author}
  {\bibfnamefont {M.}~\bibnamefont {Syv\"aj\"arvi}}, \bibinfo {author}
  {\bibfnamefont {W.~F.}\ \bibnamefont {Koehl}}, \bibinfo {author}
  {\bibfnamefont {T.}~\bibnamefont {Ohshima}}, \bibinfo {author} {\bibfnamefont
  {N.~T.}\ \bibnamefont {Son}}, \bibinfo {author} {\bibfnamefont
  {E.}~\bibnamefont {Janz\'en}}, \bibinfo {author} {\bibfnamefont
  {A.}~\bibnamefont {Gali}},\ and\ \bibinfo {author} {\bibfnamefont {D.~D.}\
  \bibnamefont {Awschalom}},\ }\bibfield  {title} {\bibinfo {title} {Isolated
  spin qubits in {SiC} with a high-fidelity infrared spin-to-photon
  interface},\ }\href@noop {} {\bibfield  {journal} {\bibinfo  {journal} {Phys.
  Rev. X}\ }\textbf {\bibinfo {volume} {7}},\ \bibinfo {pages} {021046}
  (\bibinfo {year} {2017})}\BibitemShut {NoStop}%
\bibitem [{\citenamefont {Miao}\ \emph
  {et~al.}(2019{\natexlab{a}})\citenamefont {Miao}, \citenamefont {Bourassa},
  \citenamefont {Anderson}, \citenamefont {Whiteley}, \citenamefont {Crook},
  \citenamefont {Bayliss}, \citenamefont {Wolfowicz}, \citenamefont {Thiering},
  \citenamefont {Udvarhelyi}, \citenamefont {Ivády}, \citenamefont {Abe},
  \citenamefont {Ohshima}, \citenamefont {Ádám Gali},\ and\ \citenamefont
  {Awschalom}}]{miao2019electrically}%
  \BibitemOpen
  \bibfield  {author} {\bibinfo {author} {\bibfnamefont {K.~C.}\ \bibnamefont
  {Miao}}, \bibinfo {author} {\bibfnamefont {A.}~\bibnamefont {Bourassa}},
  \bibinfo {author} {\bibfnamefont {C.~P.}\ \bibnamefont {Anderson}}, \bibinfo
  {author} {\bibfnamefont {S.~J.}\ \bibnamefont {Whiteley}}, \bibinfo {author}
  {\bibfnamefont {A.~L.}\ \bibnamefont {Crook}}, \bibinfo {author}
  {\bibfnamefont {S.~L.}\ \bibnamefont {Bayliss}}, \bibinfo {author}
  {\bibfnamefont {G.}~\bibnamefont {Wolfowicz}}, \bibinfo {author}
  {\bibfnamefont {G.}~\bibnamefont {Thiering}}, \bibinfo {author}
  {\bibfnamefont {P.}~\bibnamefont {Udvarhelyi}}, \bibinfo {author}
  {\bibfnamefont {V.}~\bibnamefont {Ivády}}, \bibinfo {author} {\bibfnamefont
  {H.}~\bibnamefont {Abe}}, \bibinfo {author} {\bibfnamefont {T.}~\bibnamefont
  {Ohshima}}, \bibinfo {author} {\bibnamefont {Ádám Gali}},\ and\ \bibinfo
  {author} {\bibfnamefont {D.~D.}\ \bibnamefont {Awschalom}},\ }\bibfield
  {title} {\bibinfo {title} {Electrically driven optical interferometry with
  spins in silicon carbide},\ }\href@noop {} {\bibfield  {journal} {\bibinfo
  {journal} {Science Advances}\ }\textbf {\bibinfo {volume} {5}},\ \bibinfo
  {pages} {eaay0527} (\bibinfo {year} {2019}{\natexlab{a}})}\BibitemShut
  {NoStop}%
\bibitem [{\citenamefont {Kiel}\ and\ \citenamefont
  {Mims}(1972)}]{PhysRevB.5.803}%
  \BibitemOpen
  \bibfield  {author} {\bibinfo {author} {\bibfnamefont {A.}~\bibnamefont
  {Kiel}}\ and\ \bibinfo {author} {\bibfnamefont {W.~B.}\ \bibnamefont
  {Mims}},\ }\bibfield  {title} {\bibinfo {title} {Linear electric field effect
  in paramagnetic resonance for {CdS}: {${\mathrm{Mn}}^{2+}$}},\ }\href
  {https://doi.org/10.1103/PhysRevB.5.803} {\bibfield  {journal} {\bibinfo
  {journal} {Phys. Rev. B}\ }\textbf {\bibinfo {volume} {5}},\ \bibinfo {pages}
  {803} (\bibinfo {year} {1972})}\BibitemShut {NoStop}%
\bibitem [{\citenamefont {Udvarhelyi}\ \emph {et~al.}(2018)\citenamefont
  {Udvarhelyi}, \citenamefont {Shkolnikov}, \citenamefont {Gali}, \citenamefont
  {Burkard},\ and\ \citenamefont {P\'alyi}}]{PhysRevB.98.075201}%
  \BibitemOpen
  \bibfield  {author} {\bibinfo {author} {\bibfnamefont {P.}~\bibnamefont
  {Udvarhelyi}}, \bibinfo {author} {\bibfnamefont {V.~O.}\ \bibnamefont
  {Shkolnikov}}, \bibinfo {author} {\bibfnamefont {A.}~\bibnamefont {Gali}},
  \bibinfo {author} {\bibfnamefont {G.}~\bibnamefont {Burkard}},\ and\ \bibinfo
  {author} {\bibfnamefont {A.}~\bibnamefont {P\'alyi}},\ }\bibfield  {title}
  {\bibinfo {title} {Spin-strain interaction in nitrogen-vacancy centers in
  diamond},\ }\href {https://doi.org/10.1103/PhysRevB.98.075201} {\bibfield
  {journal} {\bibinfo  {journal} {Phys. Rev. B}\ }\textbf {\bibinfo {volume}
  {98}},\ \bibinfo {pages} {075201} (\bibinfo {year} {2018})}\BibitemShut
  {NoStop}%
\bibitem [{\citenamefont {Jarmola}\ \emph {et~al.}(2012)\citenamefont
  {Jarmola}, \citenamefont {Acosta}, \citenamefont {Jensen}, \citenamefont
  {Chemerisov},\ and\ \citenamefont {Budker}}]{PhysRevLett.108.197601}%
  \BibitemOpen
  \bibfield  {author} {\bibinfo {author} {\bibfnamefont {A.}~\bibnamefont
  {Jarmola}}, \bibinfo {author} {\bibfnamefont {V.~M.}\ \bibnamefont {Acosta}},
  \bibinfo {author} {\bibfnamefont {K.}~\bibnamefont {Jensen}}, \bibinfo
  {author} {\bibfnamefont {S.}~\bibnamefont {Chemerisov}},\ and\ \bibinfo
  {author} {\bibfnamefont {D.}~\bibnamefont {Budker}},\ }\bibfield  {title}
  {\bibinfo {title} {Temperature- and magnetic-field-dependent longitudinal
  spin relaxation in nitrogen-vacancy ensembles in diamond},\ }\href
  {https://doi.org/10.1103/PhysRevLett.108.197601} {\bibfield  {journal}
  {\bibinfo  {journal} {Phys. Rev. Lett.}\ }\textbf {\bibinfo {volume} {108}},\
  \bibinfo {pages} {197601} (\bibinfo {year} {2012})}\BibitemShut {NoStop}%
\bibitem [{\citenamefont {Wang}\ \emph {et~al.}(2013)\citenamefont {Wang},
  \citenamefont {Shin}, \citenamefont {Avalos}, \citenamefont {Seltzer},
  \citenamefont {Budker}, \citenamefont {Pines},\ and\ \citenamefont
  {Bajaj}}]{wang2013sensitive}%
  \BibitemOpen
  \bibfield  {author} {\bibinfo {author} {\bibfnamefont {H.-J.}\ \bibnamefont
  {Wang}}, \bibinfo {author} {\bibfnamefont {C.~S.}\ \bibnamefont {Shin}},
  \bibinfo {author} {\bibfnamefont {C.~E.}\ \bibnamefont {Avalos}}, \bibinfo
  {author} {\bibfnamefont {S.~J.}\ \bibnamefont {Seltzer}}, \bibinfo {author}
  {\bibfnamefont {D.}~\bibnamefont {Budker}}, \bibinfo {author} {\bibfnamefont
  {A.}~\bibnamefont {Pines}},\ and\ \bibinfo {author} {\bibfnamefont {V.~S.}\
  \bibnamefont {Bajaj}},\ }\bibfield  {title} {\bibinfo {title} {Sensitive
  magnetic control of ensemble nuclear spin hyperpolarization in diamond},\
  }\href@noop {} {\bibfield  {journal} {\bibinfo  {journal} {Nature
  Communications}\ }\textbf {\bibinfo {volume} {4}},\ \bibinfo {pages} {1}
  (\bibinfo {year} {2013})}\BibitemShut {NoStop}%
\bibitem [{\citenamefont {Falk}\ \emph {et~al.}(2015)\citenamefont {Falk},
  \citenamefont {Klimov}, \citenamefont {Iv\'ady}, \citenamefont {Sz\'asz},
  \citenamefont {Christle}, \citenamefont {Koehl}, \citenamefont {Gali},\ and\
  \citenamefont {Awschalom}}]{PhysRevLett.114.247603}%
  \BibitemOpen
  \bibfield  {author} {\bibinfo {author} {\bibfnamefont {A.~L.}\ \bibnamefont
  {Falk}}, \bibinfo {author} {\bibfnamefont {P.~V.}\ \bibnamefont {Klimov}},
  \bibinfo {author} {\bibfnamefont {V.}~\bibnamefont {Iv\'ady}}, \bibinfo
  {author} {\bibfnamefont {K.}~\bibnamefont {Sz\'asz}}, \bibinfo {author}
  {\bibfnamefont {D.~J.}\ \bibnamefont {Christle}}, \bibinfo {author}
  {\bibfnamefont {W.~F.}\ \bibnamefont {Koehl}}, \bibinfo {author}
  {\bibfnamefont {A.}~\bibnamefont {Gali}},\ and\ \bibinfo {author}
  {\bibfnamefont {D.~D.}\ \bibnamefont {Awschalom}},\ }\bibfield  {title}
  {\bibinfo {title} {Optical polarization of nuclear spins in silicon
  carbide},\ }\href {https://doi.org/10.1103/PhysRevLett.114.247603} {\bibfield
   {journal} {\bibinfo  {journal} {Phys. Rev. Lett.}\ }\textbf {\bibinfo
  {volume} {114}},\ \bibinfo {pages} {247603} (\bibinfo {year}
  {2015})}\BibitemShut {NoStop}%
\bibitem [{\citenamefont {Wickenbrock}\ \emph {et~al.}(2016)\citenamefont
  {Wickenbrock}, \citenamefont {Zheng}, \citenamefont {Bougas}, \citenamefont
  {Leefer}, \citenamefont {Afach}, \citenamefont {Jarmola}, \citenamefont
  {Acosta},\ and\ \citenamefont {Budker}}]{doi:10.1063/1.4960171}%
  \BibitemOpen
  \bibfield  {author} {\bibinfo {author} {\bibfnamefont {A.}~\bibnamefont
  {Wickenbrock}}, \bibinfo {author} {\bibfnamefont {H.}~\bibnamefont {Zheng}},
  \bibinfo {author} {\bibfnamefont {L.}~\bibnamefont {Bougas}}, \bibinfo
  {author} {\bibfnamefont {N.}~\bibnamefont {Leefer}}, \bibinfo {author}
  {\bibfnamefont {S.}~\bibnamefont {Afach}}, \bibinfo {author} {\bibfnamefont
  {A.}~\bibnamefont {Jarmola}}, \bibinfo {author} {\bibfnamefont {V.~M.}\
  \bibnamefont {Acosta}},\ and\ \bibinfo {author} {\bibfnamefont
  {D.}~\bibnamefont {Budker}},\ }\bibfield  {title} {\bibinfo {title}
  {Microwave-free magnetometry with nitrogen-vacancy centers in diamond},\
  }\href {https://doi.org/10.1063/1.4960171} {\bibfield  {journal} {\bibinfo
  {journal} {Applied Physics Letters}\ }\textbf {\bibinfo {volume} {109}},\
  \bibinfo {pages} {053505} (\bibinfo {year} {2016})}\BibitemShut {NoStop}%
\bibitem [{\citenamefont {Broadway}\ \emph {et~al.}(2016)\citenamefont
  {Broadway}, \citenamefont {Wood}, \citenamefont {Hall}, \citenamefont
  {Stacey}, \citenamefont {Markham}, \citenamefont {Simpson}, \citenamefont
  {Tetienne},\ and\ \citenamefont {Hollenberg}}]{PhysRevApplied.6.064001}%
  \BibitemOpen
  \bibfield  {author} {\bibinfo {author} {\bibfnamefont {D.~A.}\ \bibnamefont
  {Broadway}}, \bibinfo {author} {\bibfnamefont {J.~D.~A.}\ \bibnamefont
  {Wood}}, \bibinfo {author} {\bibfnamefont {L.~T.}\ \bibnamefont {Hall}},
  \bibinfo {author} {\bibfnamefont {A.}~\bibnamefont {Stacey}}, \bibinfo
  {author} {\bibfnamefont {M.}~\bibnamefont {Markham}}, \bibinfo {author}
  {\bibfnamefont {D.~A.}\ \bibnamefont {Simpson}}, \bibinfo {author}
  {\bibfnamefont {J.-P.}\ \bibnamefont {Tetienne}},\ and\ \bibinfo {author}
  {\bibfnamefont {L.~C.~L.}\ \bibnamefont {Hollenberg}},\ }\bibfield  {title}
  {\bibinfo {title} {Anticrossing spin dynamics of diamond nitrogen-vacancy
  centers and all-optical low-frequency magnetometry},\ }\href
  {https://doi.org/10.1103/PhysRevApplied.6.064001} {\bibfield  {journal}
  {\bibinfo  {journal} {Phys. Rev. Applied}\ }\textbf {\bibinfo {volume} {6}},\
  \bibinfo {pages} {064001} (\bibinfo {year} {2016})}\BibitemShut {NoStop}%
\bibitem [{\citenamefont {Tarasenko}\ \emph {et~al.}(2018)\citenamefont
  {Tarasenko}, \citenamefont {Poshakinskiy}, \citenamefont {Simin},
  \citenamefont {Soltamov}, \citenamefont {Mokhov}, \citenamefont {Baranov},
  \citenamefont {Dyakonov},\ and\ \citenamefont
  {Astakhov}}]{https://doi.org/10.1002/pssb.201700258}%
  \BibitemOpen
  \bibfield  {author} {\bibinfo {author} {\bibfnamefont {S.~A.}\ \bibnamefont
  {Tarasenko}}, \bibinfo {author} {\bibfnamefont {A.~V.}\ \bibnamefont
  {Poshakinskiy}}, \bibinfo {author} {\bibfnamefont {D.}~\bibnamefont {Simin}},
  \bibinfo {author} {\bibfnamefont {V.~A.}\ \bibnamefont {Soltamov}}, \bibinfo
  {author} {\bibfnamefont {E.~N.}\ \bibnamefont {Mokhov}}, \bibinfo {author}
  {\bibfnamefont {P.~G.}\ \bibnamefont {Baranov}}, \bibinfo {author}
  {\bibfnamefont {V.}~\bibnamefont {Dyakonov}},\ and\ \bibinfo {author}
  {\bibfnamefont {G.~V.}\ \bibnamefont {Astakhov}},\ }\bibfield  {title}
  {\bibinfo {title} {Spin and optical properties of silicon vacancies in
  silicon carbide - a review},\ }\href
  {https://doi.org/https://doi.org/10.1002/pssb.201700258} {\bibfield
  {journal} {\bibinfo  {journal} {Physica Status Solidi (b)}\ }\textbf
  {\bibinfo {volume} {255}},\ \bibinfo {pages} {1700258} (\bibinfo {year}
  {2018})}\BibitemShut {NoStop}%
\bibitem [{\citenamefont {Auzinsh}\ \emph {et~al.}(2019)\citenamefont
  {Auzinsh}, \citenamefont {Berzins}, \citenamefont {Budker}, \citenamefont
  {Busaite}, \citenamefont {Ferber}, \citenamefont {Gahbauer}, \citenamefont
  {Lazda}, \citenamefont {Wickenbrock},\ and\ \citenamefont
  {Zheng}}]{PhysRevB.100.075204}%
  \BibitemOpen
  \bibfield  {author} {\bibinfo {author} {\bibfnamefont {M.}~\bibnamefont
  {Auzinsh}}, \bibinfo {author} {\bibfnamefont {A.}~\bibnamefont {Berzins}},
  \bibinfo {author} {\bibfnamefont {D.}~\bibnamefont {Budker}}, \bibinfo
  {author} {\bibfnamefont {L.}~\bibnamefont {Busaite}}, \bibinfo {author}
  {\bibfnamefont {R.}~\bibnamefont {Ferber}}, \bibinfo {author} {\bibfnamefont
  {F.}~\bibnamefont {Gahbauer}}, \bibinfo {author} {\bibfnamefont
  {R.}~\bibnamefont {Lazda}}, \bibinfo {author} {\bibfnamefont
  {A.}~\bibnamefont {Wickenbrock}},\ and\ \bibinfo {author} {\bibfnamefont
  {H.}~\bibnamefont {Zheng}},\ }\bibfield  {title} {\bibinfo {title} {Hyperfine
  level structure in nitrogen-vacancy centers near the ground-state level
  anticrossing},\ }\href {https://doi.org/10.1103/PhysRevB.100.075204}
  {\bibfield  {journal} {\bibinfo  {journal} {Phys. Rev. B}\ }\textbf {\bibinfo
  {volume} {100}},\ \bibinfo {pages} {075204} (\bibinfo {year}
  {2019})}\BibitemShut {NoStop}%
\bibitem [{\citenamefont {Busaite}\ \emph {et~al.}(2020)\citenamefont
  {Busaite}, \citenamefont {Lazda}, \citenamefont {Berzins}, \citenamefont
  {Auzinsh}, \citenamefont {Ferber},\ and\ \citenamefont
  {Gahbauer}}]{PhysRevB.102.224101}%
  \BibitemOpen
  \bibfield  {author} {\bibinfo {author} {\bibfnamefont {L.}~\bibnamefont
  {Busaite}}, \bibinfo {author} {\bibfnamefont {R.}~\bibnamefont {Lazda}},
  \bibinfo {author} {\bibfnamefont {A.}~\bibnamefont {Berzins}}, \bibinfo
  {author} {\bibfnamefont {M.}~\bibnamefont {Auzinsh}}, \bibinfo {author}
  {\bibfnamefont {R.}~\bibnamefont {Ferber}},\ and\ \bibinfo {author}
  {\bibfnamefont {F.}~\bibnamefont {Gahbauer}},\ }\bibfield  {title} {\bibinfo
  {title} {Dynamic $^{14}\mathrm{N}$ nuclear spin polarization in
  nitrogen-vacancy centers in diamond},\ }\href
  {https://doi.org/10.1103/PhysRevB.102.224101} {\bibfield  {journal} {\bibinfo
   {journal} {Phys. Rev. B}\ }\textbf {\bibinfo {volume} {102}},\ \bibinfo
  {pages} {224101} (\bibinfo {year} {2020})}\BibitemShut {NoStop}%
\bibitem [{\citenamefont {Zheng}\ \emph {et~al.}(2020)\citenamefont {Zheng},
  \citenamefont {Sun}, \citenamefont {Chatzidrosos}, \citenamefont {Zhang},
  \citenamefont {Nakamura}, \citenamefont {Sumiya}, \citenamefont {Ohshima},
  \citenamefont {Isoya}, \citenamefont {Wrachtrup}, \citenamefont
  {Wickenbrock},\ and\ \citenamefont {Budker}}]{PhysRevApplied.13.044023}%
  \BibitemOpen
  \bibfield  {author} {\bibinfo {author} {\bibfnamefont {H.}~\bibnamefont
  {Zheng}}, \bibinfo {author} {\bibfnamefont {Z.}~\bibnamefont {Sun}}, \bibinfo
  {author} {\bibfnamefont {G.}~\bibnamefont {Chatzidrosos}}, \bibinfo {author}
  {\bibfnamefont {C.}~\bibnamefont {Zhang}}, \bibinfo {author} {\bibfnamefont
  {K.}~\bibnamefont {Nakamura}}, \bibinfo {author} {\bibfnamefont
  {H.}~\bibnamefont {Sumiya}}, \bibinfo {author} {\bibfnamefont
  {T.}~\bibnamefont {Ohshima}}, \bibinfo {author} {\bibfnamefont
  {J.}~\bibnamefont {Isoya}}, \bibinfo {author} {\bibfnamefont
  {J.}~\bibnamefont {Wrachtrup}}, \bibinfo {author} {\bibfnamefont
  {A.}~\bibnamefont {Wickenbrock}},\ and\ \bibinfo {author} {\bibfnamefont
  {D.}~\bibnamefont {Budker}},\ }\bibfield  {title} {\bibinfo {title}
  {Microwave-free vector magnetometry with nitrogen-vacancy centers along a
  single axis in diamond},\ }\href
  {https://doi.org/10.1103/PhysRevApplied.13.044023} {\bibfield  {journal}
  {\bibinfo  {journal} {Phys. Rev. Applied}\ }\textbf {\bibinfo {volume}
  {13}},\ \bibinfo {pages} {044023} (\bibinfo {year} {2020})}\BibitemShut
  {NoStop}%
\bibitem [{\citenamefont {Iv\'ady}\ \emph {et~al.}(2021)\citenamefont
  {Iv\'ady}, \citenamefont {Zheng}, \citenamefont {Wickenbrock}, \citenamefont
  {Bougas}, \citenamefont {Chatzidrosos}, \citenamefont {Nakamura},
  \citenamefont {Sumiya}, \citenamefont {Ohshima}, \citenamefont {Isoya},
  \citenamefont {Budker}, \citenamefont {Abrikosov},\ and\ \citenamefont
  {Gali}}]{PhysRevB.103.035307}%
  \BibitemOpen
  \bibfield  {author} {\bibinfo {author} {\bibfnamefont {V.}~\bibnamefont
  {Iv\'ady}}, \bibinfo {author} {\bibfnamefont {H.}~\bibnamefont {Zheng}},
  \bibinfo {author} {\bibfnamefont {A.}~\bibnamefont {Wickenbrock}}, \bibinfo
  {author} {\bibfnamefont {L.}~\bibnamefont {Bougas}}, \bibinfo {author}
  {\bibfnamefont {G.}~\bibnamefont {Chatzidrosos}}, \bibinfo {author}
  {\bibfnamefont {K.}~\bibnamefont {Nakamura}}, \bibinfo {author}
  {\bibfnamefont {H.}~\bibnamefont {Sumiya}}, \bibinfo {author} {\bibfnamefont
  {T.}~\bibnamefont {Ohshima}}, \bibinfo {author} {\bibfnamefont
  {J.}~\bibnamefont {Isoya}}, \bibinfo {author} {\bibfnamefont
  {D.}~\bibnamefont {Budker}}, \bibinfo {author} {\bibfnamefont {I.~A.}\
  \bibnamefont {Abrikosov}},\ and\ \bibinfo {author} {\bibfnamefont
  {A.}~\bibnamefont {Gali}},\ }\bibfield  {title} {\bibinfo {title}
  {Photoluminescence at the ground-state level anticrossing of the
  nitrogen-vacancy center in diamond: A comprehensive study},\ }\href
  {https://doi.org/10.1103/PhysRevB.103.035307} {\bibfield  {journal} {\bibinfo
   {journal} {Phys. Rev. B}\ }\textbf {\bibinfo {volume} {103}},\ \bibinfo
  {pages} {035307} (\bibinfo {year} {2021})}\BibitemShut {NoStop}%
\bibitem [{\citenamefont {Chen}\ \emph {et~al.}(2020)\citenamefont {Chen},
  \citenamefont {Bhave},\ and\ \citenamefont
  {Fuchs}}]{PhysRevApplied.13.054068}%
  \BibitemOpen
  \bibfield  {author} {\bibinfo {author} {\bibfnamefont {H.~Y.}\ \bibnamefont
  {Chen}}, \bibinfo {author} {\bibfnamefont {S.~A.}\ \bibnamefont {Bhave}},\
  and\ \bibinfo {author} {\bibfnamefont {G.~D.}\ \bibnamefont {Fuchs}},\
  }\bibfield  {title} {\bibinfo {title} {Acoustically driving the
  single-quantum spin transition of diamond nitrogen-vacancy centers},\ }\href
  {https://doi.org/10.1103/PhysRevApplied.13.054068} {\bibfield  {journal}
  {\bibinfo  {journal} {Phys. Rev. Applied}\ }\textbf {\bibinfo {volume}
  {13}},\ \bibinfo {pages} {054068} (\bibinfo {year} {2020})}\BibitemShut
  {NoStop}%
\bibitem [{\citenamefont {Makhlin}\ \emph {et~al.}(2004)\citenamefont
  {Makhlin}, \citenamefont {Schön},\ and\ \citenamefont
  {Shnirman}}]{MAKHLIN2004315}%
  \BibitemOpen
  \bibfield  {author} {\bibinfo {author} {\bibfnamefont {Y.}~\bibnamefont
  {Makhlin}}, \bibinfo {author} {\bibfnamefont {G.}~\bibnamefont {Schön}},\
  and\ \bibinfo {author} {\bibfnamefont {A.}~\bibnamefont {Shnirman}},\
  }\bibfield  {title} {\bibinfo {title} {Dissipative effects in {J}osephson
  qubits},\ }\href
  {https://doi.org/https://doi.org/10.1016/j.chemphys.2003.09.025} {\bibfield
  {journal} {\bibinfo  {journal} {Chemical Physics}\ }\textbf {\bibinfo
  {volume} {296}},\ \bibinfo {pages} {315} (\bibinfo {year}
  {2004})}\BibitemShut {NoStop}%
\bibitem [{\citenamefont {Welack}\ \emph {et~al.}(2006)\citenamefont {Welack},
  \citenamefont {Schreiber},\ and\ \citenamefont
  {Kleinekathöfer}}]{doi:10.1063/1.2162537}%
  \BibitemOpen
  \bibfield  {author} {\bibinfo {author} {\bibfnamefont {S.}~\bibnamefont
  {Welack}}, \bibinfo {author} {\bibfnamefont {M.}~\bibnamefont {Schreiber}},\
  and\ \bibinfo {author} {\bibfnamefont {U.}~\bibnamefont {Kleinekathöfer}},\
  }\bibfield  {title} {\bibinfo {title} {The influence of ultrafast laser
  pulses on electron transfer in molecular wires studied by a non-markovian
  density-matrix approach},\ }\href {https://doi.org/10.1063/1.2162537}
  {\bibfield  {journal} {\bibinfo  {journal} {The Journal of Chemical Physics}\
  }\textbf {\bibinfo {volume} {124}},\ \bibinfo {pages} {044712} (\bibinfo
  {year} {2006})}\BibitemShut {NoStop}%
\bibitem [{\citenamefont {Novikov}(1965)}]{novikov1965functionals}%
  \BibitemOpen
  \bibfield  {author} {\bibinfo {author} {\bibfnamefont {E.~A.}\ \bibnamefont
  {Novikov}},\ }\bibfield  {title} {\bibinfo {title} {Functionals and the
  random-force method in turbulence theory},\ }\href@noop {} {\bibfield
  {journal} {\bibinfo  {journal} {Sov. Phys. JETP}\ }\textbf {\bibinfo {volume}
  {20}},\ \bibinfo {pages} {1290} (\bibinfo {year} {1965})}\BibitemShut
  {NoStop}%
\bibitem [{\citenamefont {Budini}(2001)}]{PhysRevA.64.052110}%
  \BibitemOpen
  \bibfield  {author} {\bibinfo {author} {\bibfnamefont {A.~A.}\ \bibnamefont
  {Budini}},\ }\bibfield  {title} {\bibinfo {title} {Quantum systems subject to
  the action of classical stochastic fields},\ }\href
  {https://doi.org/10.1103/PhysRevA.64.052110} {\bibfield  {journal} {\bibinfo
  {journal} {Phys. Rev. A}\ }\textbf {\bibinfo {volume} {64}},\ \bibinfo
  {pages} {052110} (\bibinfo {year} {2001})}\BibitemShut {NoStop}%
\bibitem [{\citenamefont {Budini}(2000)}]{PhysRevA.63.012106}%
  \BibitemOpen
  \bibfield  {author} {\bibinfo {author} {\bibfnamefont {A.~A.}\ \bibnamefont
  {Budini}},\ }\bibfield  {title} {\bibinfo {title} {Non-markovian gaussian
  dissipative stochastic wave vector},\ }\href
  {https://doi.org/10.1103/PhysRevA.63.012106} {\bibfield  {journal} {\bibinfo
  {journal} {Phys. Rev. A}\ }\textbf {\bibinfo {volume} {63}},\ \bibinfo
  {pages} {012106} (\bibinfo {year} {2000})}\BibitemShut {NoStop}%
\bibitem [{\citenamefont {Costa~Filho}()}]{costaquantum}%
  \BibitemOpen
  \bibfield  {author} {\bibinfo {author} {\bibfnamefont {J.~I.~d.}\
  \bibnamefont {Costa~Filho}},\ }\emph {\bibinfo {title} {Quantum
  non-Markovianity induced by classical stochastic noise}},\ \href@noop {}
  {Ph.D. thesis},\ \bibinfo  {school} {Universidade de S{\~a}o
  Paulo}\BibitemShut {NoStop}%
\bibitem [{\citenamefont {Costa-Filho}\ \emph {et~al.}(2017)\citenamefont
  {Costa-Filho}, \citenamefont {Lima}, \citenamefont {Paiva}, \citenamefont
  {Soares}, \citenamefont {Morgado}, \citenamefont {Franco},\ and\
  \citenamefont {Soares-Pinto}}]{PhysRevA.95.052126}%
  \BibitemOpen
  \bibfield  {author} {\bibinfo {author} {\bibfnamefont {J.~I.}\ \bibnamefont
  {Costa-Filho}}, \bibinfo {author} {\bibfnamefont {R.~B.~B.}\ \bibnamefont
  {Lima}}, \bibinfo {author} {\bibfnamefont {R.~R.}\ \bibnamefont {Paiva}},
  \bibinfo {author} {\bibfnamefont {P.~M.}\ \bibnamefont {Soares}}, \bibinfo
  {author} {\bibfnamefont {W.~A.~M.}\ \bibnamefont {Morgado}}, \bibinfo
  {author} {\bibfnamefont {R.~L.}\ \bibnamefont {Franco}},\ and\ \bibinfo
  {author} {\bibfnamefont {D.~O.}\ \bibnamefont {Soares-Pinto}},\ }\bibfield
  {title} {\bibinfo {title} {Enabling quantum non-markovian dynamics by
  injection of classical colored noise},\ }\href
  {https://doi.org/10.1103/PhysRevA.95.052126} {\bibfield  {journal} {\bibinfo
  {journal} {Phys. Rev. A}\ }\textbf {\bibinfo {volume} {95}},\ \bibinfo
  {pages} {052126} (\bibinfo {year} {2017})}\BibitemShut {NoStop}%
\bibitem [{\citenamefont {Chou}\ and\ \citenamefont
  {Gali}(2017)}]{chou2017nitrogen}%
  \BibitemOpen
  \bibfield  {author} {\bibinfo {author} {\bibfnamefont {J.-P.}\ \bibnamefont
  {Chou}}\ and\ \bibinfo {author} {\bibfnamefont {A.}~\bibnamefont {Gali}},\
  }\bibfield  {title} {\bibinfo {title} {Nitrogen-vacancy diamond sensor: novel
  diamond surfaces from ab initio simulations},\ }\href@noop {} {\bibfield
  {journal} {\bibinfo  {journal} {MRS Communications}\ }\textbf {\bibinfo
  {volume} {7}},\ \bibinfo {pages} {551} (\bibinfo {year} {2017})}\BibitemShut
  {NoStop}%
\bibitem [{\citenamefont {Safavi-Naini}\ \emph {et~al.}(2011)\citenamefont
  {Safavi-Naini}, \citenamefont {Rabl}, \citenamefont {Weck},\ and\
  \citenamefont {Sadeghpour}}]{PhysRevA.84.023412}%
  \BibitemOpen
  \bibfield  {author} {\bibinfo {author} {\bibfnamefont {A.}~\bibnamefont
  {Safavi-Naini}}, \bibinfo {author} {\bibfnamefont {P.}~\bibnamefont {Rabl}},
  \bibinfo {author} {\bibfnamefont {P.~F.}\ \bibnamefont {Weck}},\ and\
  \bibinfo {author} {\bibfnamefont {H.~R.}\ \bibnamefont {Sadeghpour}},\
  }\bibfield  {title} {\bibinfo {title} {Microscopic model of
  electric-field-noise heating in ion traps},\ }\href
  {https://doi.org/10.1103/PhysRevA.84.023412} {\bibfield  {journal} {\bibinfo
  {journal} {Phys. Rev. A}\ }\textbf {\bibinfo {volume} {84}},\ \bibinfo
  {pages} {023412} (\bibinfo {year} {2011})}\BibitemShut {NoStop}%
\bibitem [{\citenamefont {Constantin}\ \emph {et~al.}(2009)\citenamefont
  {Constantin}, \citenamefont {Yu},\ and\ \citenamefont
  {Martinis}}]{PhysRevB.79.094520}%
  \BibitemOpen
  \bibfield  {author} {\bibinfo {author} {\bibfnamefont {M.}~\bibnamefont
  {Constantin}}, \bibinfo {author} {\bibfnamefont {C.~C.}\ \bibnamefont {Yu}},\
  and\ \bibinfo {author} {\bibfnamefont {J.~M.}\ \bibnamefont {Martinis}},\
  }\bibfield  {title} {\bibinfo {title} {Saturation of two-level systems and
  charge noise in josephson junction qubits},\ }\href
  {https://doi.org/10.1103/PhysRevB.79.094520} {\bibfield  {journal} {\bibinfo
  {journal} {Phys. Rev. B}\ }\textbf {\bibinfo {volume} {79}},\ \bibinfo
  {pages} {094520} (\bibinfo {year} {2009})}\BibitemShut {NoStop}%
\bibitem [{\citenamefont {Ohresser}\ \emph {et~al.}(2005)\citenamefont
  {Ohresser}, \citenamefont {Bulou}, \citenamefont {Dhesi}, \citenamefont
  {Boeglin}, \citenamefont {Lazarovits}, \citenamefont {Gaudry}, \citenamefont
  {Chado}, \citenamefont {Faerber},\ and\ \citenamefont
  {Scheurer}}]{PhysRevLett.95.195901}%
  \BibitemOpen
  \bibfield  {author} {\bibinfo {author} {\bibfnamefont {P.}~\bibnamefont
  {Ohresser}}, \bibinfo {author} {\bibfnamefont {H.}~\bibnamefont {Bulou}},
  \bibinfo {author} {\bibfnamefont {S.~S.}\ \bibnamefont {Dhesi}}, \bibinfo
  {author} {\bibfnamefont {C.}~\bibnamefont {Boeglin}}, \bibinfo {author}
  {\bibfnamefont {B.}~\bibnamefont {Lazarovits}}, \bibinfo {author}
  {\bibfnamefont {E.}~\bibnamefont {Gaudry}}, \bibinfo {author} {\bibfnamefont
  {I.}~\bibnamefont {Chado}}, \bibinfo {author} {\bibfnamefont
  {J.}~\bibnamefont {Faerber}},\ and\ \bibinfo {author} {\bibfnamefont
  {F.}~\bibnamefont {Scheurer}},\ }\bibfield  {title} {\bibinfo {title}
  {Surface diffusion of cr adatoms on au(111) by quantum tunneling},\ }\href
  {https://doi.org/10.1103/PhysRevLett.95.195901} {\bibfield  {journal}
  {\bibinfo  {journal} {Phys. Rev. Lett.}\ }\textbf {\bibinfo {volume} {95}},\
  \bibinfo {pages} {195901} (\bibinfo {year} {2005})}\BibitemShut {NoStop}%
\bibitem [{\citenamefont {Safavi-Naini}\ \emph {et~al.}(2013)\citenamefont
  {Safavi-Naini}, \citenamefont {Kim}, \citenamefont {Weck}, \citenamefont
  {Rabl},\ and\ \citenamefont {Sadeghpour}}]{PhysRevA.87.023421}%
  \BibitemOpen
  \bibfield  {author} {\bibinfo {author} {\bibfnamefont {A.}~\bibnamefont
  {Safavi-Naini}}, \bibinfo {author} {\bibfnamefont {E.}~\bibnamefont {Kim}},
  \bibinfo {author} {\bibfnamefont {P.~F.}\ \bibnamefont {Weck}}, \bibinfo
  {author} {\bibfnamefont {P.}~\bibnamefont {Rabl}},\ and\ \bibinfo {author}
  {\bibfnamefont {H.~R.}\ \bibnamefont {Sadeghpour}},\ }\bibfield  {title}
  {\bibinfo {title} {Influence of monolayer contamination on
  electric-field-noise heating in ion traps},\ }\href
  {https://doi.org/10.1103/PhysRevA.87.023421} {\bibfield  {journal} {\bibinfo
  {journal} {Phys. Rev. A}\ }\textbf {\bibinfo {volume} {87}},\ \bibinfo
  {pages} {023421} (\bibinfo {year} {2013})}\BibitemShut {NoStop}%
\bibitem [{\citenamefont {Dutta}\ and\ \citenamefont
  {Horn}(1981)}]{RevModPhys.53.497}%
  \BibitemOpen
  \bibfield  {author} {\bibinfo {author} {\bibfnamefont {P.}~\bibnamefont
  {Dutta}}\ and\ \bibinfo {author} {\bibfnamefont {P.~M.}\ \bibnamefont
  {Horn}},\ }\bibfield  {title} {\bibinfo {title} {Low-frequency fluctuations
  in solids: $\frac{1}{f}$ noise},\ }\href
  {https://doi.org/10.1103/RevModPhys.53.497} {\bibfield  {journal} {\bibinfo
  {journal} {Rev. Mod. Phys.}\ }\textbf {\bibinfo {volume} {53}},\ \bibinfo
  {pages} {497} (\bibinfo {year} {1981})}\BibitemShut {NoStop}%
\bibitem [{\citenamefont {Kogan}(2008)}]{kogan2008electronic}%
  \BibitemOpen
  \bibfield  {author} {\bibinfo {author} {\bibfnamefont {S.}~\bibnamefont
  {Kogan}},\ }\href@noop {} {\emph {\bibinfo {title} {Electronic noise and
  fluctuations in solids}}}\ (\bibinfo  {publisher} {Cambridge University
  Press},\ \bibinfo {year} {2008})\BibitemShut {NoStop}%
\bibitem [{\citenamefont {Paladino}\ \emph {et~al.}(2014)\citenamefont
  {Paladino}, \citenamefont {Galperin}, \citenamefont {Falci},\ and\
  \citenamefont {Altshuler}}]{RevModPhys.86.361}%
  \BibitemOpen
  \bibfield  {author} {\bibinfo {author} {\bibfnamefont {E.}~\bibnamefont
  {Paladino}}, \bibinfo {author} {\bibfnamefont {Y.~M.}\ \bibnamefont
  {Galperin}}, \bibinfo {author} {\bibfnamefont {G.}~\bibnamefont {Falci}},\
  and\ \bibinfo {author} {\bibfnamefont {B.~L.}\ \bibnamefont {Altshuler}},\
  }\bibfield  {title} {\bibinfo {title} {$\mathbf{1}/\mathbf{f}$ noise:
  Implications for solid-state quantum information},\ }\href
  {https://doi.org/10.1103/RevModPhys.86.361} {\bibfield  {journal} {\bibinfo
  {journal} {Rev. Mod. Phys.}\ }\textbf {\bibinfo {volume} {86}},\ \bibinfo
  {pages} {361} (\bibinfo {year} {2014})}\BibitemShut {NoStop}%
\bibitem [{\citenamefont {Hachiya}\ \emph {et~al.}(2014)\citenamefont
  {Hachiya}, \citenamefont {Burkard},\ and\ \citenamefont
  {Egues}}]{PhysRevB.89.115427}%
  \BibitemOpen
  \bibfield  {author} {\bibinfo {author} {\bibfnamefont {M.~O.}\ \bibnamefont
  {Hachiya}}, \bibinfo {author} {\bibfnamefont {G.}~\bibnamefont {Burkard}},\
  and\ \bibinfo {author} {\bibfnamefont {J.~C.}\ \bibnamefont {Egues}},\
  }\bibfield  {title} {\bibinfo {title} {Nonmonotonic spin relaxation and
  decoherence in graphene quantum dots with spin-orbit interactions},\ }\href
  {https://doi.org/10.1103/PhysRevB.89.115427} {\bibfield  {journal} {\bibinfo
  {journal} {Phys. Rev. B}\ }\textbf {\bibinfo {volume} {89}},\ \bibinfo
  {pages} {115427} (\bibinfo {year} {2014})}\BibitemShut {NoStop}%
\bibitem [{\citenamefont {Johnson}(1928)}]{PhysRev.32.97}%
  \BibitemOpen
  \bibfield  {author} {\bibinfo {author} {\bibfnamefont {J.~B.}\ \bibnamefont
  {Johnson}},\ }\bibfield  {title} {\bibinfo {title} {Thermal agitation of
  electricity in conductors},\ }\href {https://doi.org/10.1103/PhysRev.32.97}
  {\bibfield  {journal} {\bibinfo  {journal} {Phys. Rev.}\ }\textbf {\bibinfo
  {volume} {32}},\ \bibinfo {pages} {97} (\bibinfo {year} {1928})}\BibitemShut
  {NoStop}%
\bibitem [{\citenamefont {Nyquist}(1928)}]{PhysRev.32.110}%
  \BibitemOpen
  \bibfield  {author} {\bibinfo {author} {\bibfnamefont {H.}~\bibnamefont
  {Nyquist}},\ }\bibfield  {title} {\bibinfo {title} {Thermal agitation of
  electric charge in conductors},\ }\href
  {https://doi.org/10.1103/PhysRev.32.110} {\bibfield  {journal} {\bibinfo
  {journal} {Phys. Rev.}\ }\textbf {\bibinfo {volume} {32}},\ \bibinfo {pages}
  {110} (\bibinfo {year} {1928})}\BibitemShut {NoStop}%
\bibitem [{\citenamefont {Ariyaratne}\ \emph {et~al.}(2018)\citenamefont
  {Ariyaratne}, \citenamefont {Bluvstein}, \citenamefont {Myers},\ and\
  \citenamefont {Jayich}}]{ariyaratne2018nanoscale}%
  \BibitemOpen
  \bibfield  {author} {\bibinfo {author} {\bibfnamefont {A.}~\bibnamefont
  {Ariyaratne}}, \bibinfo {author} {\bibfnamefont {D.}~\bibnamefont
  {Bluvstein}}, \bibinfo {author} {\bibfnamefont {B.~A.}\ \bibnamefont
  {Myers}},\ and\ \bibinfo {author} {\bibfnamefont {A.~C.~B.}\ \bibnamefont
  {Jayich}},\ }\bibfield  {title} {\bibinfo {title} {Nanoscale electrical
  conductivity imaging using a nitrogen-vacancy center in diamond},\ }\bibfield
   {journal} {\bibinfo  {journal} {Nature Communications}\ }\textbf {\bibinfo
  {volume} {9}},\ \href {https://doi.org/10.1038/s41467-018-04798-1}
  {10.1038/s41467-018-04798-1} (\bibinfo {year} {2018})\BibitemShut {NoStop}%
\bibitem [{\citenamefont {Son}\ \emph {et~al.}(2020)\citenamefont {Son},
  \citenamefont {Anderson}, \citenamefont {Bourassa}, \citenamefont {Miao},
  \citenamefont {Babin}, \citenamefont {Widmann}, \citenamefont {Niethammer},
  \citenamefont {Ul~Hassan}, \citenamefont {Morioka}, \citenamefont {Ivanov},
  \citenamefont {Kaiser}, \citenamefont {Wrachtrup},\ and\ \citenamefont
  {Awschalom}}]{Son2020}%
  \BibitemOpen
  \bibfield  {author} {\bibinfo {author} {\bibfnamefont {N.~T.}\ \bibnamefont
  {Son}}, \bibinfo {author} {\bibfnamefont {C.~P.}\ \bibnamefont {Anderson}},
  \bibinfo {author} {\bibfnamefont {A.}~\bibnamefont {Bourassa}}, \bibinfo
  {author} {\bibfnamefont {K.~C.}\ \bibnamefont {Miao}}, \bibinfo {author}
  {\bibfnamefont {C.}~\bibnamefont {Babin}}, \bibinfo {author} {\bibfnamefont
  {M.}~\bibnamefont {Widmann}}, \bibinfo {author} {\bibfnamefont
  {M.}~\bibnamefont {Niethammer}}, \bibinfo {author} {\bibfnamefont
  {J.}~\bibnamefont {Ul~Hassan}}, \bibinfo {author} {\bibfnamefont
  {N.}~\bibnamefont {Morioka}}, \bibinfo {author} {\bibfnamefont {I.~G.}\
  \bibnamefont {Ivanov}}, \bibinfo {author} {\bibfnamefont {F.}~\bibnamefont
  {Kaiser}}, \bibinfo {author} {\bibfnamefont {J.}~\bibnamefont {Wrachtrup}},\
  and\ \bibinfo {author} {\bibfnamefont {D.~D.}\ \bibnamefont {Awschalom}},\
  }\bibfield  {title} {\bibinfo {title} {Developing silicon carbide for quantum
  spintronics},\ }\href {https://doi.org/10.1063/5.0004454} {\bibfield
  {journal} {\bibinfo  {journal} {Applied Physics Letters}\ }\textbf {\bibinfo
  {volume} {116}},\ \bibinfo {pages} {190501} (\bibinfo {year}
  {2020})}\BibitemShut {NoStop}%
\bibitem [{\citenamefont {Stoneham}(2001)}]{Stoneham2001}%
  \BibitemOpen
  \bibfield  {author} {\bibinfo {author} {\bibfnamefont {A.~M.}\ \bibnamefont
  {Stoneham}},\ }\href@noop {} {\emph {\bibinfo {title} {Theory of Defects in
  Solids}}}\ (\bibinfo  {publisher} {Oxford University Press},\ \bibinfo
  {address} {Oxford, UK},\ \bibinfo {year} {2001})\BibitemShut {NoStop}%
\bibitem [{\citenamefont {Crook}\ \emph {et~al.}(2020)\citenamefont {Crook},
  \citenamefont {Anderson}, \citenamefont {Miao}, \citenamefont {Bourassa},
  \citenamefont {Lee}, \citenamefont {Bayliss}, \citenamefont {Bracher},
  \citenamefont {Zhang}, \citenamefont {Abe}, \citenamefont {Ohshima},
  \citenamefont {Hu},\ and\ \citenamefont {Awschalom}}]{Crook2020}%
  \BibitemOpen
  \bibfield  {author} {\bibinfo {author} {\bibfnamefont {A.~L.}\ \bibnamefont
  {Crook}}, \bibinfo {author} {\bibfnamefont {C.~P.}\ \bibnamefont {Anderson}},
  \bibinfo {author} {\bibfnamefont {K.~C.}\ \bibnamefont {Miao}}, \bibinfo
  {author} {\bibfnamefont {A.}~\bibnamefont {Bourassa}}, \bibinfo {author}
  {\bibfnamefont {H.}~\bibnamefont {Lee}}, \bibinfo {author} {\bibfnamefont
  {S.~L.}\ \bibnamefont {Bayliss}}, \bibinfo {author} {\bibfnamefont {D.~O.}\
  \bibnamefont {Bracher}}, \bibinfo {author} {\bibfnamefont {X.}~\bibnamefont
  {Zhang}}, \bibinfo {author} {\bibfnamefont {H.}~\bibnamefont {Abe}}, \bibinfo
  {author} {\bibfnamefont {T.}~\bibnamefont {Ohshima}}, \bibinfo {author}
  {\bibfnamefont {E.~L.}\ \bibnamefont {Hu}},\ and\ \bibinfo {author}
  {\bibfnamefont {D.~D.}\ \bibnamefont {Awschalom}},\ }\bibfield  {title}
  {\bibinfo {title} {Purcell enhancement of a single silicon carbide color
  center with coherent spin control},\ }\href@noop {} {\bibfield  {journal}
  {\bibinfo  {journal} {Nano Letters}\ }\textbf {\bibinfo {volume} {20}},\
  \bibinfo {pages} {3427} (\bibinfo {year} {2020})}\BibitemShut {NoStop}%
\bibitem [{\citenamefont {Gisin}\ \emph {et~al.}(2002)\citenamefont {Gisin},
  \citenamefont {Ribordy}, \citenamefont {Tittel},\ and\ \citenamefont
  {Zbinden}}]{Gisin2002}%
  \BibitemOpen
  \bibfield  {author} {\bibinfo {author} {\bibfnamefont {N.}~\bibnamefont
  {Gisin}}, \bibinfo {author} {\bibfnamefont {G.}~\bibnamefont {Ribordy}},
  \bibinfo {author} {\bibfnamefont {W.}~\bibnamefont {Tittel}},\ and\ \bibinfo
  {author} {\bibfnamefont {H.}~\bibnamefont {Zbinden}},\ }\bibfield  {title}
  {\bibinfo {title} {Quantum cryptography},\ }\href@noop {} {\bibfield
  {journal} {\bibinfo  {journal} {Rev. Mod. Phys.}\ }\textbf {\bibinfo {volume}
  {74}},\ \bibinfo {pages} {145} (\bibinfo {year} {2002})}\BibitemShut
  {NoStop}%
\bibitem [{\citenamefont {Wehner}\ \emph {et~al.}(2018)\citenamefont {Wehner},
  \citenamefont {Elkouss},\ and\ \citenamefont {Hanson}}]{Wehnereaam9288}%
  \BibitemOpen
  \bibfield  {author} {\bibinfo {author} {\bibfnamefont {S.}~\bibnamefont
  {Wehner}}, \bibinfo {author} {\bibfnamefont {D.}~\bibnamefont {Elkouss}},\
  and\ \bibinfo {author} {\bibfnamefont {R.}~\bibnamefont {Hanson}},\
  }\bibfield  {title} {\bibinfo {title} {Quantum internet: A vision for the
  road ahead},\ }\href@noop {} {\bibfield  {journal} {\bibinfo  {journal}
  {Science}\ }\textbf {\bibinfo {volume} {362}} (\bibinfo {year}
  {2018})}\BibitemShut {NoStop}%
\bibitem [{\citenamefont {Awschalom}\ \emph {et~al.}(2018)\citenamefont
  {Awschalom}, \citenamefont {Hanson}, \citenamefont {Wrachtrup},\ and\
  \citenamefont {Zhou}}]{awschalom2018quantum}%
  \BibitemOpen
  \bibfield  {author} {\bibinfo {author} {\bibfnamefont {D.~D.}\ \bibnamefont
  {Awschalom}}, \bibinfo {author} {\bibfnamefont {R.}~\bibnamefont {Hanson}},
  \bibinfo {author} {\bibfnamefont {J.}~\bibnamefont {Wrachtrup}},\ and\
  \bibinfo {author} {\bibfnamefont {B.~B.}\ \bibnamefont {Zhou}},\ }\bibfield
  {title} {\bibinfo {title} {Quantum technologies with optically interfaced
  solid-state spins},\ }\href@noop {} {\bibfield  {journal} {\bibinfo
  {journal} {Nature Photonics}\ }\textbf {\bibinfo {volume} {12}},\ \bibinfo
  {pages} {516} (\bibinfo {year} {2018})}\BibitemShut {NoStop}%
\bibitem [{\citenamefont {Hensen}\ \emph {et~al.}(2015)\citenamefont {Hensen},
  \citenamefont {Bernien}, \citenamefont {Dr{\'e}au}, \citenamefont {Reiserer},
  \citenamefont {Kalb}, \citenamefont {Blok}, \citenamefont {Ruitenberg},
  \citenamefont {Vermeulen}, \citenamefont {Schouten}, \citenamefont
  {Abell{\'a}n}, \citenamefont {Amaya}, \citenamefont {Pruneri}, \citenamefont
  {Mitchell}, \citenamefont {Markham}, \citenamefont {Twitchen}, \citenamefont
  {Elkouss}, \citenamefont {Wehner}, \citenamefont {Taminiau},\ and\
  \citenamefont {Hanson}}]{hensen2015loophole}%
  \BibitemOpen
  \bibfield  {author} {\bibinfo {author} {\bibfnamefont {B.}~\bibnamefont
  {Hensen}}, \bibinfo {author} {\bibfnamefont {H.}~\bibnamefont {Bernien}},
  \bibinfo {author} {\bibfnamefont {A.~E.}\ \bibnamefont {Dr{\'e}au}}, \bibinfo
  {author} {\bibfnamefont {A.}~\bibnamefont {Reiserer}}, \bibinfo {author}
  {\bibfnamefont {N.}~\bibnamefont {Kalb}}, \bibinfo {author} {\bibfnamefont
  {M.~S.}\ \bibnamefont {Blok}}, \bibinfo {author} {\bibfnamefont
  {J.}~\bibnamefont {Ruitenberg}}, \bibinfo {author} {\bibfnamefont {R.~F.~L.}\
  \bibnamefont {Vermeulen}}, \bibinfo {author} {\bibfnamefont {R.~N.}\
  \bibnamefont {Schouten}}, \bibinfo {author} {\bibfnamefont {C.}~\bibnamefont
  {Abell{\'a}n}}, \bibinfo {author} {\bibfnamefont {W.}~\bibnamefont {Amaya}},
  \bibinfo {author} {\bibfnamefont {V.}~\bibnamefont {Pruneri}}, \bibinfo
  {author} {\bibfnamefont {M.~W.}\ \bibnamefont {Mitchell}}, \bibinfo {author}
  {\bibfnamefont {M.}~\bibnamefont {Markham}}, \bibinfo {author} {\bibfnamefont
  {D.~J.}\ \bibnamefont {Twitchen}}, \bibinfo {author} {\bibfnamefont
  {D.}~\bibnamefont {Elkouss}}, \bibinfo {author} {\bibfnamefont
  {S.}~\bibnamefont {Wehner}}, \bibinfo {author} {\bibfnamefont {T.~H.}\
  \bibnamefont {Taminiau}},\ and\ \bibinfo {author} {\bibfnamefont
  {R.}~\bibnamefont {Hanson}},\ }\bibfield  {title} {\bibinfo {title}
  {Loophole-free bell inequality violation using electron spins separated by
  1.3 kilometres},\ }\href@noop {} {\bibfield  {journal} {\bibinfo  {journal}
  {Nature}\ }\textbf {\bibinfo {volume} {526}},\ \bibinfo {pages} {682}
  (\bibinfo {year} {2015})}\BibitemShut {NoStop}%
\bibitem [{\citenamefont {Fuchs}\ \emph {et~al.}(2011)\citenamefont {Fuchs},
  \citenamefont {Burkard}, \citenamefont {Klimov},\ and\ \citenamefont
  {Awschalom}}]{fuchs2011quantum}%
  \BibitemOpen
  \bibfield  {author} {\bibinfo {author} {\bibfnamefont {G.}~\bibnamefont
  {Fuchs}}, \bibinfo {author} {\bibfnamefont {G.}~\bibnamefont {Burkard}},
  \bibinfo {author} {\bibfnamefont {P.}~\bibnamefont {Klimov}},\ and\ \bibinfo
  {author} {\bibfnamefont {D.}~\bibnamefont {Awschalom}},\ }\bibfield  {title}
  {\bibinfo {title} {A quantum memory intrinsic to single nitrogen--vacancy
  centres in diamond},\ }\href@noop {} {\bibfield  {journal} {\bibinfo
  {journal} {Nature Physics}\ }\textbf {\bibinfo {volume} {7}},\ \bibinfo
  {pages} {789} (\bibinfo {year} {2011})}\BibitemShut {NoStop}%
\bibitem [{\citenamefont {Degen}\ \emph
  {et~al.}(2017{\natexlab{b}})\citenamefont {Degen}, \citenamefont {Reinhard},\
  and\ \citenamefont {Cappellaro}}]{RevModPhys.89.035002}%
  \BibitemOpen
  \bibfield  {author} {\bibinfo {author} {\bibfnamefont {C.~L.}\ \bibnamefont
  {Degen}}, \bibinfo {author} {\bibfnamefont {F.}~\bibnamefont {Reinhard}},\
  and\ \bibinfo {author} {\bibfnamefont {P.}~\bibnamefont {Cappellaro}},\
  }\bibfield  {title} {\bibinfo {title} {Quantum sensing},\ }\href@noop {}
  {\bibfield  {journal} {\bibinfo  {journal} {Rev. Mod. Phys.}\ }\textbf
  {\bibinfo {volume} {89}},\ \bibinfo {pages} {035002} (\bibinfo {year}
  {2017}{\natexlab{b}})}\BibitemShut {NoStop}%
\bibitem [{\citenamefont {Lowther}(1977)}]{lowther1977vacancies}%
  \BibitemOpen
  \bibfield  {author} {\bibinfo {author} {\bibfnamefont {J.}~\bibnamefont
  {Lowther}},\ }\bibfield  {title} {\bibinfo {title} {Vacancies and divacancies
  in cubic silicon carbide},\ }\href@noop {} {\bibfield  {journal} {\bibinfo
  {journal} {Journal of Physics C: Solid State Physics}\ }\textbf {\bibinfo
  {volume} {10}},\ \bibinfo {pages} {2501} (\bibinfo {year}
  {1977})}\BibitemShut {NoStop}%
\bibitem [{\citenamefont {Miao}\ \emph
  {et~al.}(2019{\natexlab{b}})\citenamefont {Miao}, \citenamefont {Bourassa},
  \citenamefont {Anderson}, \citenamefont {Whiteley}, \citenamefont {Crook},
  \citenamefont {Bayliss}, \citenamefont {Wolfowicz}, \citenamefont {Thiering},
  \citenamefont {Udvarhelyi}, \citenamefont {Iv{\'a}dy}, \citenamefont {Abe},
  \citenamefont {Ohshima}, \citenamefont {Gali},\ and\ \citenamefont
  {Awschalom}}]{miao2019electrically2}%
  \BibitemOpen
  \bibfield  {author} {\bibinfo {author} {\bibfnamefont {K.~C.}\ \bibnamefont
  {Miao}}, \bibinfo {author} {\bibfnamefont {A.}~\bibnamefont {Bourassa}},
  \bibinfo {author} {\bibfnamefont {C.~P.}\ \bibnamefont {Anderson}}, \bibinfo
  {author} {\bibfnamefont {S.~J.}\ \bibnamefont {Whiteley}}, \bibinfo {author}
  {\bibfnamefont {A.~L.}\ \bibnamefont {Crook}}, \bibinfo {author}
  {\bibfnamefont {S.~L.}\ \bibnamefont {Bayliss}}, \bibinfo {author}
  {\bibfnamefont {G.}~\bibnamefont {Wolfowicz}}, \bibinfo {author}
  {\bibfnamefont {G.}~\bibnamefont {Thiering}}, \bibinfo {author}
  {\bibfnamefont {P.}~\bibnamefont {Udvarhelyi}}, \bibinfo {author}
  {\bibfnamefont {V.}~\bibnamefont {Iv{\'a}dy}}, \bibinfo {author}
  {\bibfnamefont {H.}~\bibnamefont {Abe}}, \bibinfo {author} {\bibfnamefont
  {T.}~\bibnamefont {Ohshima}}, \bibinfo {author} {\bibfnamefont
  {{\'A}.}~\bibnamefont {Gali}},\ and\ \bibinfo {author} {\bibfnamefont
  {D.~D.}\ \bibnamefont {Awschalom}},\ }\bibfield  {title} {\bibinfo {title}
  {Electrically driven optical interferometry with spins in silicon carbide},\
  }\href@noop {} {\bibfield  {journal} {\bibinfo  {journal} {Science Advances}\
  }\textbf {\bibinfo {volume} {5}} (\bibinfo {year}
  {2019}{\natexlab{b}})}\BibitemShut {NoStop}%
\bibitem [{\citenamefont {Anderson$~$}\ and\ \citenamefont
  {Weiss}(1953)}]{anderson1953exchange}%
  \BibitemOpen
  \bibfield  {author} {\bibinfo {author} {\bibfnamefont {P.~W.}\ \bibnamefont
  {Anderson$~$}}\ and\ \bibinfo {author} {\bibfnamefont {P.~R.}\ \bibnamefont
  {Weiss}},\ }\bibfield  {title} {\bibinfo {title} {Exchange narrowing in
  paramagnetic resonance},\ }\href {https://doi.org/10.1103/RevModPhys.25.269}
  {\bibfield  {journal} {\bibinfo  {journal} {Rev. Mod. Phys.}\ }\textbf
  {\bibinfo {volume} {25}},\ \bibinfo {pages} {269} (\bibinfo {year}
  {1953})}\BibitemShut {NoStop}%
\bibitem [{\citenamefont {Anderson}(2020)}]{chris-thesis}%
  \BibitemOpen
  \bibfield  {author} {\bibinfo {author} {\bibfnamefont {C.~P.}\ \bibnamefont
  {Anderson}},\ }\href@noop {} {\bibinfo {type} {{PhD thesis}, {DOI}
  10.6082/uchicago.2263}},\ \bibinfo  {school} {The University of Chicago}
  (\bibinfo {year} {2020})\BibitemShut {NoStop}%
\bibitem [{\citenamefont {Chandrasekhar}(1943)}]{RevModPhys.15.1}%
  \BibitemOpen
  \bibfield  {author} {\bibinfo {author} {\bibfnamefont {S.}~\bibnamefont
  {Chandrasekhar}},\ }\bibfield  {title} {\bibinfo {title} {Stochastic problems
  in physics and astronomy},\ }\href {https://doi.org/10.1103/RevModPhys.15.1}
  {\bibfield  {journal} {\bibinfo  {journal} {Rev. Mod. Phys.}\ }\textbf
  {\bibinfo {volume} {15}},\ \bibinfo {pages} {1} (\bibinfo {year}
  {1943})}\BibitemShut {NoStop}%
\bibitem [{\citenamefont {Gesley}\ and\ \citenamefont
  {Swanson}(1985)}]{PhysRevB.32.7703}%
  \BibitemOpen
  \bibfield  {author} {\bibinfo {author} {\bibfnamefont {M.~A.}\ \bibnamefont
  {Gesley}}\ and\ \bibinfo {author} {\bibfnamefont {L.~W.}\ \bibnamefont
  {Swanson}},\ }\bibfield  {title} {\bibinfo {title} {Spectral analysis of
  adsorbate induced field-emission flicker noise},\ }\href
  {https://doi.org/10.1103/PhysRevB.32.7703} {\bibfield  {journal} {\bibinfo
  {journal} {Phys. Rev. B}\ }\textbf {\bibinfo {volume} {32}},\ \bibinfo
  {pages} {7703} (\bibinfo {year} {1985})}\BibitemShut {NoStop}%
\bibitem [{\citenamefont {Dubessy}\ \emph {et~al.}(2009)\citenamefont
  {Dubessy}, \citenamefont {Coudreau},\ and\ \citenamefont
  {Guidoni}}]{PhysRevA.80.031402}%
  \BibitemOpen
  \bibfield  {author} {\bibinfo {author} {\bibfnamefont {R.}~\bibnamefont
  {Dubessy}}, \bibinfo {author} {\bibfnamefont {T.}~\bibnamefont {Coudreau}},\
  and\ \bibinfo {author} {\bibfnamefont {L.}~\bibnamefont {Guidoni}},\
  }\bibfield  {title} {\bibinfo {title} {Electric field noise above surfaces: A
  model for heating-rate scaling law in ion traps},\ }\href
  {https://doi.org/10.1103/PhysRevA.80.031402} {\bibfield  {journal} {\bibinfo
  {journal} {Phys. Rev. A}\ }\textbf {\bibinfo {volume} {80}},\ \bibinfo
  {pages} {031402} (\bibinfo {year} {2009})}\BibitemShut {NoStop}%
\bibitem [{\citenamefont {Low}\ \emph {et~al.}(2011)\citenamefont {Low},
  \citenamefont {Herskind},\ and\ \citenamefont {Chuang}}]{PhysRevA.84.053425}%
  \BibitemOpen
  \bibfield  {author} {\bibinfo {author} {\bibfnamefont {G.~H.}\ \bibnamefont
  {Low}}, \bibinfo {author} {\bibfnamefont {P.~F.}\ \bibnamefont {Herskind}},\
  and\ \bibinfo {author} {\bibfnamefont {I.~L.}\ \bibnamefont {Chuang}},\
  }\bibfield  {title} {\bibinfo {title} {Finite-geometry models of electric
  field noise from patch potentials in ion traps},\ }\href
  {https://doi.org/10.1103/PhysRevA.84.053425} {\bibfield  {journal} {\bibinfo
  {journal} {Phys. Rev. A}\ }\textbf {\bibinfo {volume} {84}},\ \bibinfo
  {pages} {053425} (\bibinfo {year} {2011})}\BibitemShut {NoStop}%
\bibitem [{\citenamefont {Lai}\ \emph {et~al.}(2018)\citenamefont {Lai},
  \citenamefont {Lin}, \citenamefont {Twamley},\ and\ \citenamefont
  {Goan}}]{PhysRevA.97.052303}%
  \BibitemOpen
  \bibfield  {author} {\bibinfo {author} {\bibfnamefont {Y.-Y.}\ \bibnamefont
  {Lai}}, \bibinfo {author} {\bibfnamefont {G.-D.}\ \bibnamefont {Lin}},
  \bibinfo {author} {\bibfnamefont {J.}~\bibnamefont {Twamley}},\ and\ \bibinfo
  {author} {\bibfnamefont {H.-S.}\ \bibnamefont {Goan}},\ }\bibfield  {title}
  {\bibinfo {title} {Single-nitrogen-vacancy-center quantum memory for a
  superconducting flux qubit mediated by a ferromagnet},\ }\href@noop {}
  {\bibfield  {journal} {\bibinfo  {journal} {Phys. Rev. A}\ }\textbf {\bibinfo
  {volume} {97}},\ \bibinfo {pages} {052303} (\bibinfo {year}
  {2018})}\BibitemShut {NoStop}%
\bibitem [{\citenamefont {M\"uhlherr}\ \emph
  {et~al.}(2019{\natexlab{b}})\citenamefont {M\"uhlherr}, \citenamefont
  {Shkolnikov},\ and\ \citenamefont {Burkard}}]{PhysRevB.99.195413}%
  \BibitemOpen
  \bibfield  {author} {\bibinfo {author} {\bibfnamefont {C.}~\bibnamefont
  {M\"uhlherr}}, \bibinfo {author} {\bibfnamefont {V.~O.}\ \bibnamefont
  {Shkolnikov}},\ and\ \bibinfo {author} {\bibfnamefont {G.}~\bibnamefont
  {Burkard}},\ }\bibfield  {title} {\bibinfo {title} {Magnetic resonance in
  defect spins mediated by spin waves},\ }\href@noop {} {\bibfield  {journal}
  {\bibinfo  {journal} {Phys. Rev. B}\ }\textbf {\bibinfo {volume} {99}},\
  \bibinfo {pages} {195413} (\bibinfo {year} {2019}{\natexlab{b}})}\BibitemShut
  {NoStop}%
\bibitem [{\citenamefont {Gonzalez-Ballestero}\ \emph
  {et~al.}(2020)\citenamefont {Gonzalez-Ballestero}, \citenamefont {van~der
  Sar},\ and\ \citenamefont {Romero-Isart}}]{ballestero}%
  \BibitemOpen
  \bibfield  {author} {\bibinfo {author} {\bibfnamefont {C.}~\bibnamefont
  {Gonzalez-Ballestero}}, \bibinfo {author} {\bibfnamefont {T.}~\bibnamefont
  {van~der Sar}},\ and\ \bibinfo {author} {\bibfnamefont {O.}~\bibnamefont
  {Romero-Isart}},\ }\bibfield  {title} {\bibinfo {title} {Towards a quantum
  interface between spin waves and paramagnetic spin baths},\ }\href@noop {}
  {\bibfield  {journal} {\bibinfo  {journal} {arXiv:2012.00540}\ } (\bibinfo
  {year} {2020})}\BibitemShut {NoStop}%
\bibitem [{\citenamefont {Bourassa}\ \emph {et~al.}(2020)\citenamefont
  {Bourassa}, \citenamefont {Anderson}, \citenamefont {Miao}, \citenamefont
  {Onizhuk}, \citenamefont {Ma}, \citenamefont {Crook}, \citenamefont {Abe},
  \citenamefont {Ul-Hassan}, \citenamefont {Ohshima}, \citenamefont {Son},
  \citenamefont {Galli},\ and\ \citenamefont
  {Awschalom}}]{bourassa2020entanglement}%
  \BibitemOpen
  \bibfield  {author} {\bibinfo {author} {\bibfnamefont {A.}~\bibnamefont
  {Bourassa}}, \bibinfo {author} {\bibfnamefont {C.~P.}\ \bibnamefont
  {Anderson}}, \bibinfo {author} {\bibfnamefont {K.~C.}\ \bibnamefont {Miao}},
  \bibinfo {author} {\bibfnamefont {M.}~\bibnamefont {Onizhuk}}, \bibinfo
  {author} {\bibfnamefont {H.}~\bibnamefont {Ma}}, \bibinfo {author}
  {\bibfnamefont {A.~L.}\ \bibnamefont {Crook}}, \bibinfo {author}
  {\bibfnamefont {H.}~\bibnamefont {Abe}}, \bibinfo {author} {\bibfnamefont
  {J.}~\bibnamefont {Ul-Hassan}}, \bibinfo {author} {\bibfnamefont
  {T.}~\bibnamefont {Ohshima}}, \bibinfo {author} {\bibfnamefont {N.~T.}\
  \bibnamefont {Son}}, \bibinfo {author} {\bibfnamefont {G.}~\bibnamefont
  {Galli}},\ and\ \bibinfo {author} {\bibfnamefont {D.~D.}\ \bibnamefont
  {Awschalom}},\ }\bibfield  {title} {\bibinfo {title} {Entanglement and
  control of single nuclear spins in isotopically engineered silicon carbide},\
  }\href@noop {} {\bibfield  {journal} {\bibinfo  {journal} {Nature Materials}\
  }\textbf {\bibinfo {volume} {19}},\ \bibinfo {pages} {1319} (\bibinfo {year}
  {2020})}\BibitemShut {NoStop}%
\bibitem [{\citenamefont {Iwata}\ \emph {et~al.}(2000)\citenamefont {Iwata},
  \citenamefont {Itoh},\ and\ \citenamefont {Pensl}}]{mobility-4HSiC}%
  \BibitemOpen
  \bibfield  {author} {\bibinfo {author} {\bibfnamefont {H.}~\bibnamefont
  {Iwata}}, \bibinfo {author} {\bibfnamefont {K.~M.}\ \bibnamefont {Itoh}},\
  and\ \bibinfo {author} {\bibfnamefont {G.}~\bibnamefont {Pensl}},\ }\bibfield
   {title} {\bibinfo {title} {Theory of the anisotropy of the electron hall
  mobility in $n$--type 4$\rm{H}$-- and 6$\rm{H}$--$\rm{SiC}$},\ }\href@noop {}
  {\bibfield  {journal} {\bibinfo  {journal} {Journal of Applied Physics}\
  }\textbf {\bibinfo {volume} {88}},\ \bibinfo {pages} {1956} (\bibinfo {year}
  {2000})}\BibitemShut {NoStop}%
\bibitem [{\citenamefont {Falk}\ \emph {et~al.}(2014)\citenamefont {Falk},
  \citenamefont {Klimov}, \citenamefont {Buckley}, \citenamefont {Iv\'ady},
  \citenamefont {Abrikosov}, \citenamefont {Calusine}, \citenamefont {Koehl},
  \citenamefont {Gali},\ and\ \citenamefont {Awschalom}}]{falko2014}%
  \BibitemOpen
  \bibfield  {author} {\bibinfo {author} {\bibfnamefont {A.~L.}\ \bibnamefont
  {Falk}}, \bibinfo {author} {\bibfnamefont {P.~V.}\ \bibnamefont {Klimov}},
  \bibinfo {author} {\bibfnamefont {B.~B.}\ \bibnamefont {Buckley}}, \bibinfo
  {author} {\bibfnamefont {V.}~\bibnamefont {Iv\'ady}}, \bibinfo {author}
  {\bibfnamefont {I.~A.}\ \bibnamefont {Abrikosov}}, \bibinfo {author}
  {\bibfnamefont {G.}~\bibnamefont {Calusine}}, \bibinfo {author}
  {\bibfnamefont {W.~F.}\ \bibnamefont {Koehl}}, \bibinfo {author}
  {\bibfnamefont {A.}~\bibnamefont {Gali}},\ and\ \bibinfo {author}
  {\bibfnamefont {D.~D.}\ \bibnamefont {Awschalom}},\ }\bibfield  {title}
  {\bibinfo {title} {Optical polarization of nuclear spins in silicon
  carbide},\ }\href@noop {} {\bibfield  {journal} {\bibinfo  {journal} {Phys.
  Rev. Lett.}\ }\textbf {\bibinfo {volume} {112}},\ \bibinfo {pages} {187601}
  (\bibinfo {year} {2014})}\BibitemShut {NoStop}%
\bibitem [{\citenamefont {Robledo}\ \emph {et~al.}(2011)\citenamefont
  {Robledo}, \citenamefont {Childress}, \citenamefont {Bernien}, \citenamefont
  {Hensen}, \citenamefont {Alkemade},\ and\ \citenamefont
  {Hanson}}]{robledo2011high}%
  \BibitemOpen
  \bibfield  {author} {\bibinfo {author} {\bibfnamefont {L.}~\bibnamefont
  {Robledo}}, \bibinfo {author} {\bibfnamefont {L.}~\bibnamefont {Childress}},
  \bibinfo {author} {\bibfnamefont {H.}~\bibnamefont {Bernien}}, \bibinfo
  {author} {\bibfnamefont {B.}~\bibnamefont {Hensen}}, \bibinfo {author}
  {\bibfnamefont {P.~F.}\ \bibnamefont {Alkemade}},\ and\ \bibinfo {author}
  {\bibfnamefont {R.}~\bibnamefont {Hanson}},\ }\bibfield  {title} {\bibinfo
  {title} {High-fidelity projective read-out of a solid-state spin quantum
  register},\ }\href@noop {} {\bibfield  {journal} {\bibinfo  {journal}
  {Nature}\ }\textbf {\bibinfo {volume} {477}},\ \bibinfo {pages} {574}
  (\bibinfo {year} {2011})}\BibitemShut {NoStop}%
\bibitem [{\citenamefont {McMillan}\ \emph {et~al.}(2020)\citenamefont
  {McMillan}, \citenamefont {Harmon},\ and\ \citenamefont
  {Flatt\'e}}]{PhysRevLett.125.257203}%
  \BibitemOpen
  \bibfield  {author} {\bibinfo {author} {\bibfnamefont {S.~R.}\ \bibnamefont
  {McMillan}}, \bibinfo {author} {\bibfnamefont {N.~J.}\ \bibnamefont
  {Harmon}},\ and\ \bibinfo {author} {\bibfnamefont {M.~E.}\ \bibnamefont
  {Flatt\'e}},\ }\bibfield  {title} {\bibinfo {title} {Image of dynamic local
  exchange interactions in the dc magnetoresistance of spin-polarized current
  through a dopant},\ }\href {https://doi.org/10.1103/PhysRevLett.125.257203}
  {\bibfield  {journal} {\bibinfo  {journal} {Phys. Rev. Lett.}\ }\textbf
  {\bibinfo {volume} {125}},\ \bibinfo {pages} {257203} (\bibinfo {year}
  {2020})}\BibitemShut {NoStop}%
\end{thebibliography}%

\end{document}